\documentclass[useAMS,usenatbib]{mn2e}

\usepackage[colorlinks]{hyperref}
\usepackage{amssymb}
\usepackage{amsmath}
\usepackage{graphicx}
\usepackage[table, dvipsnames]{xcolor}
\usepackage{xspace}
\usepackage{subfig}
\usepackage{array}
\usepackage{multirow}
\DeclareUnicodeCharacter{2212}{\textendash}
\usepackage{url}
\usepackage[colorinlistoftodos]{todonotes}

\def\spi{\textit{INTEGRAL/SPI}\xspace}

\begin{document}
\title[Constraints on Decaying DM from INTEGRAL/SPI]{New Constraints on Decaying Dark Matter from INTEGRAL/SPI}
\author[Fischer S. et al.]{S. Fischer$^{1}$\thanks{E-mail: fischer@astro.uni-tuebingen.de,\newline 98fischersimon@gmail.com}, 
          D. Malyshev$^{1}$, L. Ducci$^{1,2}$, A. Santangelo$^{1}$
          \\
$^{1}$ Institut f{\"u}r Astronomie und Astrophysik T{\"u}bingen, Universit{\"a}t T{\"u}bingen, Sand 1, D-72076 T{\"u}bingen, Germany \\
$^{2}$ ISDC Data Center for Astrophysics, Universit\'e de Gen\`eve, 16 chemin d'\'Ecogia, 1290, Versoix, Switzerland \\
}

\date{Received $<$date$>$  ; in original form  $<$date$>$ }

\label{firstpage}
\pagerange{\pageref{firstpage}--\pageref{lastpage}} 
\pubyear{2022}

\maketitle

\begin{abstract}
Based on almost 20~years of data collected by the high-resolution spectrometer SPI on board the \textit{International Gamma-Ray Astrophysics Laboratory (INTEGRAL)} we present constraints on a decaying dark matter particle undergoing a decay into two bodies, at least one of which is a photon, manifesting itself via a narrow line-like spectral feature. Our ON-OFF type analysis of the Milky Way observations allowed us to constrain the lifetime to be  $\gtrsim 10^{20}-10^{21}$~yrs for DM particles with masses $40\,\text{keV}\,<\,M_{\text{DM}}\,<\,14\,\text{MeV}$. Within this mass range our analysis also reveals 32 line-like features detected at $\geq 3\sigma$ significance, 29 of which coincide with known instrumental and astrophysical lines. In particular, we report on the detection of the electron-positron annihilation (511~keV) and $^{26}$Al (1809~keV) lines with spatial profiles consistent with previous results in the literature. For the particular case of the sterile neutrino DM we report the limits on the mixing angle as a function of sterile neutrino mass. We discuss the dominant impact of systematic uncertainties connected to the strongly time-variable INTEGRAL/SPI instrumental background as well as the ones connected to the uncertainties of MW DM density profile measurements on the derived results.
\end{abstract}


\section{Introduction}
There is a substantial amount of evidence in favor of the existence of dark matter (DM) in the Universe. The most prominent evidences for DM originate from observations of rotation curves of galaxies \citep{Trimble_1987,Ashman_1992,Sofue_2001}, cosmic microwave background fluctuations \citep{Planck_2016,Dvorkin_2022} and Bullet clusters \citep{Clowe:2006eq}.
The possibility of a Standard Model (SM) particle being DM has been ruled already out for a long time~\citep[see e.g.][for recent reviews]{pdg18,boyarsky18,profumo19}. Extensions of the SM proposed several well-motivated DM-candidate particles. The suggested mass range of such particles extends from smaller than $\mu$eV (axions and axion-like particles) through keV (sterile neutrinos) to GeV-TeV (weakly interacting massive particles (WIMPs)) and even to $>10^{13}$\,GeV (WIMPZILLAs)~\citep[for a review of all candidates see e.g.][]{Feng_2017}.

In what follows we consider the general keV-MeV mass scale DM which could decay with an emission of one or more photons. A typical example of such a DM particle is the right-handed (sterile) neutrino~\citep[see][for a recent review]{boyarsky18}. Such a neutrino interacts with the rest of the matter only via a mixing with SM (active) neutrinos and can therefore survive cosmologically for a long time. The sterile neutrino could also be produced in the early Universe with the correct abundances~\citep{dodelson94, shi99}. The mass of the sterile neutrino (like any fermionic dark matter) is constrained to be higher than $M_{\text{DM}}\gtrsim 1$~keV. Such a bound arises from limits imposed by the Pauli exclusion principle. Specifically, the phase space density of the DM particles in the halos of dwarf spheroidal galaxies cannot exceed the fundamental limits imposed by the exclusion principle and the initial phase space density at the moment of production of the DM in the Early Universe~\citep{tremainegunn,tg1,gorbunov08,savchenko19}.

The (lightest) sterile neutrino particle $N$ could decay into a photon and SM neutrino $N\rightarrow\nu+\gamma$. Consequently, the decay signal is a narrow line at energy $E=M_{\text{DM}}/2$ with the strength of the signal being determined by the sterile-active neutrino mixing angle $\theta$.

High values of $\theta$ are forbidden as the abundance of sterile neutrinos produced in the Early Universe with such mixing angles would exceed the observed DM density in the present day~\citep{dodelson94,numsm1,boyarsky18}. Additional upper limits on $\theta$ originate from non-detection of the line-like feature in multiple DM-dominated objects with the current generation of instruments~\citep{boyarsky18}. Such searches in particular were performed in the keV-MeV band with XMM-Newton~\citep{2014PhRvD..90j3506M,2016MNRAS.460.1390R,Foster_2021}, NuSTAR~\citep{2016PhRvD..94l3504N,2019PhRvD..99h3005N, Roach_2022}, \spi~\citep{we_spi, laha20, calore22} and are suggested to be performed with future missions, such as Athena, eXTP and THESEUS in~\citet{we_ath,we_extp,we_ths}. In this work we update such searches with almost 20 years of \spi data.\\

\noindent This paper is organized as follows: In section \ref{sec:dm_signal} we review the expected signal from decaying DM as detected by SPI. In section \ref{ch:Instrument} we review the instrument itself, the used data set and analysis methods. In section \ref{sec:dm_lines} we present the results of the analysis and discuss the possibility to discriminate the lines with DM-admixture from purely instrumental-origin line-candidates adopting off-GC angular profile studies. 
Finally, we summarise the results and discuss them in section \ref{sec:results_discussion}.

\section{DM decay signal}
\label{sec:dm_signal}
We consider the signal from a decay of a dark matter particle with mass $M_{\text{DM}}$ over two massless particles, the typical example of which could be a decay of the sterile (right-handed) neutrino. For such a type of decay the expected photon signal is a narrow line at energy $E=M_{\text{DM}}/2$, with the flux:
\begin{align}
 &   \Phi (E, \phi) = \frac{dN}{dt dA dE} = k_x \frac{\Gamma_{\text{DM}}}{4 \pi M_{\text{DM}}} \delta (E-M_{\text{DM}}/2) S_{\text{DM}}(\phi) \nonumber \\
 & S_{\text{DM}}(\phi) = \int\limits_{l.o.s.}\int\limits_{\Omega_{\text{FoV}}} \rho_{\text{DM}}(r,l,\phi) d\ell d\Omega.
 \label{eq:dm_decay_signal_general}
\end{align}
Here $k_x$ is the number of photons emitted per decay, $S_{\text{DM}}$ is the dark matter column density (``$D$-factor'') of the DM in the field of view (FoV) of the instrument $\Omega_{\text{FoV}}$ and $\Gamma_{\text{DM}}$ is the decay width of the DM particle.   
Generally speaking, the signal discussed above is given by Eq.~\ref{eq:dm_decay_signal_general} and can be thought of as a product of the spatial/instrument-dependent part $S_{\text{DM}}$ and the spectral/particle-physics determined part $\Gamma_{\text{DM}}\delta(E-M_{\text{DM}}/2)$.

\subsection{Spatial part of the signal}

The $S_{\text{DM}}$-component of the expected signal is given by the integral of the DM density $\rho_{\text{DM}}$ over the line of sight $\ell$ and the instrument's FoV $\Omega$. For the observation of decaying DM in the Milky Way (MW) galaxy with a narrow field instrument (such that the integral $\rho_{\text{DM}}d\ell$ does not vary significantly within the FoV) and for radially-symmetric DM distributions, the DM column density is
\begin{align}
 &    S_{\text{DM}}(\phi) = \Omega_{\text{FoV}} \int\limits_0^{\infty} dz \rho_{\text{DM}}\left(\sqrt{r_{\odot}^2 -2zr_{\odot} \cos(\phi) + z^2}\right), \nonumber \\
 & \cos(\phi) = \cos(b)\cos(l),
 \label{eq:Sdm}
\end{align}
which is a function of a single angle $\phi$ -- the angle between the center of observation and the direction to the Galactic Center (GC)~\citep[see e.g.][]{we_spi}. Here, $(l,b)$ represent the galactic coordinates of the center of observation and $r_\odot\simeq 8.5$~kpc stands for the distance Sun-GC. 

\begin{figure}
    \centering
    \includegraphics[clip=true,width=1\columnwidth]{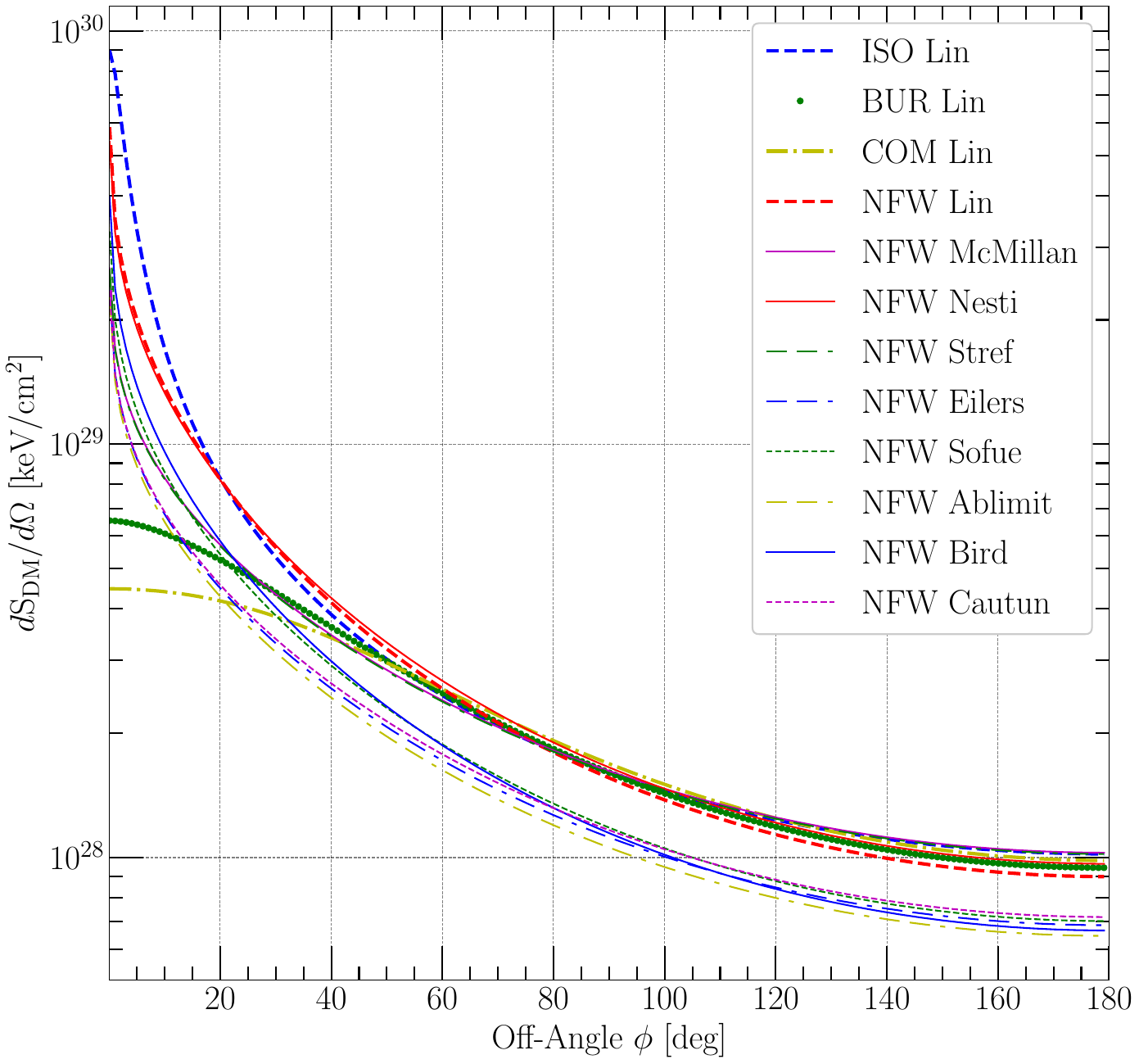}
    \centering
    \caption{Comparison of the surface brightness of DM $dS_{\text{DM}}/d\Omega$ from Eq.~\ref{eq:dm_decay_signal_general} along different off-angles for all DM density profiles. The parameters used can be found in Tab.~\ref{tab:models} and represent the best fit parameters for the DM distribution in the Milky Way by different authors. As a reference profile for later constraints we will pick the widely used parameters obtained by \citet{Cautun_2020}.}
     \label{fig:profiles}
\end{figure}

We note that the DM column density, representing the total DM mass in the FoV, is a function of the considered DM density profile $\rho_{\text{DM}}$ and is consequently a  subject to statistical and systematic uncertainties connected to the poor knowledge of the DM distribution in our Galaxy. In order to marginalize over possible sources of uncertainty, we consider several DM density profiles of the MW with recently reported parameters from the literature. Namely, we focus on the following DM density profiles:\\
the Navarro-Frank-White (NFW) profile \citep{NFW1996}
\begin{equation}
    \rho_{\text{NFW}} = \frac{\rho_0}{\frac{r}{r_0}\left(1+\frac{r}{r_0}\right)^2},
    \label{eq:NFW_profile}
\end{equation}
the Burkert profile~\citep{Burkert_1995}
\begin{equation}
    \rho_{\text{BUR}} = \frac{\rho_0 r_0^3}{(r+r_0)(r^2 + r_0^2)},
\end{equation}
the Pseudo-isothermal profile\citep{Jimenez_2003}
\begin{equation}
    \rho_{\text{ISO}} = \frac{\rho_0}{1+\left(\frac{r}{r_0}\right)^2},
\end{equation}
and the Core-modified profile \citep{Brownstein_2009}
\begin{equation}
    \rho_{\text{COM}} = \frac{\rho_0r_0^3}{r^3+r_0^3}.
\end{equation}
The parameters of the described profiles, namely the density and radius $\rho_0$ and $r_0$, the type of profile and a reference are summarised in Tab.~\ref{tab:models}. The corresponding column densities $S_{\text{DM}}(\phi)$ for the profiles parameters reported in the table are shown in Fig.~\ref{fig:profiles}.

One can see that the differences in DM column densities could reach an order of magnitude close to the Galactic Center for certain DM density profiles. In what follows we adopt the NFW DM-density profile with parameters reported in~\citet{Cautun_2020} as a reference profile. We use the rest of the considered profiles to estimate the level of systematic uncertainty connected to the insufficient knowledge of the DM distribution in our Galaxy. The corresponding NFW profile is marked in Tab.~\ref{tab:models} with a $^\dagger$ symbol.

\subsection{Spectral part of the signal}
As discussed above for the decay of a DM particle with mass $M_{\text{DM}}$ into two massless bodies, the spectral signal is expected to be a narrow line with intensity $\propto k_x\Gamma_{\text{DM}}\delta(E-M_{\text{DM}}/2)$. For the case of the sterile neutrino decay into an active neutrino and a photon $k_x=1$ the decay width $\Gamma_{\text{DM}}$ is given by~\citep{Palomares_Ruiz_2010, Barger_1995}:
\begin{equation}
    \Gamma_{\text{DM}}\,\approx\,1.3\times 10^{-32} \left[ \frac{\sin^2(2\theta)}{10^{-10}} \right] \left[ \frac{M_{\text{DM}}}{1\text{keV}} \right]^5 \frac{1}{\text{s}},
    \label{eq:gamma_DM}
\end{equation}
where $\theta$ is the mixing angle of the sterile neutrino.

To summarise, the expected flux from the DM-decay signal represents a narrow line at energy $E=M_{\text{DM}}/2$ with the intensity
\begin{align}
& \Phi^E(\phi) \equiv \int E\Phi(E,\phi)dE,\nonumber\\ 
& \Phi^E(\phi) = 2\cdot 10^{-10}k_x \left(\frac{\Gamma_{\text{DM}}}{10^{-20}\mbox{yr}^{-1}}\right)\left(\frac{S_{\text{DM}}}{10^{28}\frac{\mbox{keV}}{\mbox{cm}^2}}\right)\quad\frac{\mbox{erg}}{\mbox{cm}^2\mbox{s}  }, \nonumber \\
& \Phi^E(\phi) = 8.3\cdot 10^{-15}\left[ \frac{\sin^2(2\theta)}{10^{-10}} \right] \left[ \frac{M_{\text{DM}}}{1\text{keV}} \right]^5\left(\frac{S_{\text{DM}}}{10^{28}\frac{\mbox{keV}}{\mbox{cm}^2}}\right)\quad\frac{\mbox{erg}}{\mbox{cm}^2\mbox{s}  }.
\label{eq:signal_final}
\end{align}

\def\arraystretch{1.5}
\begin{table*}
\begin{center}

        \begin{tabular}{>{\centering\arraybackslash}m{3in}|llll}
        
            \hline
            \hline

          \multicolumn{1}{c}{\textbf{Reference}} & \multicolumn{1}{c}{$\rho_0$ [$ 10^7 \,M_{\odot}/\text{kpc}^3$]} & \multicolumn{1}{c}{$r_0$ [kpc]} & \multicolumn{1}{c}{\textbf{Profile}}\\
          \hline \hline

          \citet{Bird_2022} & \multicolumn{1}{c}{$3.24 \pm 1.25$} & \multicolumn{1}{c}{$8.75\pm1.57$} & \multicolumn{1}{c}{NFW} \\\hline \hline
          
          \citet{Cautun_2020}$^\dagger$ & \multicolumn{1}{c}{$1.0^{+0.68}_{-0.55}$} & \multicolumn{1}{c}{$15.56^{+2.2}_{-2.0}$} & \multicolumn{1}{c}{NFW$^\dagger$} \\\hline \hline

          \citet{Ablimit_2020} & \multicolumn{1}{c}{$1.05 \pm 0.12$} & \multicolumn{1}{c}{$14.45\pm0.46$} & \multicolumn{1}{c}{NFW} \\\hline \hline

          \citet{Sofue_2020} & \multicolumn{1}{c}{$2.07 \pm 0.1$} & \multicolumn{1}{c}{$10.94\pm1.05$} & \multicolumn{1}{c}{NFW} \\\hline \hline
          
        \citet{HaiNan2019} & \multicolumn{1}{c}{$5.21 \pm 0.58$} & \multicolumn{1}{c}{$7.8\pm0.4$} & \multicolumn{1}{c}{BUR} \\\cline{2-4}
        
              & \multicolumn{1}{c}{$2.0 \pm 0.19$} & \multicolumn{1}{c}{$9.6\pm0.4$} & \multicolumn{1}{c}{COM} \\\cline{2-4}
             
              & \multicolumn{1}{c}{$821\pm 1090$} & \multicolumn{1}{c}{$0.3\pm0.2$} & \multicolumn{1}{c}{ISO} \\\cline{2-4} 
             
              & \multicolumn{1}{c}{$5.22 \pm 0.92$ } & \multicolumn{1}{c}{$8.1\pm0.7$} & \multicolumn{1}{c}{NFW} \\\hline\hline
            
         \citet{Eilers_2019} & \multicolumn{1}{c}{$1.06 \pm 0.09$} & \multicolumn{1}{c}{$14.8\pm0.4$} & \multicolumn{1}{c}{NFW} \\\hline \hline

         \citet{Stref_2017} & \multicolumn{1}{c}{$0.84$} & \multicolumn{1}{c}{$20.2$} & \multicolumn{1}{c}{NFW} \\\cline{2-4}
         
             & \multicolumn{1}{c}{$0.84$} & \multicolumn{1}{c}{$19.6$} & \multicolumn{1}{c}{NFW} \\ \hline \hline
            
        \citet{McMillan_2016} & \multicolumn{1}{c}{$0.854$} & \multicolumn{1}{c}{$19.6$} & \multicolumn{1}{c}{NFW} \\\hline \hline

            \citet{Nesti_2013} & \multicolumn{1}{c}{$4.13^{+6.2}_{-1.6}$} & \multicolumn{1}{c}{$9.26^{+5.3}_{-4.2}$} & \multicolumn{1}{c}{NFW} \\\cline{2-4}
            
                 & \multicolumn{1}{c}{$1.4^{+2.9}_{-0.93}$} & \multicolumn{1}{c}{$16.1^{+17}_{-7.8}$} & \multicolumn{1}{c}{NFW} \\

            \hline
            \hline
            
            \citet{McMillan_2011} & \multicolumn{1}{c}{$0.846$} & \multicolumn{1}{c}{$20.2$} & \multicolumn{1}{c}{NFW} \\\hline

            \hline\hline
        \end{tabular}
    \caption{Collection of DM profiles with the parameters for the Milky Way reported in the literature. The table summarises the reference, the values of characteristic radius $r_0$ and density $\rho_0$ and the types of the corresponding DM profiles. Where applicable we refer to the NFW profile from~\citet{Cautun_2020} (marked with $^{\dagger}$) as to the ``reference DM-density'' profile.}
    \label{tab:models}
\end{center}
\end{table*}

\section{Instrumentation and data analysis} \label{ch:Instrument}
The analysis performed in this work is based on publicly available INTEGRAL/SPI data taken during about 20~years of instrumental operation. In this section we briefly describe the basic characteristics of the INTEGRAL/SPI instrument, describe the used data set and our approach to its analysis.
\subsection{SPI spectrometer}\label{subsec:SPI_Spectrometer}
SPI is a spectrometer, coded-mask instrument onboard of the ``INTernational Gamma-Ray Astrophysics Laboratory''~\citep{Winkler_2003,kuulkers21} that has been operational since 2002. SPI operates in the energy band of $20$\,keV -- 8~MeV and is characterised by an excellent energy resolution $\Delta E/E \sim 1/500$~\citep{refId0}. More precisely, the energy-dependence of the full-width half-maximum (FWHM) energy resolution can be approximated as~\citep{Roques2003}:
\begin{equation}
    \frac{\Delta E}{\mbox{1\,keV}} = 1.54 + 4.6\cdot10^{-3} \sqrt{\frac{E}{\mbox{1\,keV}}} + 6.0\cdot10^{-4} \left(\frac{E}{\mbox{1\,keV}}\right).
    \label{eq:eres}
\end{equation}

The satellite orbits Earth on $\sim 3$-day period, highly eccentric orbit which is partially located in Earth’s radiation belts. The regular crossings of the belts lead to a strong irradiation of the satellite by high-energy charged particles and consecutively to a strong time-variable instrumental background.

Contrary to focusing-optics instruments like e.g. XMM-Newton, \spi has limited imaging capabilities due to its coded-mask aperture. For the analysis presented below, similarly to~\citet{we_spi}, we explicitly treat the instrument as a collimator with $\sim 17.5^\circ$-radius (partially coded) FoV. Such a FoV corresponds to $\Omega_{\text{FoV}}\approx 0.29$~sr. 

\begin{figure}
    \centering
    \includegraphics[clip=true,width=1\columnwidth]{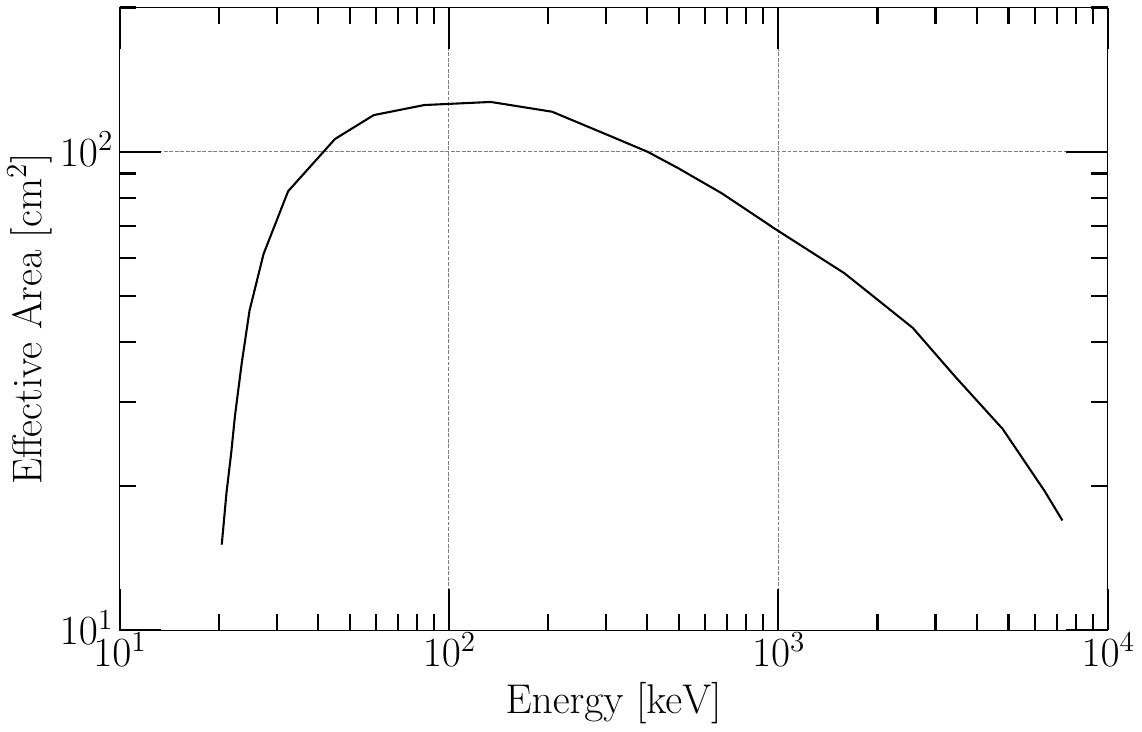}
    \centering
    \caption{The dependence of the on-axis effective area $A_{\text{eff,on}}$ of the SPI spectrometer on the energy of incoming photons.}
     \label{fig:EffArea}
\end{figure}

The on-axis effective area $A_{\text{eff,on}}$ of \spi is energy dependent and reaches $\sim 100$~cm$^2$, see Fig.~\ref{fig:EffArea}. For the analysis of the signal that covers the whole FoV of the instrument, the off-axis dependence of $A_{\text{eff}}$ should be taken into account, which leads to somewhat lower values of the effective area for extended sources~\citep{we_spi}:

\begin{align}
& A_{\text{eff,ext}}(E) = \frac{1}{\Omega_{\text{FoV}} }\int\limits_{\text{FoV}} d\Omega A_{\text{eff}}(E,\Omega)  \nonumber \\
& \approx 0.165\left(\frac{E}{\text{1\,keV}}\right)^{0.11}A_{\text{eff,ext}}.
    \label{eq_A_eff}
\end{align}

\subsection{Data set and data analysis}
In what presented below we utilise the data collected by \spi during its operation since 2002. The data set covers almost 19 years (02-28-2003 to 02-11-2021, revolutions: 46-2330) of publicly available observations\footnote{The \spi observations are available from the \href{https://www.isdc.unige.ch/integral/archive}{isdc website}.} taken in normal operational mode (\texttt{spimode=41} data selection flag) with at least 100~s \texttt{good\_spi} exposure. We additionally excluded observations performed during periods of SPI annealing phases\footnote{The list of annealing phases is available at the \href{https://www.isdc.unige.ch/integral/download/osa/doc/10.1/osa_um_spi/node70.html}{isdc website}.}, i.e. time periods during which the  SPI detectors are heated to recover from radiative damage \citep{SPI_User_Manual}. 

For each of the selected observations, we extracted the detected photons from the \texttt{spi-oper.fits} files ( \texttt{SPI.-OPSD-ALL} + \texttt{SPI.-OSGL-ALL} extensions). The main part of the electronics of SPI is the Analogue Front End Electronics (AFEE) where signals are amplified and filtered. The energy range 400--2000~keV  ($650-2200$~keV before 13 May 2012 or before Rev. 1170) is known to be affected by the AFEE artifacts~\citep{refId0,Jourdain2009}. In this energy range we explicitly used only photon events detected by the Pulse Shape discriminator (PSD) which are stored in \texttt{SPI.-OPSD-ALL} of the \texttt{spi-oper.fits} files. To take into account the reduced SPI efficiency in this energy band we additionally scaled down the effective area by a factor of 0.85~\citep{Jourdain2009}.

Based on the extracted photon list mentioned above, we produced a spectrum $dN/(dEdtdA)$ for each observation using the effective area given by Eq.~\ref{eq_A_eff} and the exposure time of the corresponding observation. We binned
the spectra into narrow energy bins of width $\Delta E / 3$, see Eq.~\ref{eq:eres}. We note that some of the energy bins host a low ($N<25$) number of photons. In order to properly handle the Poisson statistics in these bins we adopted $\Delta N=\sqrt{N+\frac{1}{4}} +1 $~\citep{Barlow:2004mi} for the statistical uncertainty. For the energy bins with $N\geq25$ photons we employed the standard Gaussian statistical uncertainty $\Delta N=\sqrt{N}$.

Given that the DM surface brightness increases towards the direction of the Galactic Center and an a priori independence of the instrumental background on the sky-position, we applied an ``ON-OFF'' technique aiming to detect the signal from the decaying DM and to eliminate the strong contribution of the instrumental background. Namely, we split the entire data set into two groups: ``ON'' (centered closer than $15^\circ$ from the GC) and ``OFF'' (centered further than $120^\circ$ from the GC). 
To minimize the effects of the time-dependent variability of the instrumental background, we selected one observation from the OFF data set for every observation from the ON data set, taken not later than 20 days apart.

Such a pairing and the consequent subselection allow for an accurate control of the time-variable
\spi instrumental background. As shown in~\citet{Teegarden_2006} for observations separated in time by not more than 20~days, the level of instrumental background is in a good correlation with the strength of a strong instrumental Germanium line at 198~keV. In this case the strength of the line can be used to rescale the \spi instrumental background. The rescaled spectra of the OFF-group observations were then treated as a background and subtracted from the respective ON-group spectra.
Consequently, the expected strength of the signal in our ON-OFF type analysis is given by (see Eq.~\ref{eq:signal_final}):
\begin{align}
 & \Phi^E(\phi) = 2\cdot 10^{-10}k_x \left(\frac{\Gamma_{\text{DM}}}{10^{-20}\mbox{yr}^{-1}}\right)\times \nonumber \\
 &\times \frac{\langle S_{\text{DM}}(\text{ON})\rangle - \langle S_{\text{DM}}(\text{OFF})\rangle }{10^{28}\mbox{keV/cm}^2}\quad\mbox{erg/cm$^2$/s},   \nonumber \\\nonumber \\
 & \langle S_{\text{DM}}(\text{ON})\rangle = \sum\limits_i S_{\text{DM}}(\text{ON}_i)T_{\text{ON,}i} / \sum\limits_i T_{\text{ON,}i}, \nonumber \\
 & \langle S_{\text{DM}}(\text{OFF})\rangle = \sum\limits_i \alpha_i S_{\text{DM}}({\text{OFF}_i})T_{\text{OFF,}i} / \sum\limits_i T_{\text{OFF,}i}, \nonumber \\
 & \alpha_i = f_{\text{ON,}i}(198\,\mbox{keV}) / f_{\text{OFF,}i}(198\,\mbox{keV}).
 \label{eq:on_off_flux_gamma}
\end{align}

Here, index $i$ numbers the pairs of ON/OFF observations and $f_{\text{ON,}i}(198\,\mbox{keV})$, $ f_{\text{OFF,}i}(198\,\mbox{keV})$ stand for the flux in the 198~keV line in ON and OFF observations respectively. The mean $\langle S_{\text{DM}}(\text{ON})\rangle$ and $\langle S_{\text{DM}}(\text{OFF})\rangle$ values correspond to the exposure-time weighted averages of the DM column densities in the ON and OFF regions. We note also that all rescaling coefficients $\alpha_i$ are close to 1 with a standard deviation of merely $\sim$ 11\%.

Such an approach resulted in 9162 observational pairs with a total duration for the ON data set of 20~Ms and 24~Ms for the OFF data set. The mean values $\langle S_{\text{DM}}(\text{ON})\rangle$ and $\langle S_{\text{DM}}(\text{OFF})\rangle$ of the DM column densities for ON and OFF data sets are summarised in Tab.~\ref{tab:S_ON-V}. 

\subsection{Residual background and systematics treatment} \label{subsec:Residual_and_systematics}
\begin{figure}
    \centering
    
    \includegraphics[width=\columnwidth]{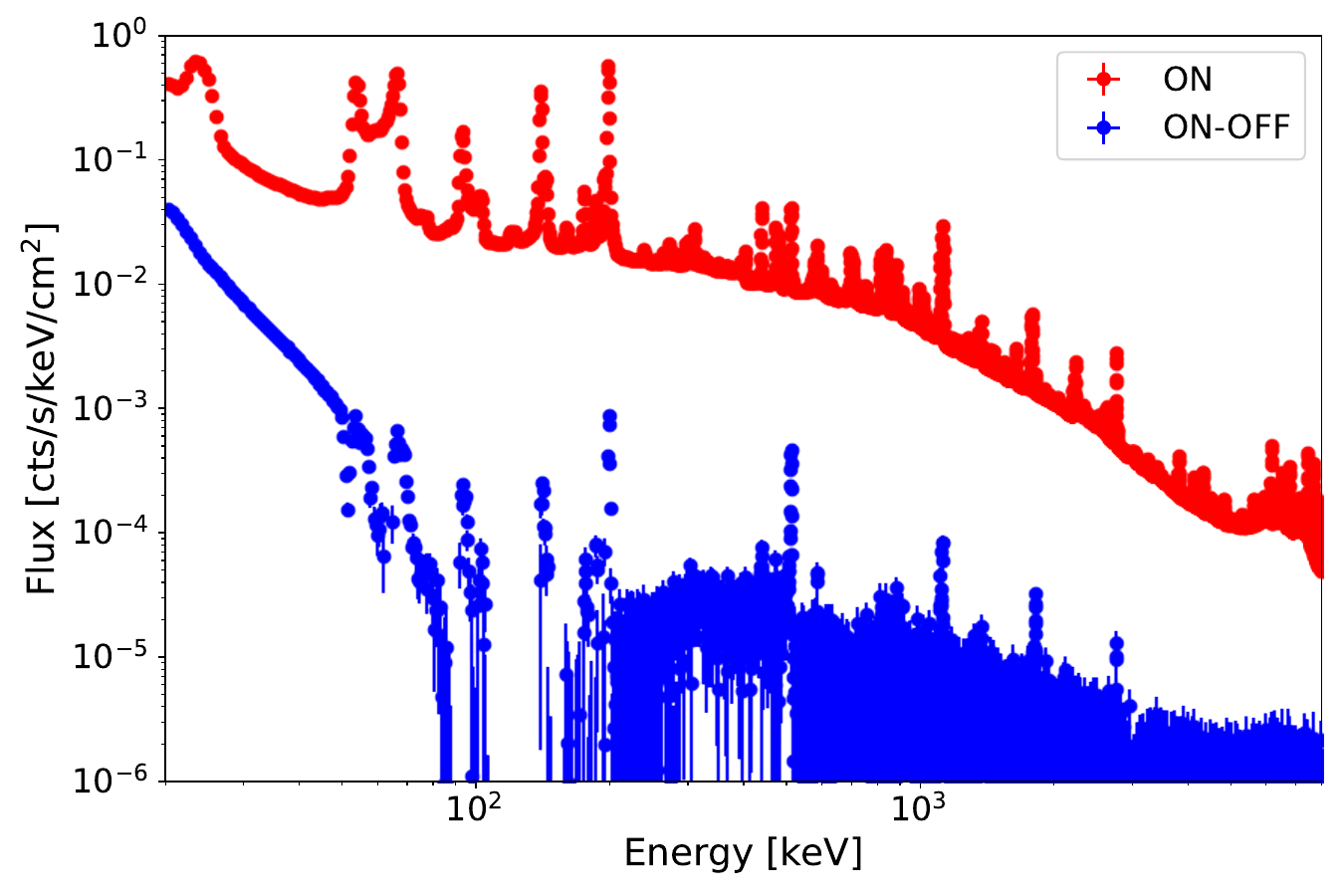}\\
    \includegraphics[width=\columnwidth]{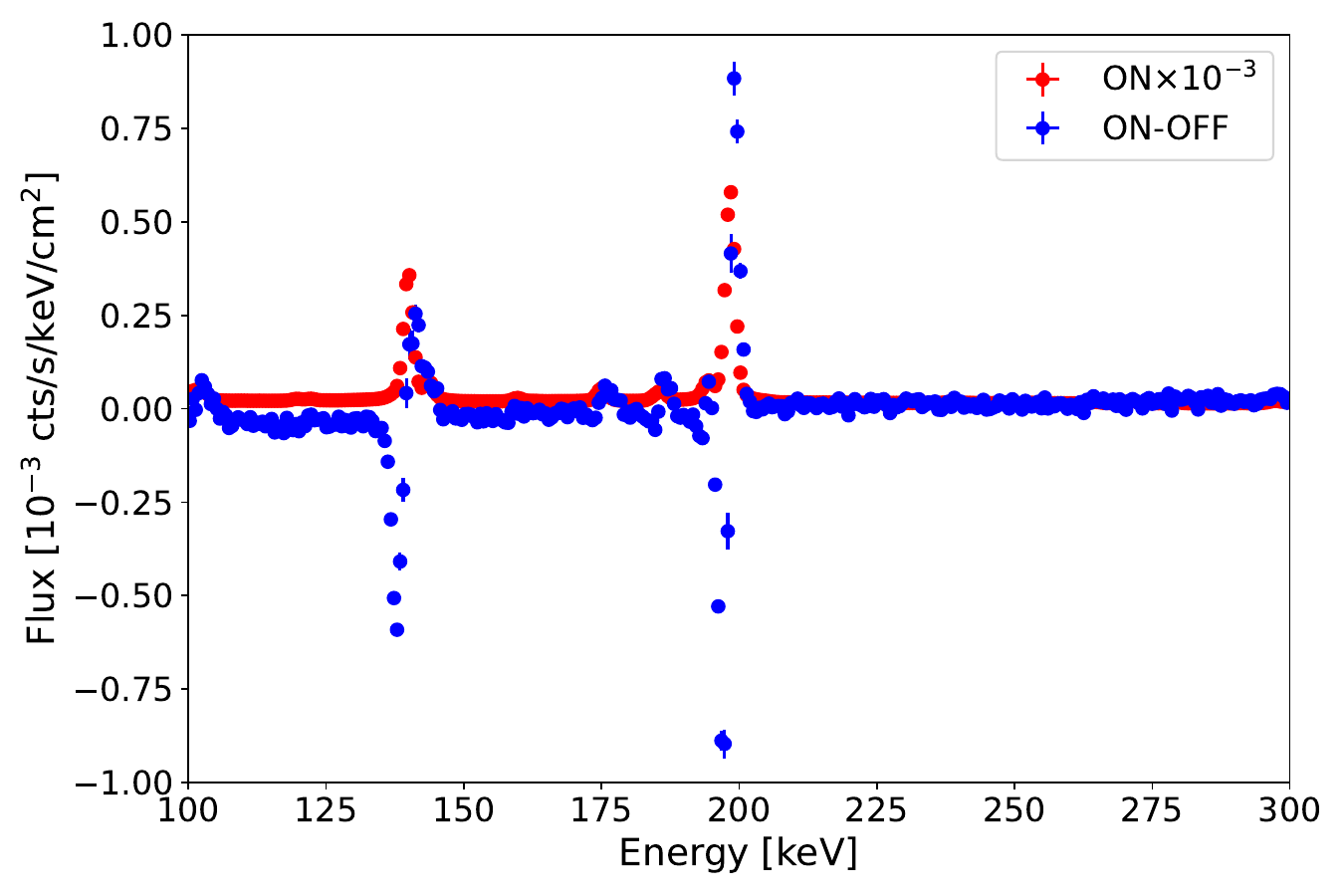}
    \centering
    \caption{Top: ON spectrum (red) as compared to the ON-OFF (blue). Bottom: a zoom of the top panel to the vicinity of the 198~keV line used in the subtraction for renormalizations of the ON spectra. The wiggle-like residuals are used to estimate the level of systematic uncertainty, see section \ref{subsec:Residual_and_systematics} for the details. }
     \label{fig:ON_vs_ON-OFF}
\end{figure}

The ON-OFF procedure described above allowed us to significantly reduce the flux level observed in the ON group of observations by a factor of ~ $10^2-10^3$, see Fig.~\ref{fig:ON_vs_ON-OFF} (upper panel), by strongly suppressing the time-variable instrumental background of SPI.

\begin{figure*}
    \centering
    \includegraphics[clip=true,width=1\columnwidth]{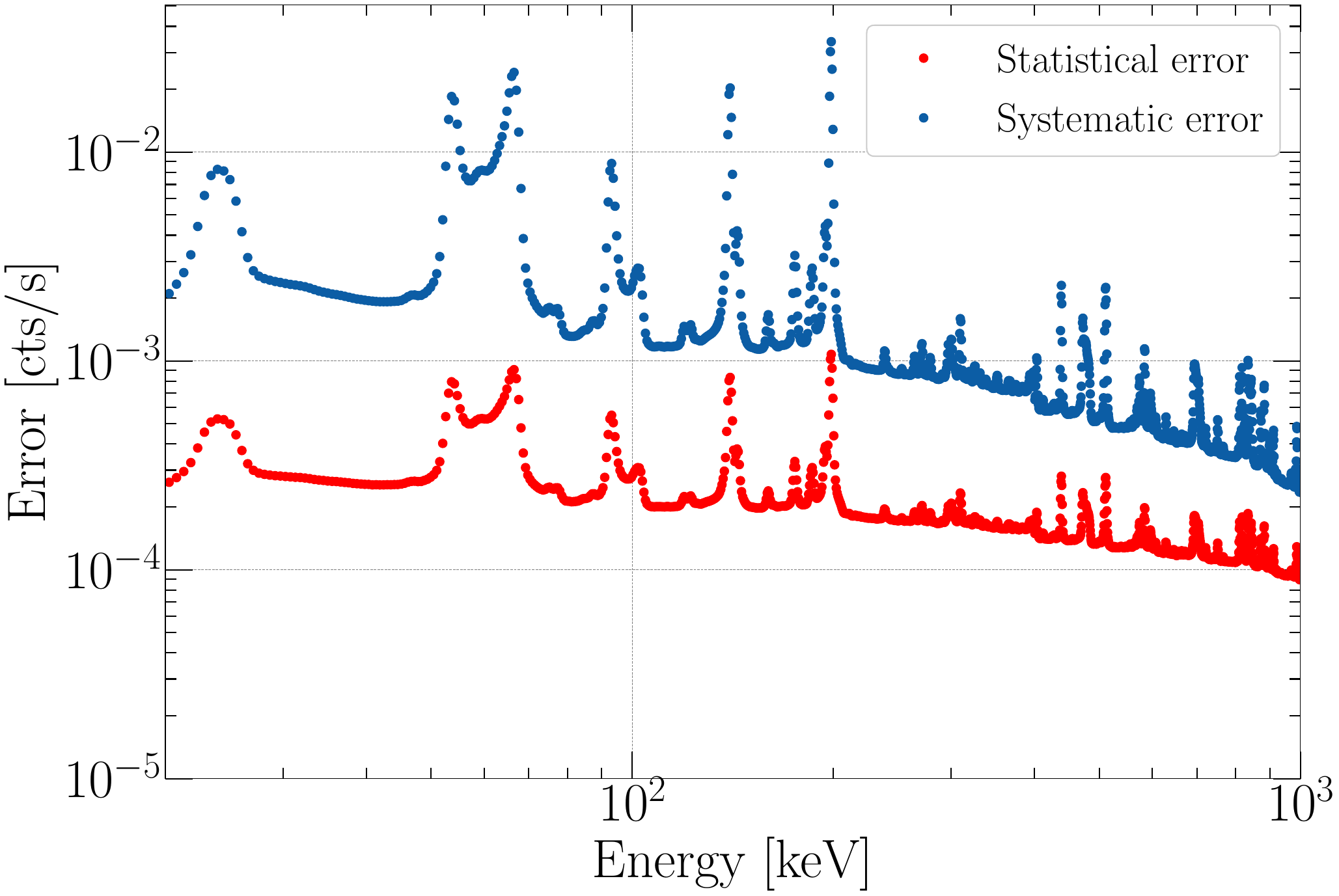}
    \includegraphics[clip=true,width=1\columnwidth]{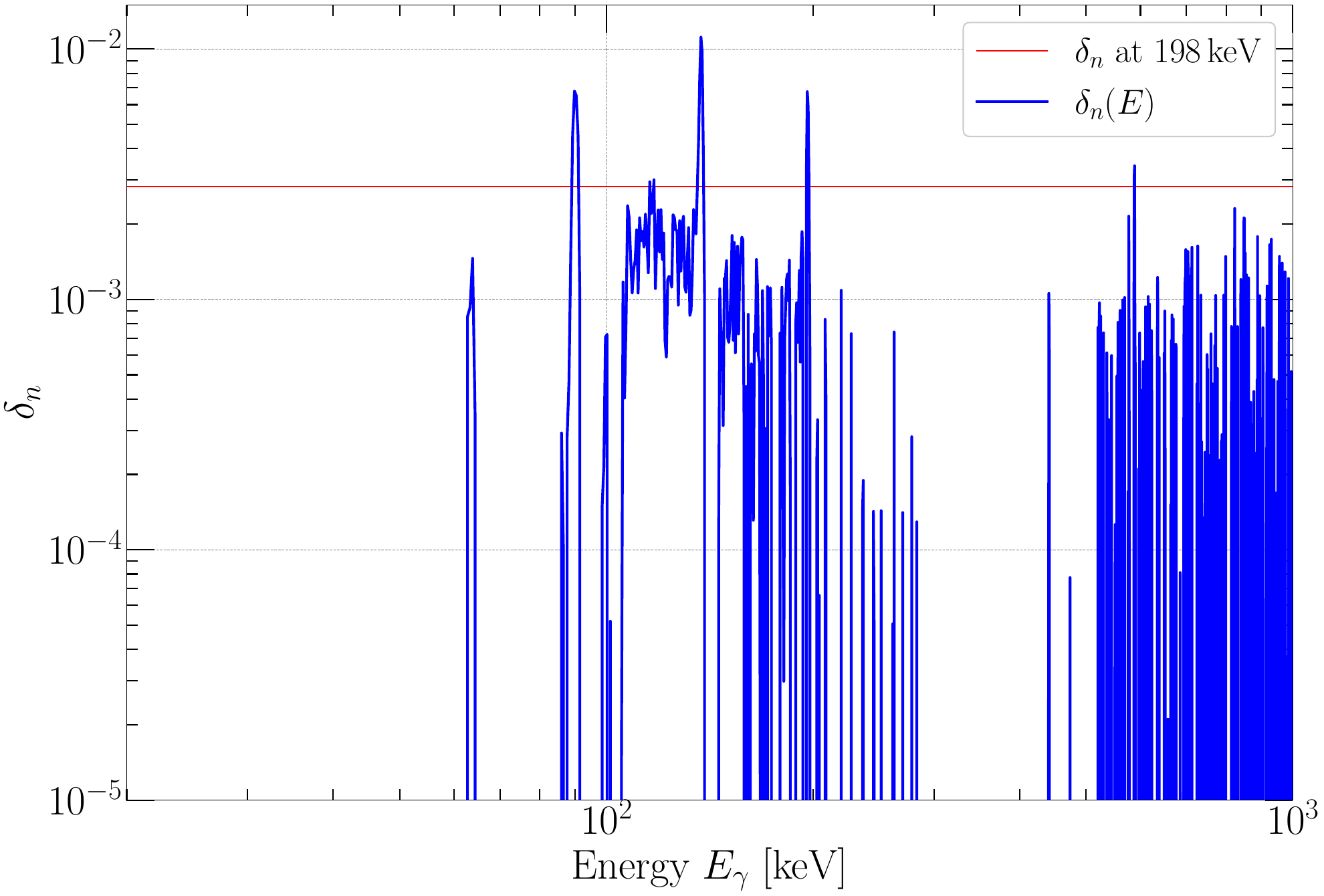}
    \centering
    \caption{Left panel: Systematic (blue) and statistical (red) uncertainties as a function of energy. The level of systematic uncertainty was estimated to be $\delta_n\approx 2.8\cdot 10^{-3}$, see text for the details. Right panel: The level of systematic uncertainty $\delta_n$ derived from the presence of negative residuals close to the 198~keV Ge line as compared to the level of systematics estimated from negative residuals at other energies.}
     \label{fig:sys_error_198_vs_stat}
\end{figure*}

One can see that close to the 198~keV line the subtraction is not perfect and results in a wiggle-like structure, see Fig.~\ref{fig:ON_vs_ON-OFF}, bottom panel. The negative wing of this structure, however, allows to estimate the level of systematic uncertainty connected to the selected method of modelling and subtracting the instrumental background. This was achieved by adding a systematic error at a fractional ON-flux level which allows to make this negative residual insignificant at $1\sigma$ level. We found such a systematic uncertainty to be
\begin{equation}
    \delta_n = \frac{\Delta \Phi^E}{\Phi^E_{\text{ON}}}\approx 2.8\cdot10^{-3}.
    \label{eq:delta_n_1}
\end{equation}
At almost all energies the systematics at $198\,\text{keV}$ exceed the others, see Fig.~\ref{fig:sys_error_198_vs_stat} (right panel).
It should be noted that this level of systematic uncertainty substantially exceeds the one of the statistical uncertainty at all energies, see left panel of Fig.~\ref{fig:sys_error_198_vs_stat}. 

In addition to the time-variable instrumental SPI background, the selected approach of using SPI as a collimator does not allow the contribution from astrophysical sources to be cancelled out. Closer to the GC the concentration of such sources results in an overall positive residual ON-OFF spectrum. 
For searching a narrow line-like signal from DM decay, however, one could adopt any smooth (i.e. not hosting lines) model of the astrophysical emission. Such a model would present a continuum-like signal from the joint contribution of all astrophysical sources. To build such a model we used the ``sliding window'' method~\citep{we_spi}. Namely, in every energy bin we replaced the flux $f(E)$ with
\begin{align}
& \text{Sliding}(f(E), \text{size}) = \langle f(E) \rangle_{[-\text{size}; +\text{size}]}\quad - \nonumber \\
& -\frac{1}{2}\left(\langle f(E) \rangle_{[-2\text{size}; -\text{size}]} + \langle f(E) \rangle_{[\text{size}; +2\text{size}]}\right) \nonumber,
\end{align}
where $ \langle f(E) \rangle_{[E_1; E_2]}$ stands for the total flux integrated (summed) between energies $[E-E_1 ; E+E_2]$ (see Eq.~\ref{eq:signal_final}); $\text{size}$ -- for the characteristic size of the sliding window and $\Delta E$ -- for the SPI energy resolution, Eq.~\ref{eq:eres}. In our analysis we used $\text{size}=2/3\,\Delta E$. Such an approach allows the subtraction of a half-sum of the adjacent energy intervals from any energy bin and to effectively eliminate the smooth (non-line-like) background. At the positions of lines such a method allows the subtraction of the pedestal emission estimated from the vicinity around the line energies. 

We note also that after applying the sliding window method to the spectrum its points lose their statistical independence. However, for broad enough sizes ($\gtrsim \Delta E$) in the sliding windows method, the significance of the signal at fixed energy is just given by the significance (above 0) of the given point. That is, no additional summation is needed to account for the presence of the signal in several nearby energy bins.

\begin{figure}
    \centering
    \includegraphics[clip=true,width=1\columnwidth]{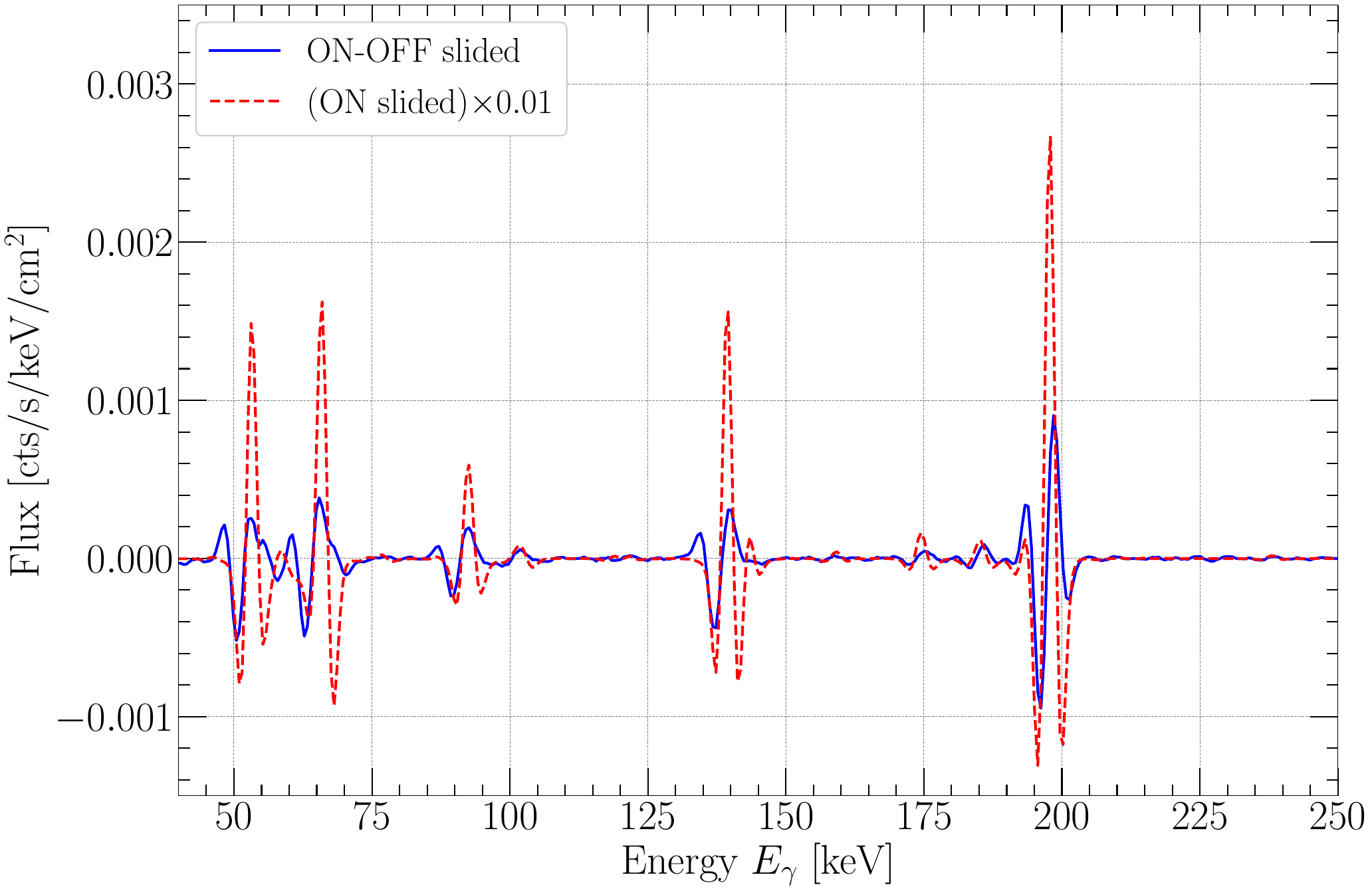}
    \centering
    \caption{Comparison of the ON–OFF spectrum and the ON spectrum scaled by $0.01$, both after applying the sliding window method. The instrumental lines cancel better than $1\%$.}
     \label{fig:wiggles}
\end{figure}

To the left and right of the positions of strong spectral lines the sliding window method is known to create wiggle-like negative residuals, see Fig.~\ref{fig:wiggles}. Such wiggles are connected to over-subtraction of the estimated background level when the left/right wing of the sliding window coincides with the position of the line. The discussed wiggles present an additional source of systematic uncertainty. Namely, at energies with the negative flux after sliding window we added a systematic uncertainty equal to the absolute value of the flux. 

\begin{figure}
    \centering
    \includegraphics[clip=true,width=1\columnwidth]{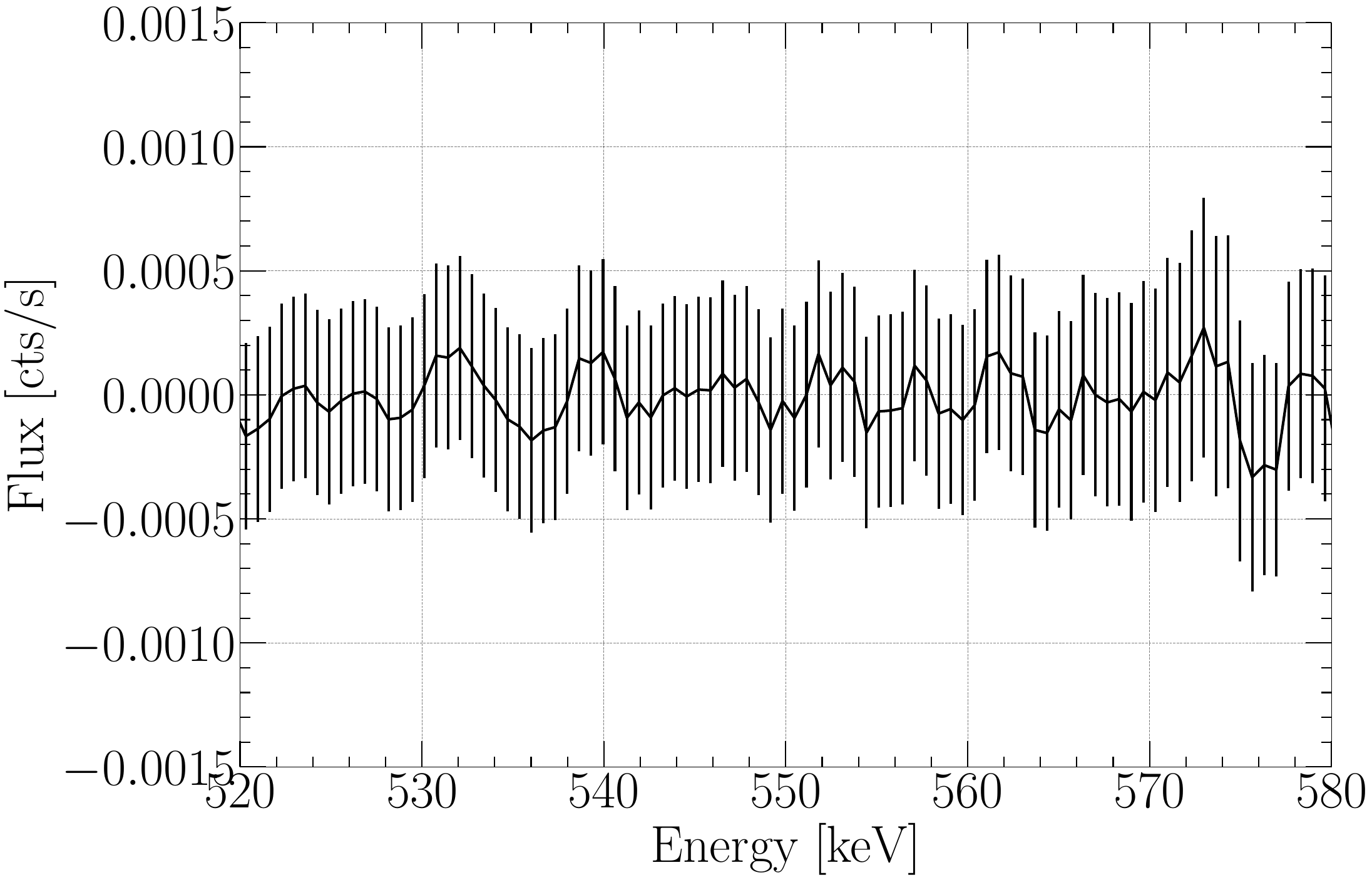}
    \centering
    \caption{Subtracted flux after sliding window method is applied in a region far from any lines.}
     \label{fig:error_dom}
\end{figure}

To summarise, in what follows and where applicable, we added a statistical uncertainty, a systematic uncertainty connected to the instrumental background modelling (see Eq.~\ref{eq:delta_n_1}) and a systematic uncertainty connected to wiggles due to the sliding window method in quadratures. The discussed uncertainties result in an ON-OFF flux (after sliding window) at $1\,\sigma$ level consistent with zero in regions far from positions of strong instrumental lines, see Fig.~\ref{fig:error_dom}.

\begin{table*}
    \centering
    \begin{tabular}{|c|c|c|c|c|}
  \hline
  
  Profile & Reference & $dS_{\text{ON}}/d\Omega$ [$10^{28}\frac{\text{keV}}{\text{cm}^2}$] & $dS_{\text{OFF}}/d\Omega$   [$10^{28}\frac{\text{keV}}{\text{cm}^2}$]& $S_{\text{ON}}/S_{\text{OFF}}$ \\
  \hline \hline
    ISO & \citet{HaiNan2019} & 40.25 & 1.07 & 37.8 \\ \hline
    \multirow{8}{*}{NFW}
         & \citet{HaiNan2019} & 22.04 & 0.95 & 23.21 \\ 
         
         & \citet{Nesti_2013} & 20.09 & 1.02 & 20.53\\
         
         & \citet{Bird_2022} & 15.19 & 7.03 & 21.62\\

         & \citet{Sofue_2020} & 13.10 & 0.739 & 17.72\\ 
         
         & \citet{Ablimit_2020} & 9.96 & 0.68 & 14.08\\ 
         
         & \citet{Eilers_2019} & 9.95 & 0.72 & 13.81\\ 
         
         & \citet{Cautun_2020}$^\dagger$ & 10.0 & 0.75 & 13.28\\
         
         & \citet{McMillan_2011} & 11.45 & 1.02 & 11.19\\

         & \citet{Stref_2017} & 11.70 & 1.07 & 10.96\\ \hline

    BUR & \citet{HaiNan2019} & 6.33 & 0.997 & 6.34 \\ \hline
    
    COM & \citet{HaiNan2019} & 4.44 & 1.04 & 4.27 \\ \hline

\end{tabular}

    \caption{Time averaged DM column densities $dS_{\text{DM}}/d\Omega$ for the ON and OFF regions as described in Eq.~\ref{eq:on_off_flux_gamma} using the virial parameters listed in Tab.~\ref{tab:models}, ordered in descending order by $S_{\text{ON}}/S_{\text{OFF}}$. Here, $d\Omega\approx0.29 $ for SPI's FoV (see section \ref{subsec:SPI_Spectrometer}). It is evident that for all models (see Tab.~\ref{tab:models}) the fraction $S_{\text{ON}}/S_{\text{OFF}}$ is greater than $4$ and for NFW significantly larger than the one of BUR and COM models.} 
    \label{tab:S_ON-V}
\end{table*}

\section{DM-decay line candidates and constraints}
\label{sec:dm_lines}
\subsection{Line search and decaying DM constraints}
In the ON-OFF spectrum discussed above (after the sliding window method was applied) we performed a search for narrow line-like features above zero background. 

The positions of $3\sigma$ (statistical) significant lines  are summarised in Tab.~\ref{tab:lines}. In the same table we also provide possible identifications of the lines with known instrumental lines listed in \citet{Weidenspointner2003, Diehl_2018}. 

At all energies we also constructed a $3\sigma$ (stat+syst) upper limit on the residual flux defined as a sum of the mean ON-OFF flux and three (statistical+systematic) uncertainties. Using Eq.~\ref{eq:on_off_flux_gamma} (see also Eq.~\ref{eq:signal_final}), we converted this flux limit into constraints on the lifetime $\tau = 1/\Gamma_{\text{DM}}$ of decaying DM (for $k_x=1$) and similarly into constraints on the mixing angle $\sin^2(2\theta)$ for sterile neutrinos. These constraints based on the reference DM density profile~\citep{Cautun_2020} of the MW are shown in Fig.~\ref{fig:lifetime_full_comparison} (DM decay lifetime) and Fig.~\ref{fig:mixing_full_comparison} (sterile neutrino mixing angle). The shaded/hatched regions correspond to the full range of limits derived for other DM density profiles listed in Tab.~\ref{tab:models}, i.e. to systematic uncertainties connected with poor knowledge of the DM distribution in the MW.

\subsection{Off-GC angle analysis} \label{subsec:Off_Angle_analysis}
The analysis presented above revealed 32 line-candidates detected with a statistical significance of $\gtrsim 3\sigma$, see Tab.~\ref{tab:lines}. Although most of the lines coincide with known \spi instrumental lines, three of them do not have a clear instrumental counterpart.
In this subsection we discuss the off-GC angle dependency of the flux in selected line candidates, which can help to further discriminate between the purely instrumental lines and lines of potential DM-decay origin.

We note that the strength of any line originating from a potential DM-decay is proportional to the DM column density in the observed region, see Eq.~\ref{eq:signal_final}, Eq.~\ref{eq:Sdm} and consequently depends on the DM density profile, see also Fig.~\ref{fig:profiles}. As discussed above this results in a higher expected DM decay signal from the inner regions of the Galaxy, which should gradually decrease to the MW outskirts. The angular, off-Galactic Center dependency of the expected signal strength can be used to discriminate between such a signal and another signals of instrumental or astrophysical origin.

\def\arraystretch{1.5}
\setlength\tabcolsep{2pt}
\begin{table*}

    \begin{minipage}{.5\linewidth}
    \begin{center}

    \resizebox{0.95\textwidth}{!}{\begin{tabular}{|c|c|c|c|c|}
      \hline
      \textbf{E} [keV] & \textbf{Sign} $\sigma$ (stat) & \textbf{Sign} $\sigma$ (stat+sys) & $\Delta E$ [keV] & \textbf{Identification}\\
      \hline \hline
      22.4 & 23.4 & 0 & 1.58 & $^{71\text{m}}\text{ Ge(IT)}^{71} \text{Ge}$\\
      \hline
      47.8 &	16 & 0 &	1.6 & $^{2380} \text{U series}$\\
      \hline
      53.1 & 9.7 & 0 &	1.6 & $^{73m} \text{Ge complex}$\\
      \hline
      60.6 &	7.4 & 0 &	1.61 & $^{73m} \text{ Ge complex, sum peak}$\\
      \hline
      65.98 &	12.2 & 0 &	1.62 & $^{73m} \text{ Ge complex, sum peak}$\\
      \hline
      86.57 &	9.1 & 0 & 	1.63 & $\text{Various isotopes}$\\
      \hline
      92.57 &	13 & 0 &	1.64 & $^{67m} \text{Zn complex}$\\
      \hline
      101.88 &	6.31 & 0 &	1.65 & $^{67m} \text{Zn complex}$\\
      \hline
      140.11 &	17 & 0 &	1.68 & $^{75m} \text{ Ge complex}$\\
      \hline
      175.07 &	5.63 & 0 &	1.71 & $^{71} \text{As(EC)}^{71} \text{Ge}$\\
      \hline
      185.9 &	11.27 & 0 &	1.71 & $^{71} \text{As(EC)}^{71} \text{Ge} + \text{K}$\\
      \hline
      193.34 &	29.42 &	0 & 1.72 & $^{71m} \text{ Ge complex}$\\
      \hline
      438.09 &	5.7	& 0 & 1.9 & $^{69m} \text{Zn(IT)}^{69} \text{Zn}$\\
      \hline
      511.0 &	35.22 & 4.2 &	1.95 & $\text{Positron annihilation}$\\
      \hline
      583.63 &	7.72 & 0 &	2 & $^{208} \text{Tl}(\beta^{−}) ^{208} \text{Pb}$\\
      \hline
      699.25 &	3.02 & 0 &	2.08 & $^{205}\text{Bi}(\text{EC})^{205}\text{Pb} + \text{L} ^{\dagger}$\\ 
      \hline

      \hline
    \end{tabular}}
  \end{center}
    \end{minipage}%
    \hfill
    \begin{minipage}{.5\linewidth}
      \begin{center}
    
    \resizebox{0.95\textwidth}{!}{\begin{tabular}{|c|c|c|c|c|}
      \hline
      \textbf{E} [keV] & \textbf{Sign} $\sigma$ (stat) & \textbf{Sign} $\sigma$ (stat+sys) & $\Delta E$ [keV] & \textbf{Identification}\\
      \hline \hline
      810.82 &	3.34 & 0 &	2.16 & $^{58}\text{ Co(EC)}^{58} \text{Fe}$ \\
      \hline
      835.36 &	3.09 & 0 &	2.17 & $^{54} \text{Mn(EC)}^{54} \text{Cr + L }$\\
      \hline
      880.56 &	5.08 & 0 &	2.2 & $^{206} \text{Bi(EC)}^{206} \text{Pb}$\\
      \hline
      910.86 &	3.94 & 0 &	2.23 & $^{232}\text{ Th series}$\\
      \hline
      1105.47 &	4.51 & 0 &	2.34 & $^{69}\text{Ge}(\text{EC})^{69}\text{Ga} ^{\dagger}$\\
      \hline
      1114.9 &	9.28 & 0 &	2.36 & $^{65} \text{Zn(EC)}^{65} \text{Cu}$ \\
      \hline
      1135.43 &	3.10 & 0 &	2.38 & $^{77}\text{Ge}(\beta^{-})^{77}\text{As} ^{\dagger}$ \\
      \hline
      1172.09 &	4.1 & 0 &	2.40 & $^{60} \text{Co}(\beta^{-})^{60} \text{Ni}$\\
      \hline
      1320.61 &	3.07 & 0 &	2.5 & $\text{?}$\\
      \hline
      1344.84 &	3.23 & 0 &	2.52 & $^{96}\text{Ge(EC)}^{69}\text{Ge} + \text{K} ^{\dagger}$\\
      \hline
      1355.76 &	3.15 & 0 &	2.52 & $^{65}\text{Ga(EC)}^{65}\text{Zn} + ^{19}O(\beta^{-})^{\dagger} $\\ 
      \hline
      1368.34 &	3.56 & 0 &	2.53 & $^{24} \text{Na}(\beta^{-}) ^{24} \text{Mg}$\\
      \hline
      1471.82 &	3.07 & 0 &	2.6 & $\text{?}$\\
      \hline
      1809.0 &	9.41 & 3.4 &	2.82 & $^{26} \text{Mg} ^{26} \text{Na}(\beta^{-})$\\
      \hline
      2403.35 &	3.11 & 0 &	3.21 & $\text{?}$\\
      \hline
      2749.42 &	5.03 & 0 &	3.43 & $^{66}\text{ Ga(EC)}^{66} \text{Zn}$\\
      \hline

      \hline
      
    \end{tabular}}
  \end{center}
  
    \end{minipage} 
    \caption{Residual lines. In total $32$ lines were detected with statistical error only, $2$ with statistical + systematic uncertainties. Lines with an identification within $\Delta E$ were listed by \citet{Weidenspointner2003}, lines with a $^{\dagger}$ are listed exclusively by \citet{Diehl_2018}. Lines with ``?" are not listed or identified by either of the latter catalogues.} 
    \label{tab:lines}
    \vspace{2mm}
\end{table*}

For the off-GC angle analysis we utilised the entire available data set of \spi without any selection criteria regarding the off-GC angle (as was done for the ON-OFF spectrum construction). However, we explicitly removed all ScWs taken during high solar activity periods (based on reports in the INTEGRAL Operation Reporting\footnote{INTEGRAL Operation Reporting on \href{https://www.isdc.unige.ch/integral/operations/reports.cgi?endRev=1}{isdc website}.}). In particular, we ignored the data taken during Rev. 283--325 and Rev. 1850-1911. The final data set contained 124769 ScWs with a total exposure of $\sim 300$\,Msec. 

In order to control the time-variable instrumental background of SPI for the off-GC angle analysis we split the whole data over bunches of data taken between particular annealing phases where the SPI instrumental background remains relatively stable \citep{Siegert_2019}. In each of the 32 periods, we determined the time-averaged flux in the 198~keV line and, similar to the ON-OFF analysis, renormalized all spectra taken during this period to match the same flux in this line. Consequently, we built the off-GC angle profile for all other lines listed in Tab.~\ref{tab:lines} during the corresponding periods.

To compensate for a possible time-variability of the level of the instrumental background's continuum between the selected periods we explicitly subtracted a constant from every off-GC angle profile to eliminate the (weighted) mean signal at angles $\phi>60^\circ$. We note that for the considered DM profiles the expected strength of a DM-decay signal at $\phi>60^\circ$ is smaller than the signal from the GC direction by a factor of at least $\sim 4$. Therefore, the subtraction of the constant cannot fully eliminate the DM decay signal as a function of the off-GC angle.

The usage of the 198~keV line for renormalizing the spectra does not allow the use of this line for the estimation of systematics as it was done in the ON-OFF analysis. In order to estimate the level of systematic uncertainties in this case we utilised the second brightest instrumental line at 138~keV. Namely, we added systematic uncertainty at a level sufficient to make the flux in this background subtracted line consistent with zero at $1\sigma$ level in each of the considered time periods and off-GC angles. Thus, the systematic error for the 138~keV line is given by the residual flux $f_R(\phi)$ at each angle:
\begin{equation}
    \sigma_{\text{sys}}(138\,\text{keV},\phi) = f_R(138\,\text{keV},\phi).
\end{equation}
In order to account for different fluxes and, therefore, different background variations within a line, we scale the systematic error for each candidate by the respective background, i.e. the mean background signal $\Bar{f_R}$ at angles $\phi>60^\circ$:
\begin{align}
    \delta n_{138}(E) =& \frac{\Bar{f_R}(E,\phi>60^\circ)}{\Bar{f_R}(138\,\text{keV},\phi>60^\circ)}.
\end{align} 
On average, this yields $\delta n_{138} \sim 5.8\cdot 10^{-2}$ over all energies, a factor of 20 higher than in the energy-independent estimation in the ON-OFF analysis, see Eq.~\ref{eq:delta_n_1}. Similarly to the ON-OFF systematics we then estimate the systematic uncertainty by $\delta n_{138}(E) \times f_R(E,\phi)$.\\

\noindent After such a subtraction we have calculated the weighted (with stat+sys uncertainties) mean of the individual off-angle profiles for the discussed line-candidates, see Fig.~\ref{fig:offangle_lines}.

We found that for the majority of the lines the obtained off-GC angle profile is consistent with a constant and only in few cases the profiles demonstrate a clear excess towards the GC. To verify this statement we performed a formal fit of every profile with a model that is a sum of a constant and DM column density for the reference NFW profile (Eq.~\ref{eq:Sdm} with mater density from Eq.~\ref{eq:NFW_profile}) with a free normalisation: 
\begin{align}
\Phi^E(\phi) = c+a*S_{\text{DM}}(\phi).
\label{eq:off_gc_profile_model}
\end{align}
Such a model includes the cases of a line with a purely instrumental angular behaviour (a and c consistent with 0) and a line with the contributions from decaying DM ($a>0$).
Tab.~\ref{tab:line_results_test} summarises the results of the fit: line energy, $\chi^2$ for the model with $a$ fixed to 0 and the $\Delta\chi^2$ of the model with a free parameter $a$ and the model with $a$ fixed to 0. The number of degrees of freedom (NDF) for $a$ free is 34 due to 36 data points and two parameters, while for $a=0$ we have NDF$=35$.
We found that only for $511$ keV and $1809$ keV the fit with non-zero $a$ is $>2\sigma$ preferable over a fit with a constant model ($\Delta\chi^2>4$) and is not excluded by either a high false-fit probability or high reduced $\chi^2$.

\section{Results and Discussion}
\label{sec:results_discussion}
In this work we present the results of an ON-OFF type analysis based on almost 20~years of \spi data aiming to search for a narrow line-like signal from a decaying dark matter particle. Our analysis revealed the presence of 32 line-candidates detected in the data with statistical significance of $>3\sigma$, see Tab.~\ref{tab:lines}.
Among these lines 29 coincide with known \spi instrumental lines; 2 of these also coincide with lines of known astrophysical origin and 3 do not have clear associations with known instrumental or astrophysical lines. Our analysis also allowed us to put strong constraints on the lifetime of the decaying DM and, in particular, constrain the parameters of the sterile neutrino DM candidate. Below we briefly discuss and summarise the derived results.

\subsection{Line candidates} \label{subsec:line_candidates}
Among the detected 32 lines, 29 coincide by energy with known \spi instrumental lines. We note that some lines, e.g. the~511~keV line, are of ``mixed'' origin. The flux in these lines is produced by a sum of astrophysical and instrumental lines coinciding in energy. In order to check the possibility of an admixture of a signal from DM decay at energies coinciding with known instrumental lines we performed the off-GC angle analysis as described above. In this analysis we built the flux in the discussed lines as a function of the angle with respect to the direction of the GC. In case a line contains some contribution from decaying DM, one would expect that its brightness would increase close to the GC. For purely instrumental lines one can expect a behaviour independent of the off-GC angle. 

The off-GC angle dependency for all detected lines is shown in Fig.~\ref{fig:offangle_lines}. We note that the majority of the lines (21 out of 32) demonstrate an off-GC angle dependency consistent with a constant at $<3\sigma$ level, see Tab.~\ref{tab:lines}. We further note that for all 3 line-candidates without a clear instrumental/astrophysical counterpart, the off-GC angle profile is also consistent with a constant. This rules out the possibility of these lines to host a significant contribution from decaying dark matter.

Among the remaining lines which do not demonstrate consistency with a constant off-GC angle profile, two lines (511~keV and 1809~keV) are statistically consistent with the ``DM decay profile'' model, that is, Eq.~\ref{eq:off_gc_profile_model} with non-zero normalization $a$ of the DM-related part of the model. This clearly indicates that the brightness of these lines gradually decrease with off-GC angle and that these lines have a substantial part of astrophysical emission from the GC region.

The 511~keV electron-positron annihilation line originating from the vicinity of the GC has been detected by GRS/SMM more than 30 years ago~\citep{share90}. The spatial profile of this line was studied with CGRO/OSSE~\citep{cheng97,purcell97} and later in more detail with \spi~\citep{jurgen03,jurgen05}. The off-GC angle profile of the 511~keV line derived in this work is shown in Fig.~\ref{fig:511_off_angle}. We found that the profile is well-described also with a simple Gaussian centered at $\phi=0$ and FWHM of $\sim 37^\circ$, consistent with previous reports of~\citet{jurgen05}.

The $^{26}\text{Al}$ decay, 1809~keV line was the first astronomical gamma-ray line from radioactive decay discovered in our Galaxy~\citep{mahoney82, mahoney84} with HEAO-3 observations. Later the distribution of this line was studied with COMPTEL/EGRET, RHESSI and later with \spi by~\citet{PRANTZOS199599,smith03,Schanne_2007,Bouchet_2015}.

The $^{26}\text{Al}$ line is produced via $\beta^+$ decay into the excited state of $^{26}\text{Mg}$ and consequent transition of $^{26}\text{Mg}$ to the ground state~\citep[see e.g.][]{Isern_2021}. The short half-life time time of $\sim 10^6$~yr of $^{26}\text{Al}$ makes this line a tracer of the recent nucleosynthesis activity. The line is believed to originate from environments hosting Type II supernovae, novae, and the winds of Wolf-Rayet and asymptotic giant branch stars~\citep[see e.g.][]{smith03,Isern_2021}.

\begin{figure}
    \centering
    \includegraphics[clip=true,width=1\columnwidth]{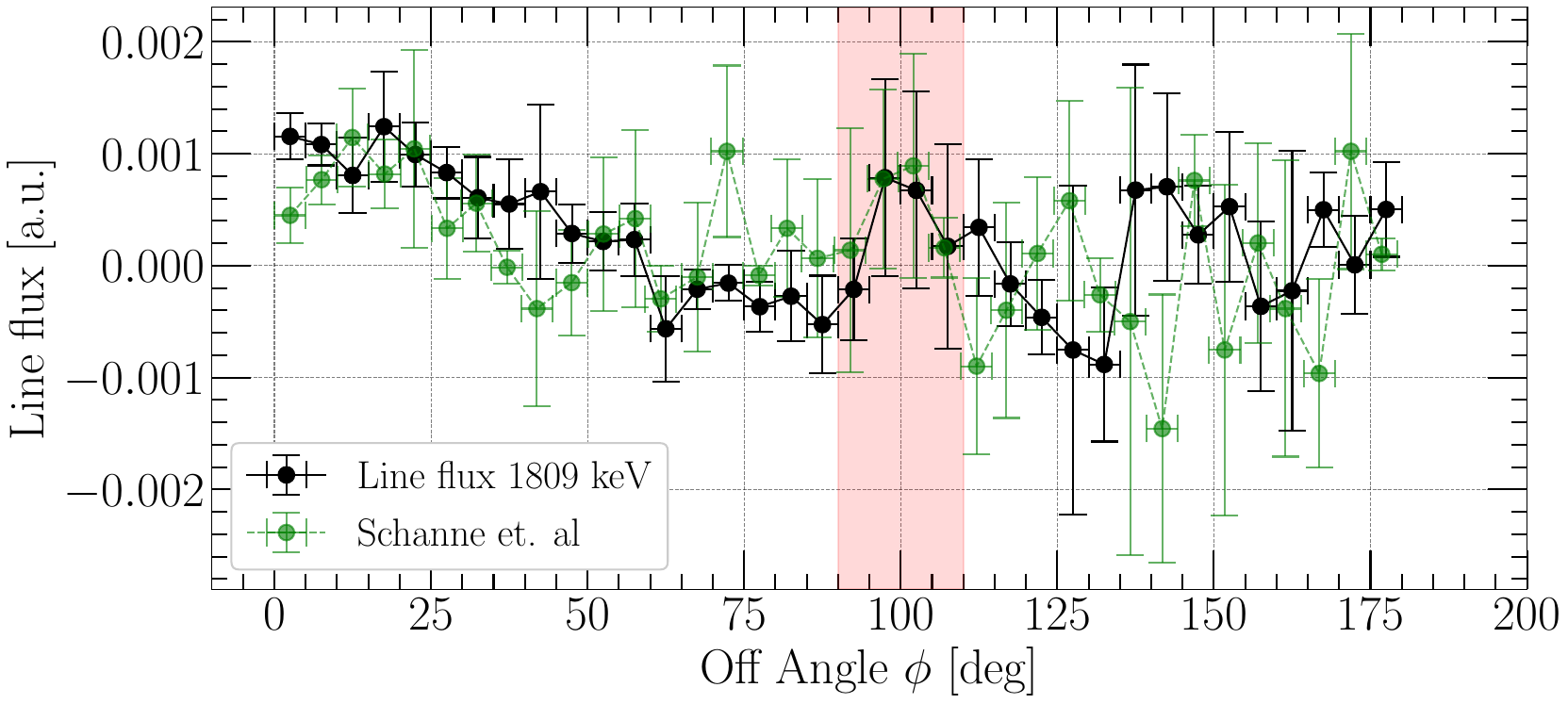}
    \centering
    \caption{Dependence of the background subtracted intensity of the $1809\,\text{keV}$ line of our analysis (black solid line) and properly scaled and background subtracted results of \citet{Schanne_2007} who also used SPI data (green dashed line). The red shaded area shows the Cygnus/Vela region which is known to have a high concentration of $^{26}\text{Al}$ and the mapping of which was the goal of the latter mentioned investigation. We observe almost perfect agreement in this region.}
     \label{fig:SPI_1809}
\end{figure}

The results for the $^{26}\text{Al}$ line profile obtained in this work are in good agreement with previous works, see Fig.~\ref{fig:SPI_1809}. In particular, we report on observations of the increased $^{26}\text{Al}$ line flux close to the GC and Cygnus/Vela regions at $\phi\sim 100^\circ$ in agreement with the previous \spi results~\citep[see e.g.][]{wang09,Bouchet_2015, Schanne_2007, jurgen04}. 

All remaining 9 detected lines with off-GC angle profile not consistent with a constant have known \spi instrumental line counterparts. Contrary to the lines of known astrophysical origin (511~keV and 1809~keV), the off-GC angle profiles for these are not consistent either with the ``DM decay profile'' model, Eq.~\ref{eq:off_gc_profile_model}. Still, for 8 of these lines the fit with the model (Eq.~\ref{eq:off_gc_profile_model}) results in statistically significant improvement of the  best-fit $\chi^2$. We argue that this could be an indication of either an underestimated level of systematic uncertainty for these lines, or a presence of the admixture of a signal that is of astrophysical/dark matter decay origin.

\subsection{Constraints on decaying DM parameters} \label{subsec:Constraints}
Within our analysis we derived constraints on the level of the residual flux in ON-OFF \spi observations. Using Eq.~\ref{eq:on_off_flux_gamma} we converted the derived constraints to constraints on the DM decay width $\Gamma_{\text{DM}}$ and its lifetime $\tau = 1/\Gamma_{\text{DM}}$ for $k_x=1$. The $3\sigma$ (statistical + systematic) constraints for the reference DM profile~\citep{Cautun_2020} on the lifetime are shown in Fig.~\ref{fig:lifetime_full_comparison} and compared to previous results from NuSTAR~\citep{Roach_2022}, \spi ~\citep{we_spi} and COMPTEL~\citep{2013JHEP...11..193E}.

The constraints on the sterile neutrino mixing angle $\sin^2(2\theta)$ are shown in Fig.~\ref{fig:mixing_full_comparison} for a reference DM density profile and compared to constraints from different instruments. In both Figures the shaded/hatched regions correspond to the full range of limits derived for DM density profiles listed in Tab.~\ref{tab:models}, i.e. to systematic uncertainties connected with poor knowledge of the DM distribution in the MW.

The derived constraints on the lifetime of the decaying DM and on the mixing angle of the sterile neutrino are by up to an order of magnitude better than the previously reported constraints~\citep{we_spi, laha20} and by an order of magnitude worse than ones reported in~\citet{calore22}. This difference could be explained by a different (agnostic) approach to the modelling of the \spi instrumental/astrophysical background contrary to a template-based approach of~\citet{calore22}. We argue, however, that the approach adopted in this paper allows a clear estimation of the systematic uncertainty connected to the \spi instrumental background variations which was not discussed in~\citet{calore22}.

The improvement in comparison to~\citet{we_spi} mainly originates from the higher DM column density considered in this work. In particular, conservative DM profile of the MW considered by~~\citet{we_spi} resulted in $S_{ON}-S_{OFF}\lesssim 10^{28}$~keV/cm$^2$ which is by more than an order of magnitude less than typical values for  $S_{ON}-S_{OFF}$ considered in this work, see Tab.~\ref{tab:S_ON-V}. 

On the other hand, we note that the level of systematic uncertainty derived in this work exceeds the level reported in~\citet{we_spi} by a factor of $\sim 3$ . This is connected to a larger data set considered in this work which, different from~\citet{we_spi}, covers time periods of strongly time-variable \spi instrumental background. Additional systematic uncertainty could arise from the different regions of the sky observed by \spi available to~\citet{we_spi}. The increased number of ON-GC observations (20~Ms current work vs. 12~Ms for~\citet{we_spi}) results in a somewhat increased number of Galactic sources of astrophysical emission in the FoV of \spi and consequently also impacts the level of systematics. 

We argue that controlling the level of systematics is a key issue for any type of DM searches with instrumental-background dominated instruments. Only with the next generation of instruments like eXTP, THESEUS, AMEGO or GECCO~\citep{we_extp,we_ths,ray21,coogan21} this issue may be resolved.

\section*{Acknowledgements}
The authors acknowledge support by the state of Baden-W\"urttemberg through bwHPC. DM work was supported by DLR through grant 50OR2104 and by DFG through the
grant MA 7807/2-1. The presented work is based on observations with INTEGRAL, an ESA project with instruments and a science data center funded by ESA member states.
\section*{Data Availability}
\begin{figure}
    \centering
    \includegraphics[clip=true,width=1\columnwidth]{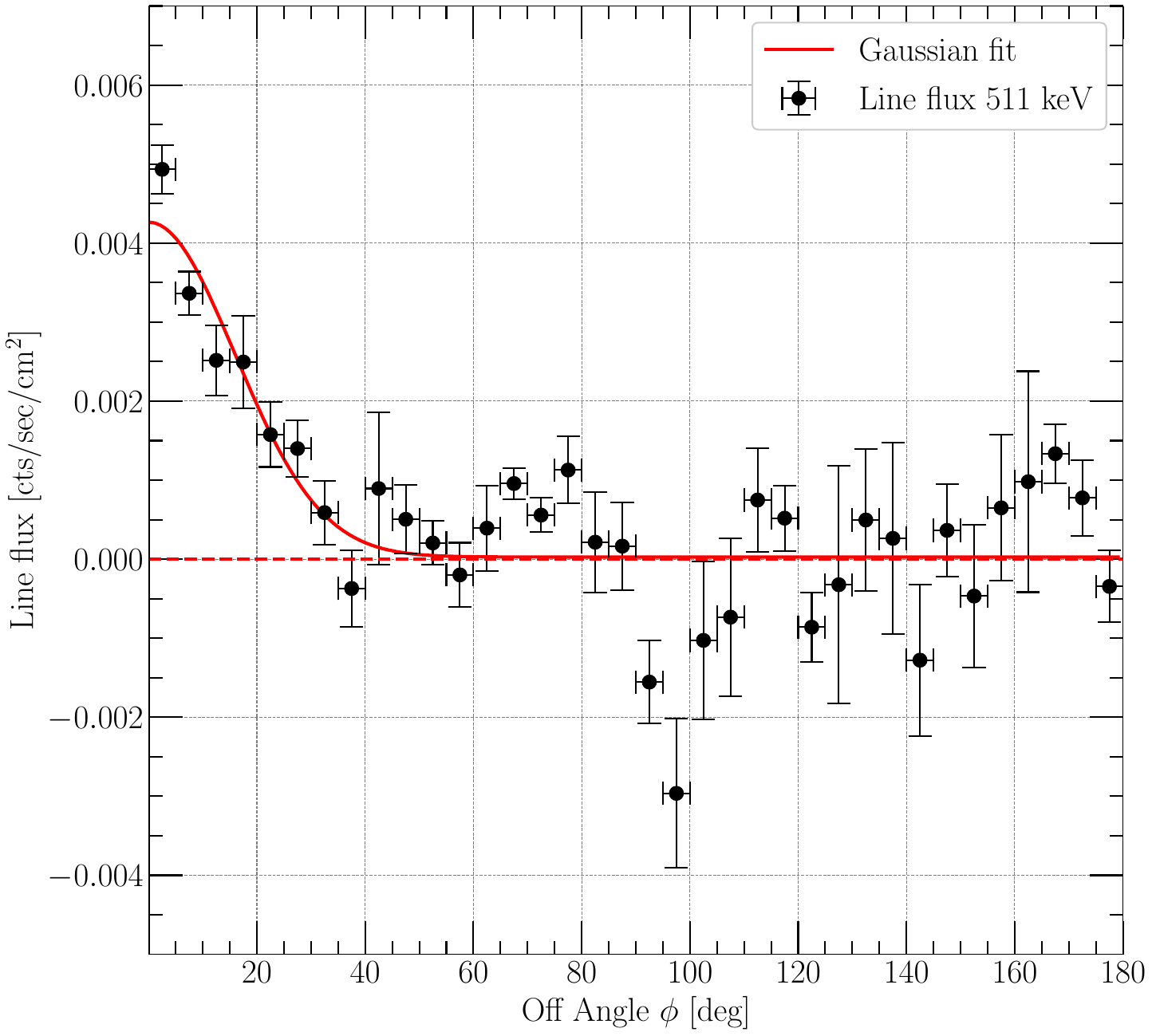}
    \centering
    \caption{Dependence of the background subtracted intensity of the $511\,\text{keV}$ positron annihilation line on the off-angle $\phi$ from the GC. Error bars are $1\,\sigma$ and the fit is Gaussian of the form $\text{const}+Ne^{-\phi^2/(2\rho^2)}$.}
     \label{fig:511_off_angle}
\end{figure}

\begin{figure*}
  \centering
  \begin{center}
  \includegraphics[clip=true,width=1\linewidth]{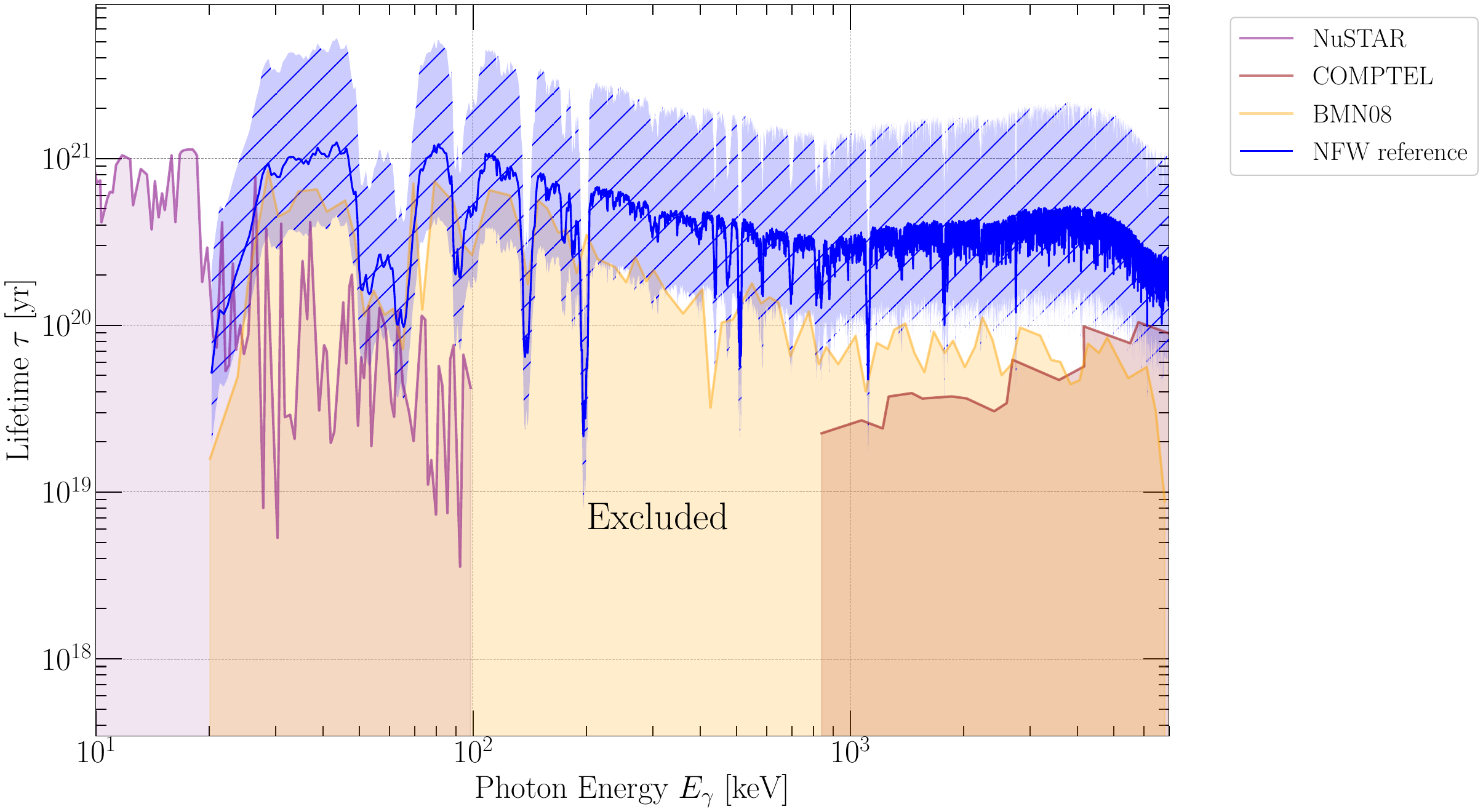}
  \caption{Complete comparison of lower limits on the lifetime $\tau$ of any unstable DM particle decaying into one photon to recent results. The region \textit{below} each curve is excluded. The blue hatched region shows the uncertainty due to different DM profiles with lowest constraints for the COM and highest for the ISO model (see Tab.~\ref{tab:models}). The blue solid line represents constraints based on NFW-parameters of \citet{Cautun_2020} for reference. Model-specific constraints for different DM profiles or parameters can be obtained by scaling the reference appropriately by $(S_{\text{ON}}-S_{\text{OFF}})/(S_{\text{ON,REF}}-S_{\text{OFF,REF}})$ listed in Tab.~\ref{tab:S_ON-V}. The other constraints shown were obtained by data from NuSTAR \citep[maximum of][]{2019PhRvD..99h3005N,2020PhRvD.101j3011R,Roach_2022}, \spi \citep{we_spi} and COMPTEL \citep[][scaled by a factor of 1/2 due to a decay into 1 photon, see Eq.~\ref{eq:gamma_DM} with $k_x=1$]{2013JHEP...11..193E}.} 
  \label{fig:lifetime_full_comparison}
  \end{center}
\end{figure*}

\begin{figure*}
  \centering
  \begin{center}
  \includegraphics[clip=true,width=1\linewidth]{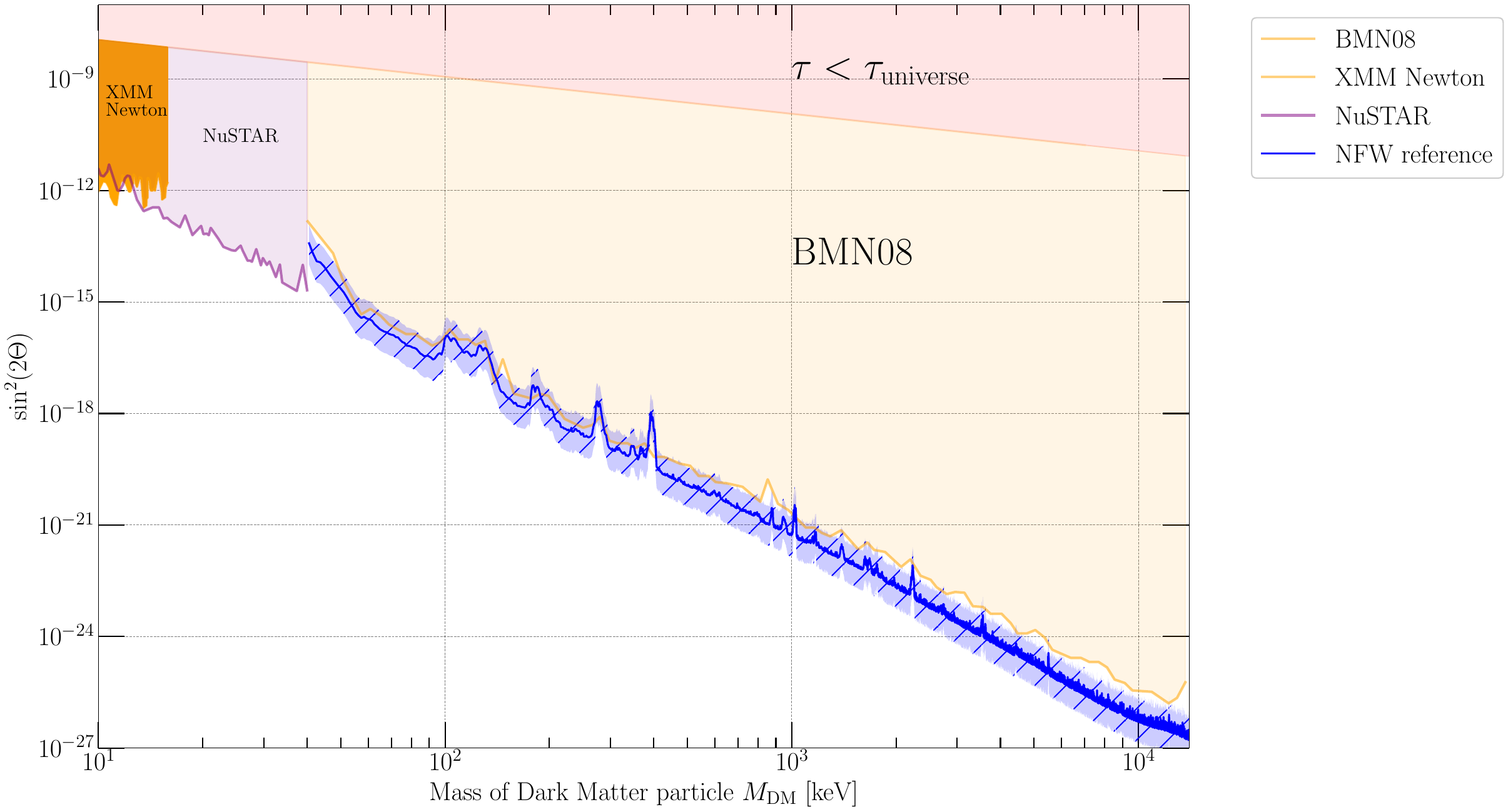}
  \caption{Complete comparison of upper limits on the mixing angle $\theta$ of the sterile neutrino decaying into two bodies, one of which is a photon.
  The region \textit{above} each curve is excluded. The blue hatched region shows the uncertainty due to different DM profiles with highest constraints for the COM and lowest for the ISO model (see Tab.~\ref{tab:models}). The blue solid line represents constraints based on NFW-parameters of \citet{Cautun_2020} for reference. Model-specific constraints for different DM profiles or parameters can be obtained by scaling the reference appropriately by $(S_{\text{ON}}-S_{\text{OFF}})/(S_{\text{ON,REF}}-S_{\text{OFF,REF}})$ listed in Tab.~\ref{tab:S_ON-V}. The other constraints shown are based on XMM-Newton \citep{Dessert_2020,Foster_2021}, INTEGRAL (called BMN08) \citep{we_spi} and NuSTAR \citep[maximum of][]{2016PhRvD..94l3504N,Perez_2017,2019PhRvD..99h3005N,2020PhRvD.101j3011R,Roach_2022}. The constraints $\tau<\tau_{\text{universe}}$ \citep{Fernandez_Martinez_2021} are of fundamental nature.}
  \label{fig:mixing_full_comparison}
  
  \end{center}
\end{figure*}

\begin{center}
\begin{figure*}
\captionsetup[subfigure]{labelformat=empty}

\begin{tabular}{cccc}

\subfloat[$910.86\,\text{keV}$]{\includegraphics[width = 1.7in]{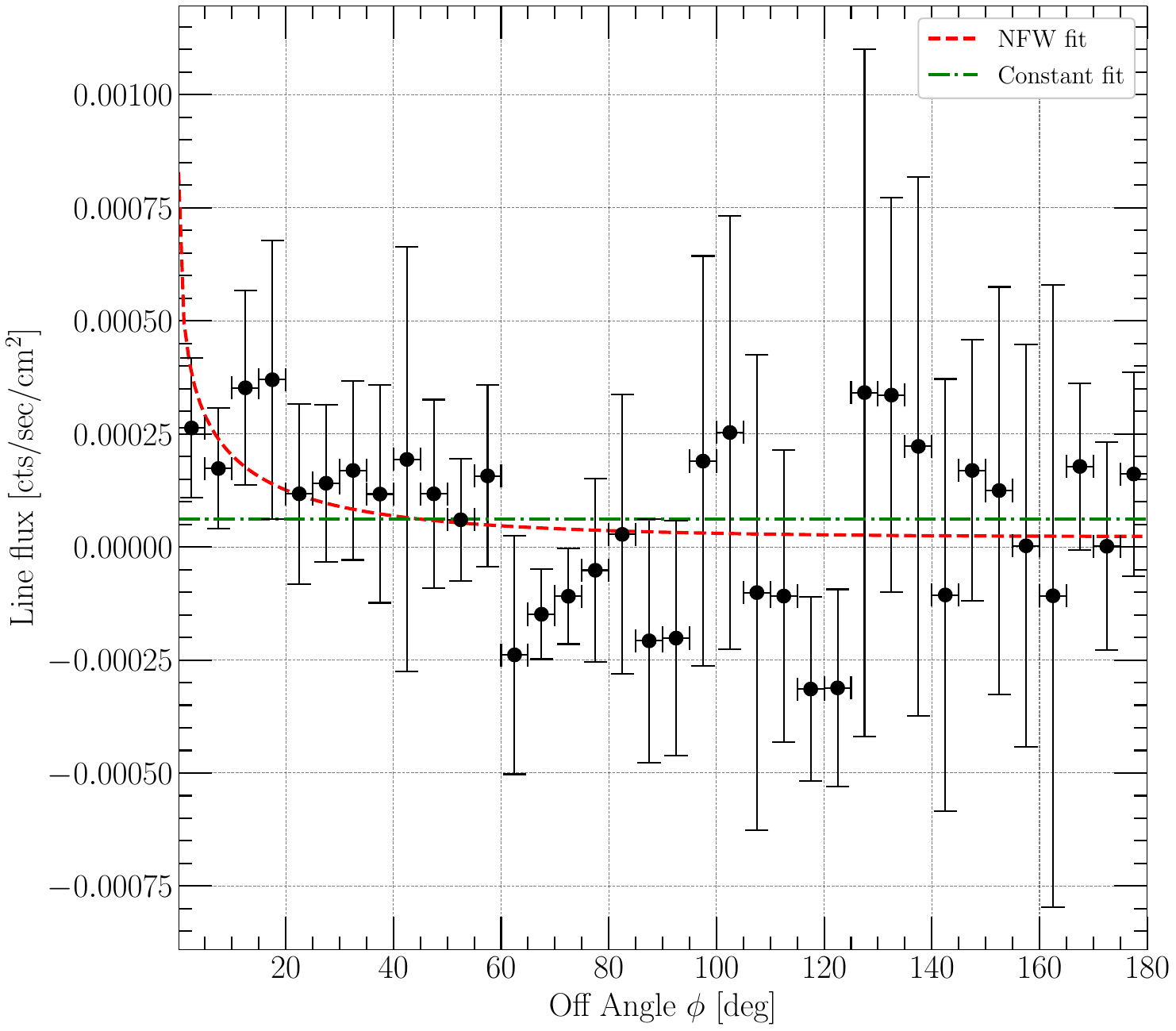}} &
\subfloat[$1105.47\,\text{keV}$]{\includegraphics[width = 1.7in]{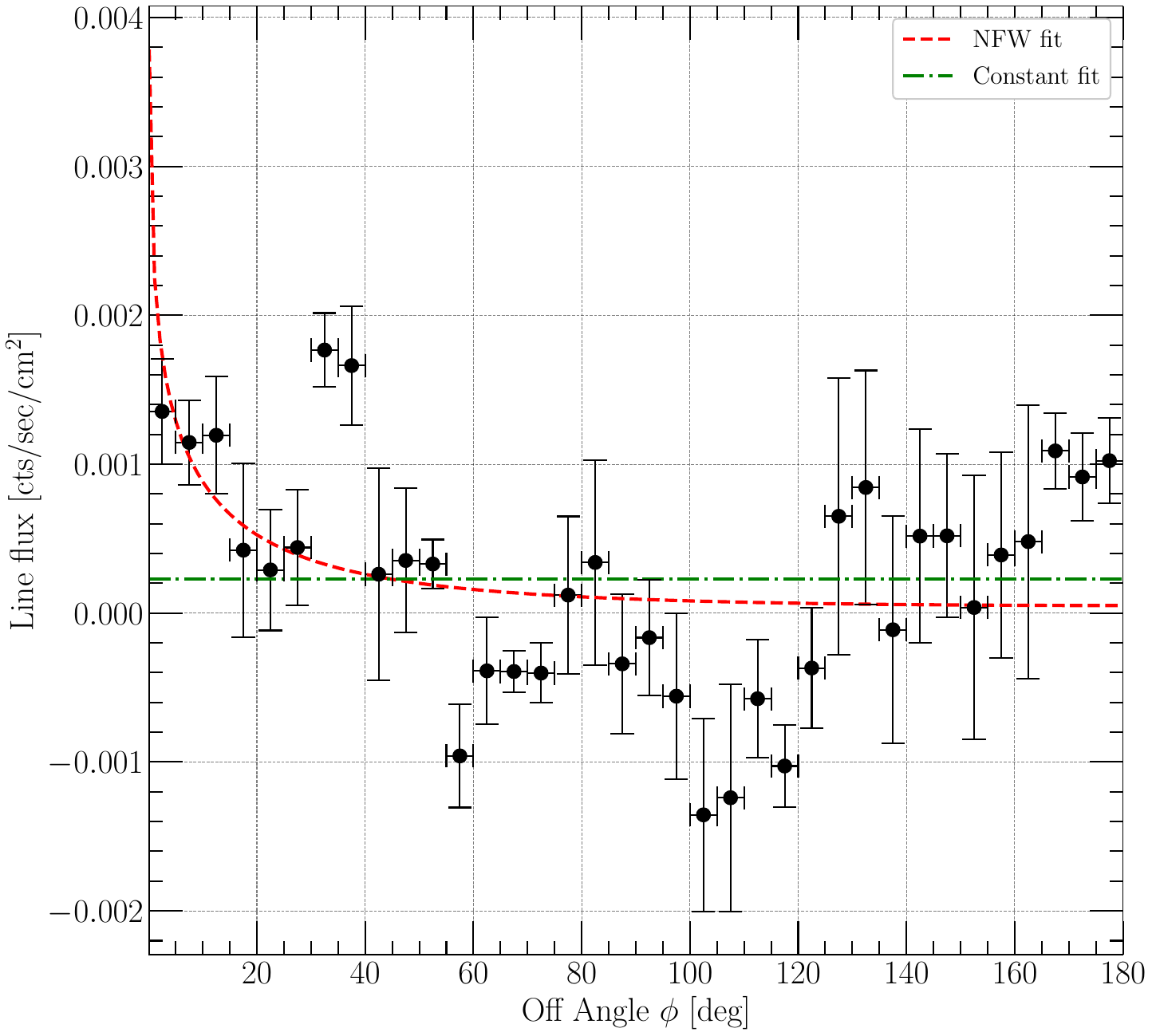}} &
\subfloat[$1114.9\,\text{keV}$]{\includegraphics[width = 1.7in]{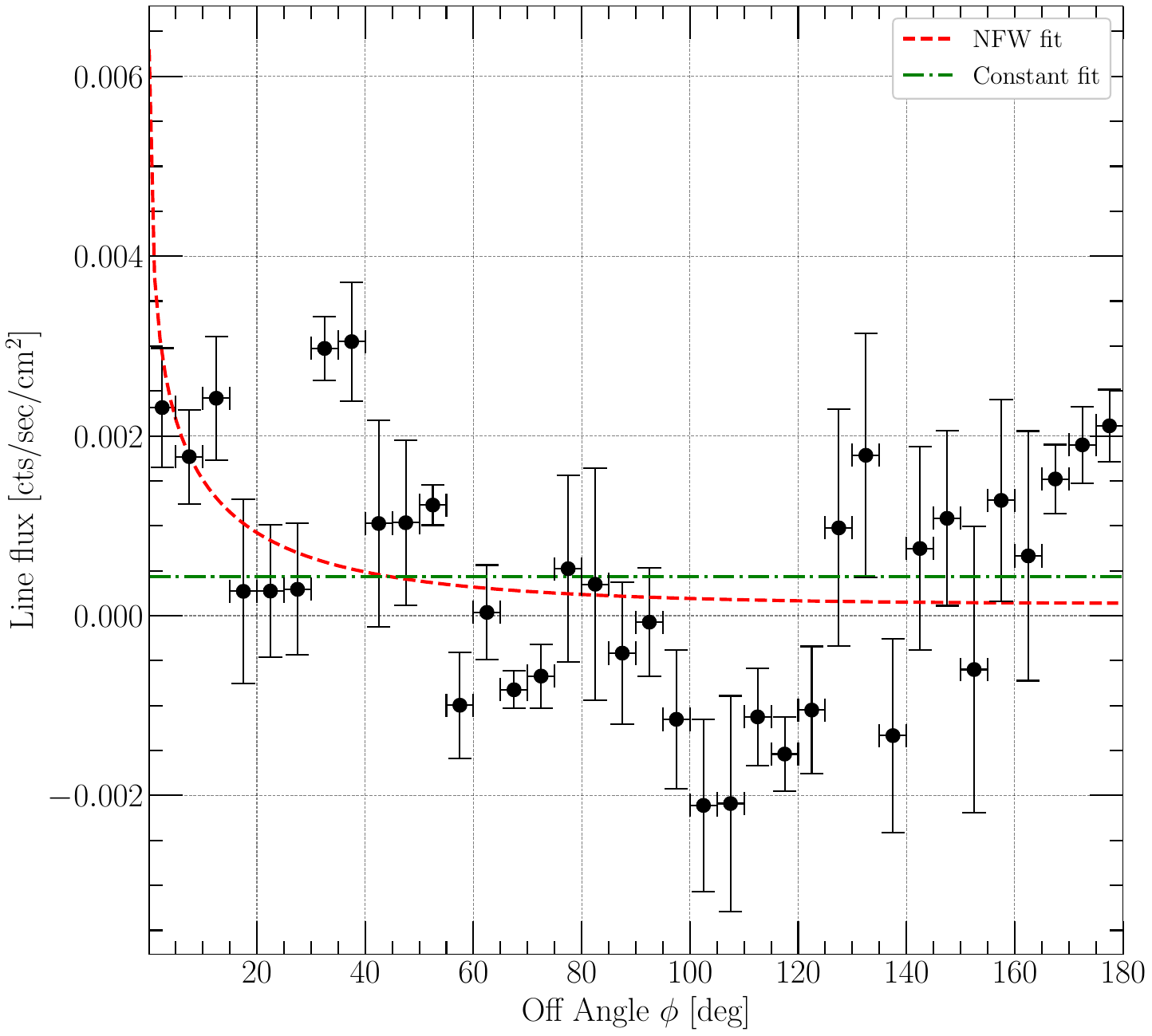}} &
\subfloat[$1135.43\,\text{keV}$]{\includegraphics[width = 1.7in]{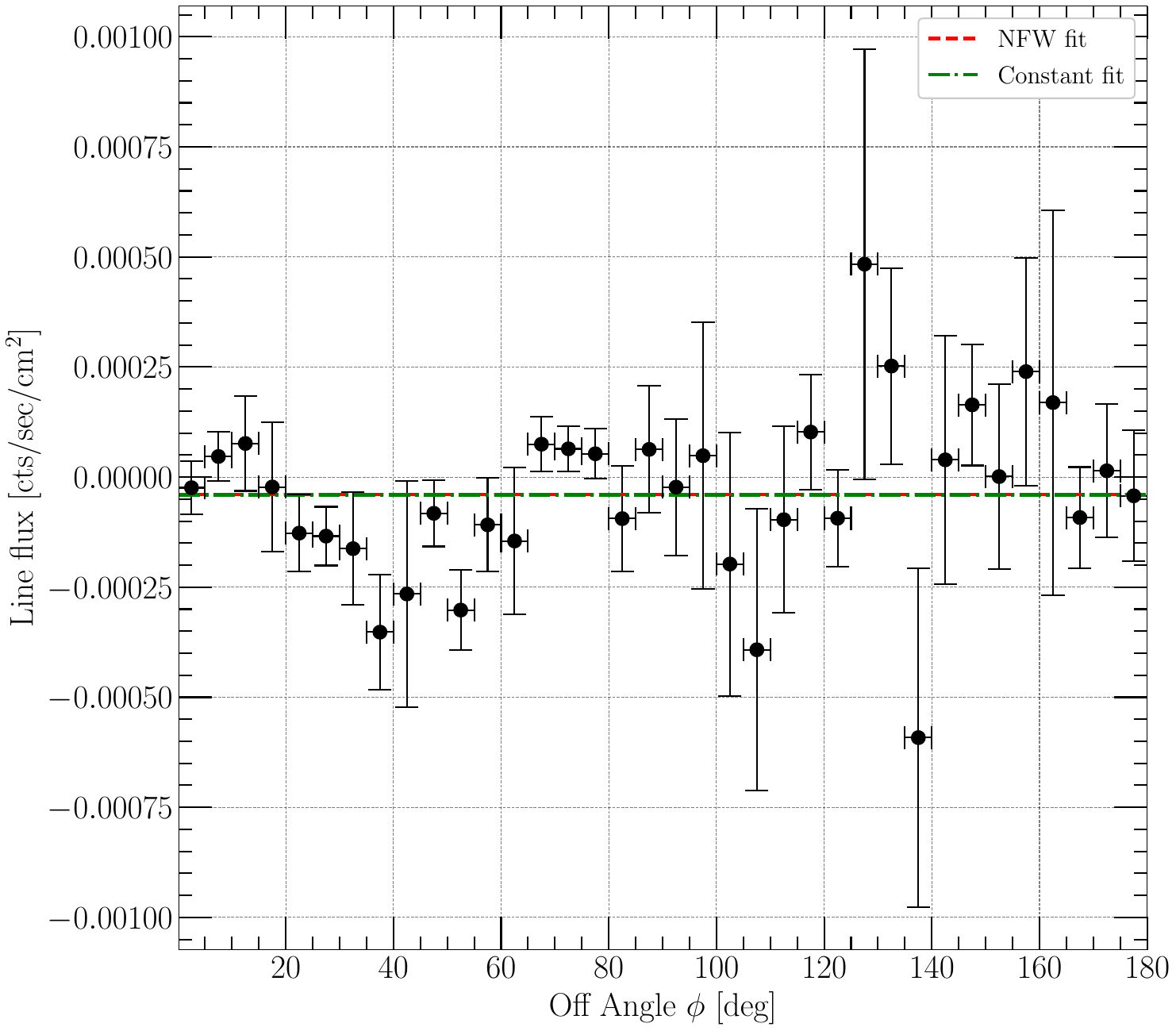}} \\
\subfloat[$1172.09\,\text{keV}$]{\includegraphics[width = 1.7in]{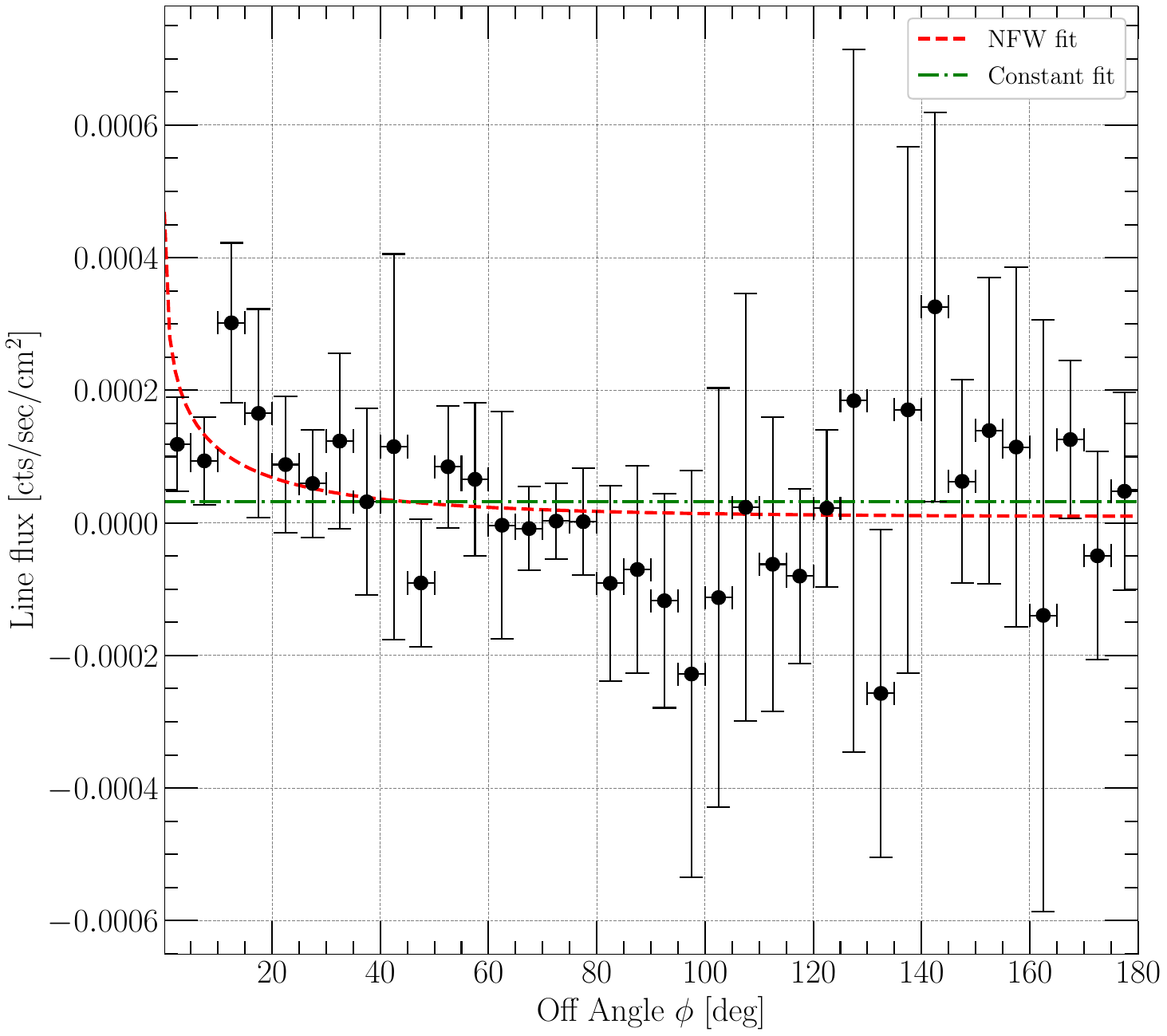}} &
\subfloat[$1320.61\,\text{keV}$]{\includegraphics[width = 1.7in]{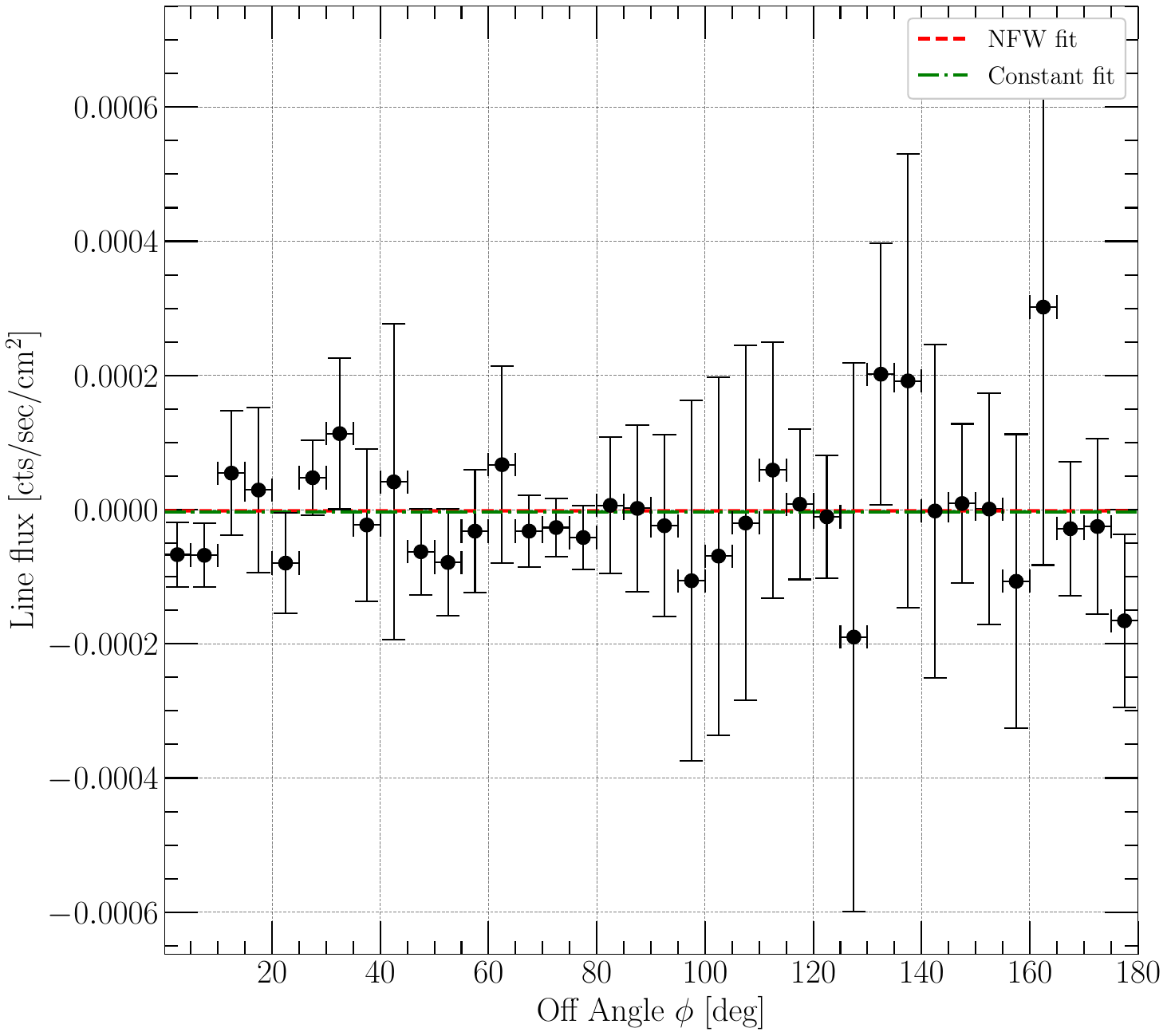}} &
\subfloat[$1344.84\,\text{keV}$]{\includegraphics[width = 1.7in]{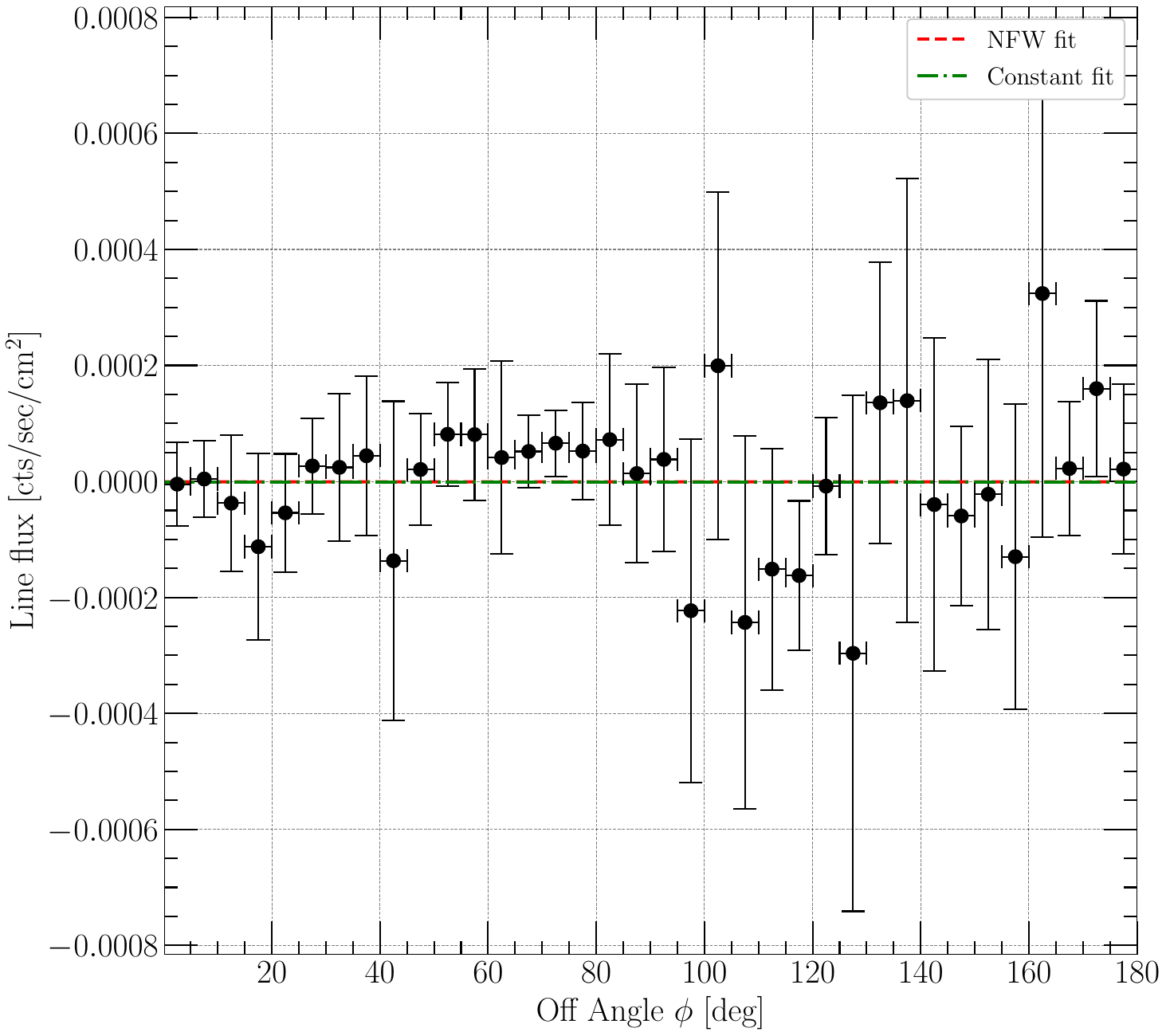}} &
\subfloat[$1355.76\,\text{keV}$]{\includegraphics[width = 1.7in]{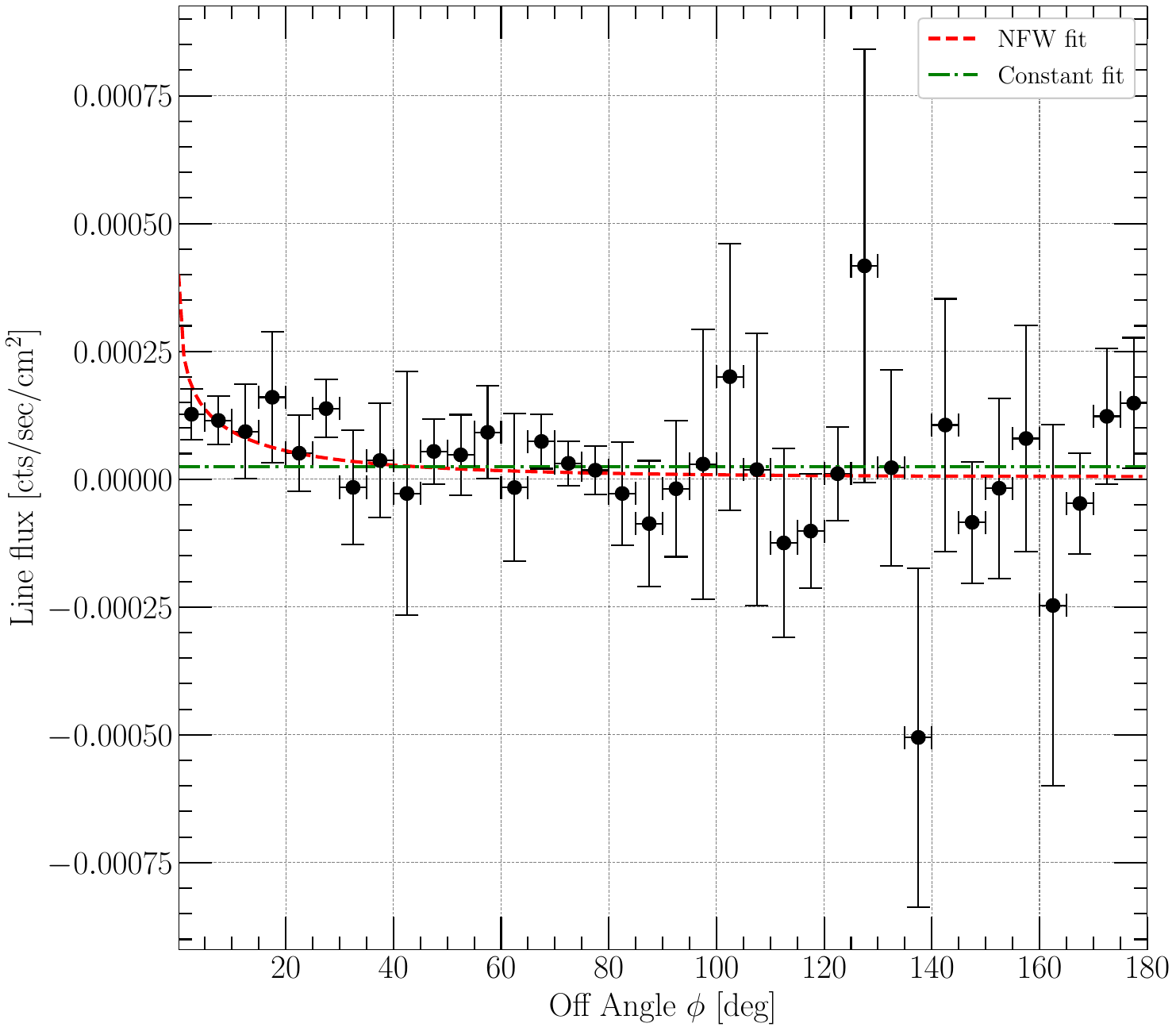}} \\
\subfloat[$1368.34\,\text{keV}$]{\includegraphics[width = 1.7in]{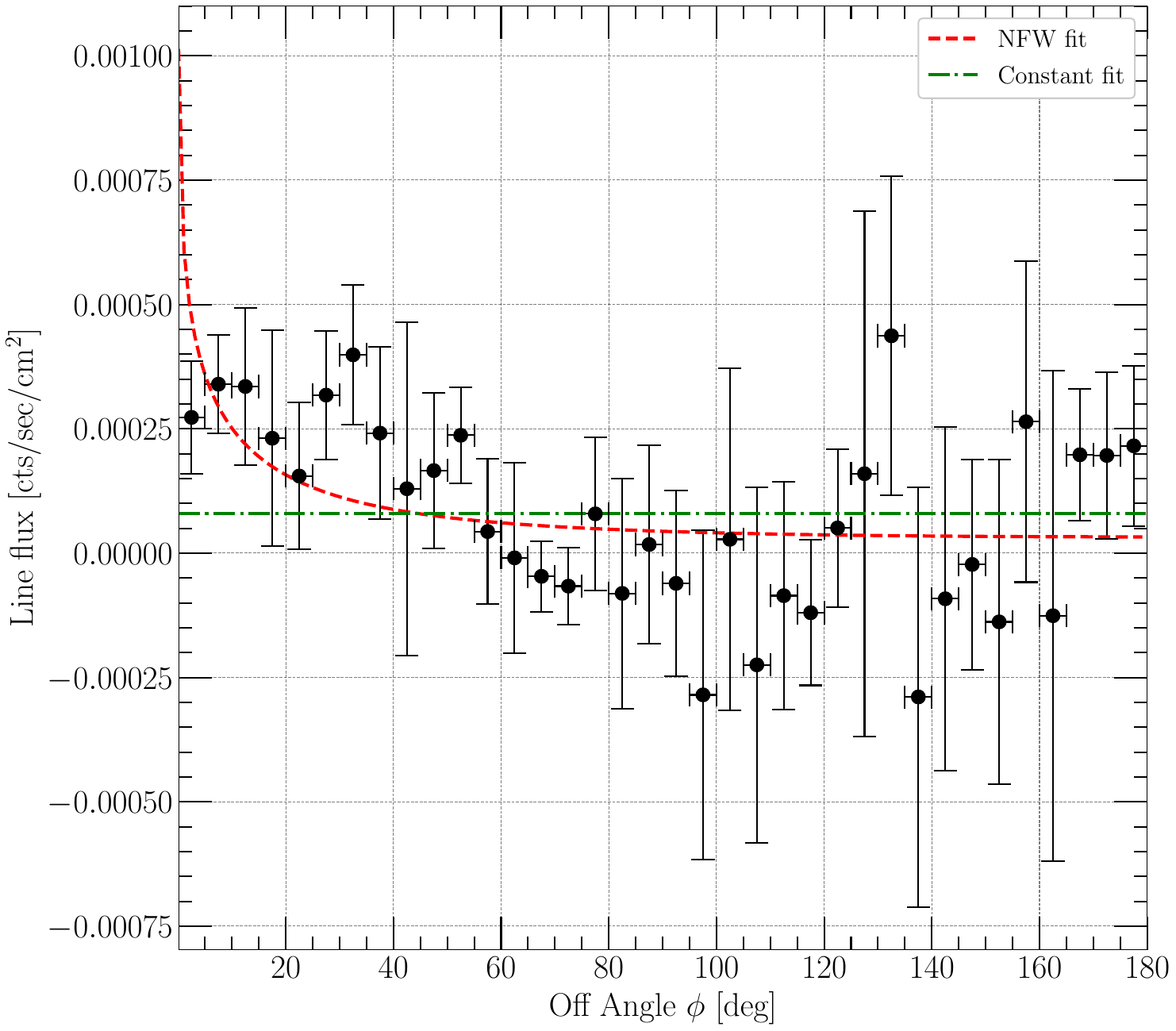}} &
\subfloat[$1471.82\,\text{keV}$]{\includegraphics[width = 1.7in]{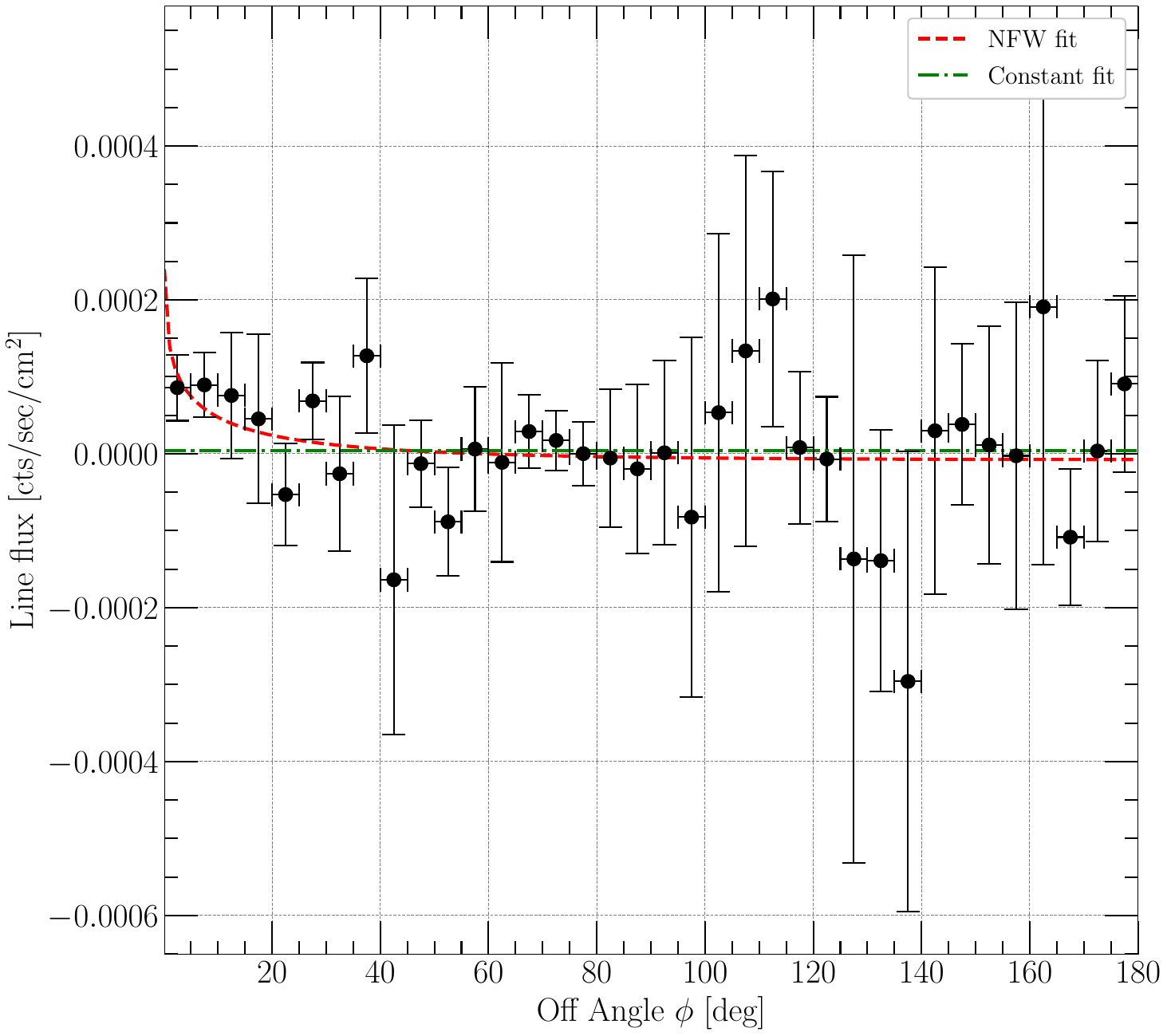}} &
\subfloat[$1809\,\text{keV}$]{\includegraphics[width = 1.7in]{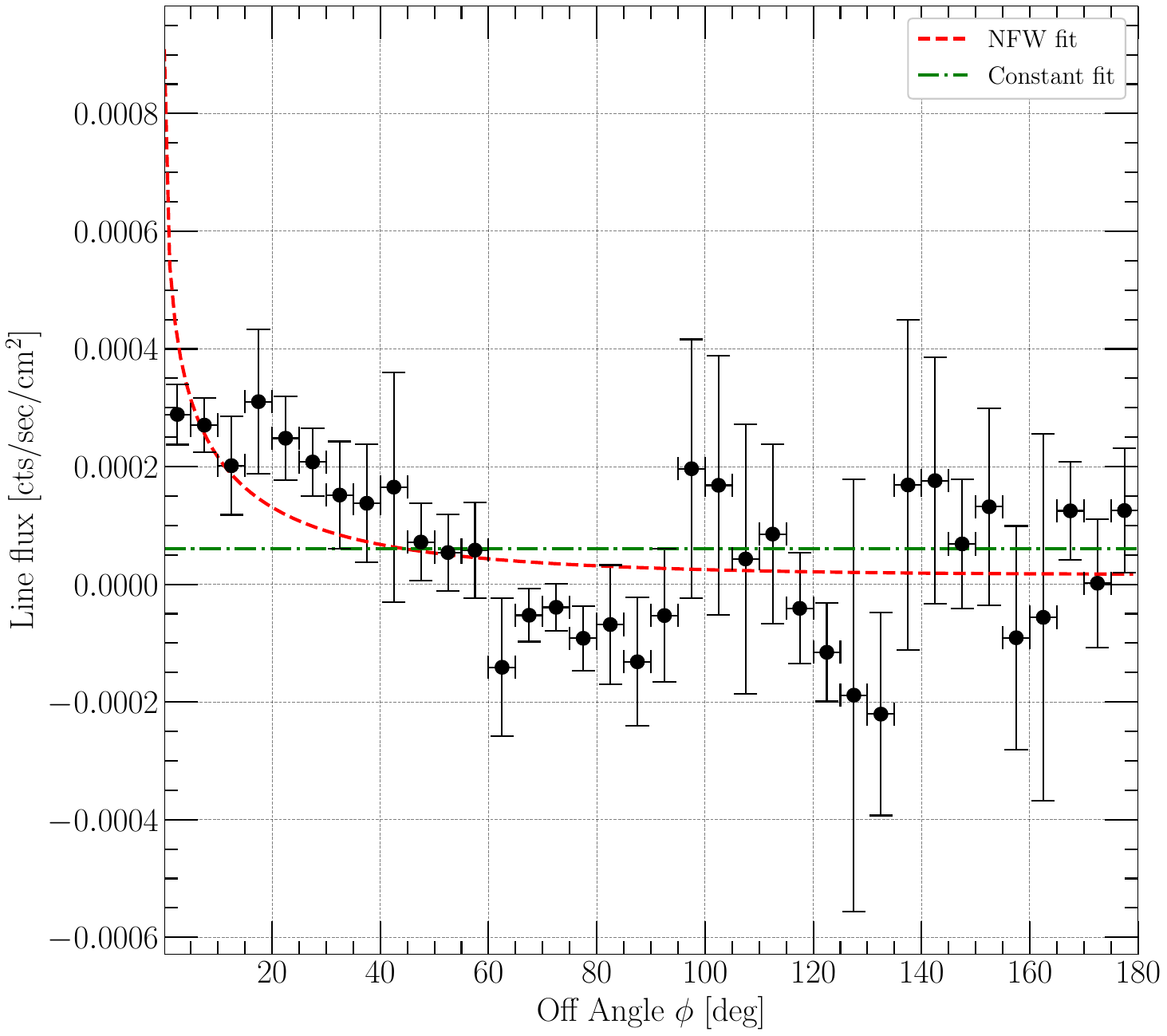}} &
\subfloat[$2403.35\,\text{keV}$]{\includegraphics[width = 1.7in]{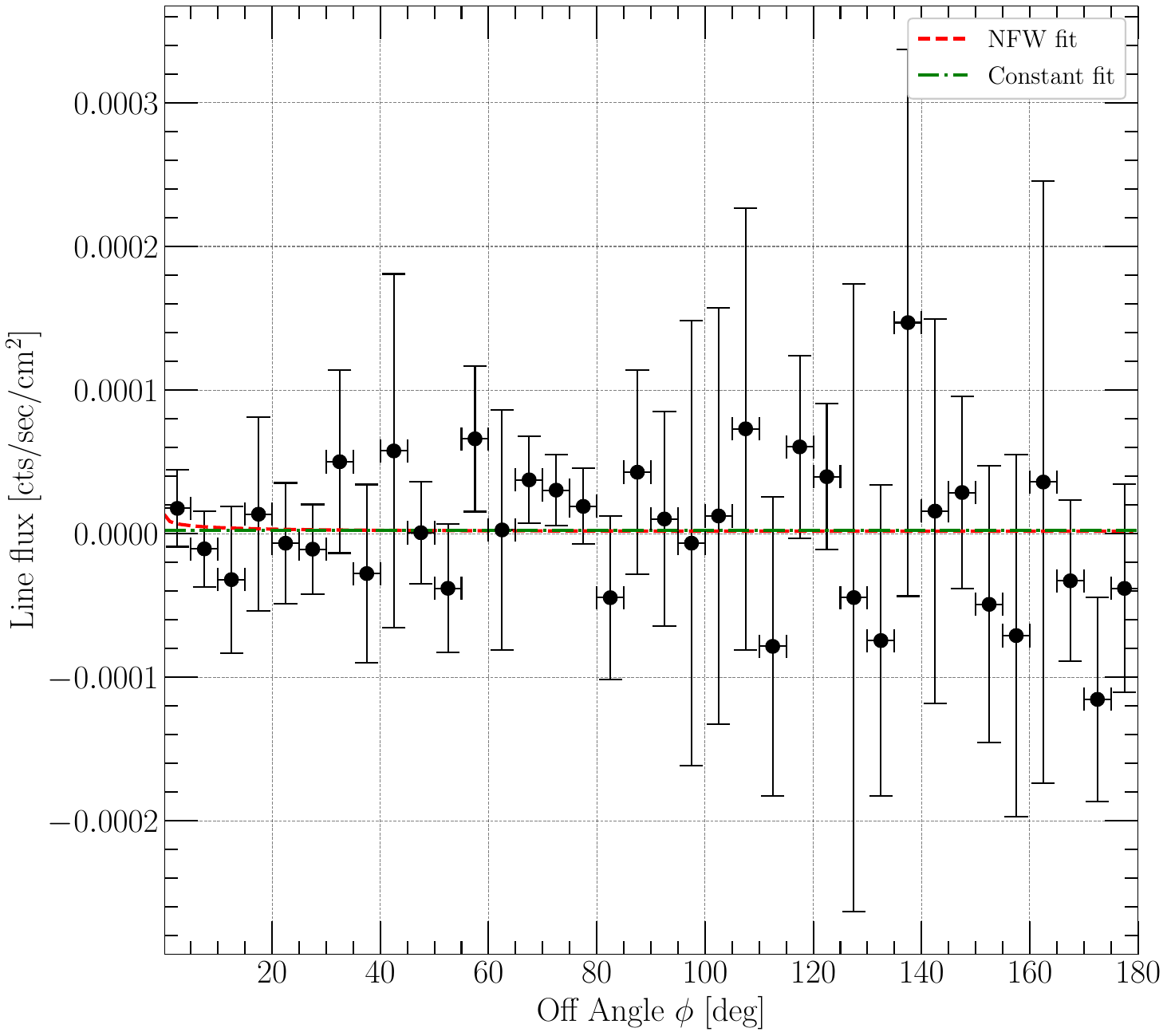}} \\
\subfloat[$2749.42\,\text{keV}$]{\includegraphics[width = 1.7in]{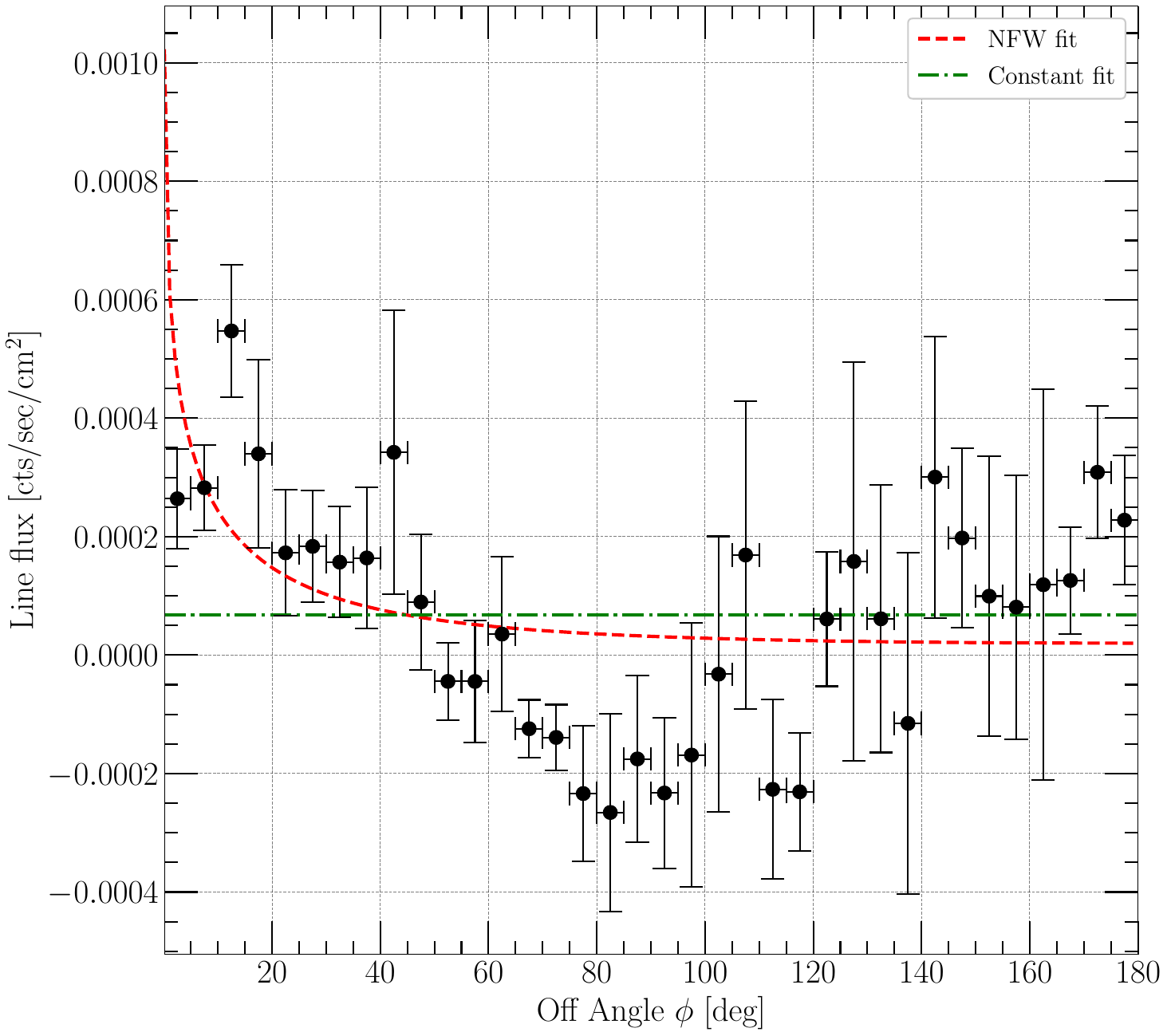}}

\end{tabular}

\caption{Off-angle profiles of all candidates detected and presented in Tab.~\ref{tab:line_results_test}. The black dots with error bars show the background subtracted line fluxes, in green dashed lines the best fit constant and in red with solid lines the best NFW fit. Note that the flux in $198\,\text{keV}$ is constant since we used it as a normalization tracer and $138\,\text{keV}$ is constant within $1\sigma$ (for details see section \ref{subsec:Off_Angle_analysis}). $511\,\text{keV}$ follows the expected shape with a drop in intensity with the galactic off-angle $\phi$ (see section \ref{subsec:line_candidates}, in particular Fig.~\ref{fig:511_off_angle}).}
\label{fig:offangle_lines}
\end{figure*}
\end{center}

The data for selected figures are publicly available on GitHub\footnote{See \href{https://github.com/fischers98/Twenty-years-of-Dark-Matter-Searches-with-INTEGRAL-SPI}{GitHub} of the author.}, namely for Fig.~\ref{fig:EffArea}, Fig.~\ref{fig:511_off_angle}, as well as constraints in Fig.~\ref{fig:lifetime_full_comparison} and Fig.~\ref{fig:mixing_full_comparison} for all models and parameters listed in Tab.~\ref{tab:S_ON-V}.\\ 

\noindent Further data underlying this article will be shared on reasonable request to the corresponding author.\\


\bibliographystyle{mn2e}
\def\aj{AJ}%
\def\actaa{Acta Astron.}%
\def\araa{ARA\&A}%
\def\apj{ApJ}%
\def\apjl{ApJL}%
\def\apjs{ApJS}%
\def\ao{Appl.~Opt.}%
\def\apss{Ap\&SS}%
\def\aap{A\&A}%
\def\aapr{A\&A~Rev.}%
\def\aaps{A\&AS}%
\def\azh{AZh}%
\def\baas{BAAS}%
\def\bac{Bull. astr. Inst. Czechosl.}%
\def\caa{Chinese Astron. Astrophys.}%
\def\cjaa{Chinese J. Astron. Astrophys.}%
\def\icarus{Icarus}%
\def\jcap{J. Cosmology Astropart. Phys.}%
\def\jrasc{JRASC}%
\def\mnras{MNRAS}%
\def\memras{MmRAS}%
\def\na{New A}%
\def\nar{New A Rev.}%
\def\pasa{PASA}%
\def\pra{Phys.~Rev.~A}%
\def\prb{Phys.~Rev.~B}%
\def\prc{Phys.~Rev.~C}%
\def\prd{Phys.~Rev.~D}%
\def\pre{Phys.~Rev.~E}%
\def\prl{Phys.~Rev.~Lett.}%
\def\pasp{PASP}%
\def\pasj{PASJ}%
\def\qjras{QJRAS}%
\def\rmxaa{Rev. Mexicana Astron. Astrofis.}%
\def\skytel{S\&T}%
\def\solphys{Sol.~Phys.}%
\def\sovast{Soviet~Ast.}%
\def\ssr{Space~Sci.~Rev.}%
\def\zap{ZAp}%
\def\nat{Nature}%
\def\iaucirc{IAU~Circ.}%
\def\aplett{Astrophys.~Lett.}%
\def\apspr{Astrophys.~Space~Phys.~Res.}%
\def\bain{Bull.~Astron.~Inst.~Netherlands}%
\def\fcp{Fund.~Cosmic~Phys.}%
\def\gca{Geochim.~Cosmochim.~Acta}%
\def\grl{Geophys.~Res.~Lett.}%
\def\jcp{J.~Chem.~Phys.}%
\def\jgr{J.~Geophys.~Res.}%
\def\jqsrt{J.~Quant.~Spec.~Radiat.~Transf.}%
\def\memsai{Mem.~Soc.~Astron.~Italiana}%
\def\nphysa{Nucl.~Phys.~A}%
\def\physrep{Phys.~Rep.}%
\def\physscr{Phys.~Scr}%
\def\planss{Planet.~Space~Sci.}%
\def\procspie{Proc.~SPIE}%
\let\astap=\aap
\let\apjlett=\apjl
\let\apjsupp=\apjs
\let\applopt=\ao
\bibliography{bibliography}

\newpage

\def\arraystretch{1}
\setlength\tabcolsep{2pt}

\def\arraystretch{1}
\setlength\tabcolsep{2pt}

\newpage

\def\arraystretch{1}
\begin{table*}
     \centering
     \begin{tabular}{|c||c|c|c||c|c|c||c|}
   \hline
  
   \textbf{E} [keV] & $\chi^2_{\text{Const}}$ & Red.$\chi^2_{\text{Const}}$ & $P_{\text{Const}}$ [\%] & $\chi^2_{\text{NFW}}$ & Red. $\chi^2_{\text{NFW}}$ & $P_{\text{NFW}}$ [\%] & $\Delta \chi^2$\\
   \hline \hline
   
   22.4&93.93&2.68&0.0&93.93&2.76&0.0&0\\
   \hline 47.8 & 42.61&1.22&17.64&37.38&1.1&31.66&5.23\\
   \hline53.1&53.06&1.52&2.58&52.01&1.53&2.48&1.05\\
   \rowcolor{lightgray}\hline60.6&989.88&28.28&0.0&965.06&28.38&0.0&24.82\\
   \hline65.98& 44.41&1.27&13.23&44.4&1.31&10.93&0.01\\
   \rowcolor{lightgray}\hline86.57& 555.41&15.87&0.0&368.88&10.85&0.0&186.53\\
   \hline92.57& 60.19&1.72&0.51&58.41&1.72&0.57&1.78\\
   \hline 101.88& 10.54&0.3&100.0&10.54&0.31&100.0&0\\
   \hline140.11& 19.68&0.56&98.28&19.68&0.58&97.32&0\\
   \hline175.07& 36.68&1.05&39.07&36.25&1.07&36.41&0.43\\
   \hline185.9&15.42&0.44&99.83&15.41&0.45&99.74&0.01\\
   \rowcolor{lightgray}\hline193.34&356.57&10.19&0.0&246.81&7.26&0.0&109.76\\
   \hline438.09&27.08&0.77&82.83&27.04&0.8&79.6&0.04\\
   \rowcolor{gray}\hline511.0$^{*}$&92.81&2.65&0.0&44.35&1.3&11.01&48.46\\
   \hline583.63&55.7&1.59&1.452&36.82&1.08&33.97&18.88\\
   \hline699.253&12.51&0.36&99.98&10.94&0.32&99.99&1.57\\
   \rowcolor{lightgray}\hline810.82&127.33&3.64&0.0&101.44&2.98&0.0&25.89\\
   \rowcolor{lightgray}\hline835.36&128.47&3.67&0.0&110.69&3.26&0.0&17.78\\
   \rowcolor{lightgray}\hline880.56&101.88&2.91&0.0&72.53&2.13&0.01&29.35\\
   \hline910.86&25.58&0.73&87.79&19.52&0.57&97.78&6.06\\
   \rowcolor{lightgray}\hline1105.47&190.43&5.44&0.0&152.76&4.49&0.0&37.67\\
   \rowcolor{lightgray}\hline1114.9&251.52&7.19&0.0&219.56&6.46&0.0&31.96\\
   \hline1135.43&46.58&1.33&9.13&46.5&1.37&7.48&0.08\\
   \hline1172.09&19.72&0.56&98.25&15.73&0.46&99.68&3.99\\
   \hline1320.61$^{\dagger}$&14.9&0.43&99.88&14.9&0.44&99.76&0\\
   \hline1344.84&12.52&0.36&99.98&12.4&0.36&99.97&0.12\\
   \hline1355.76&26.63&0.76&84.41&17.47&0.51&99.15&9.16\\
   \hline1368.34&43.42&1.24&15.54&29.86&0.88&67.07&13.56\\
   \hline1471.82$^{\dagger}$&21.97&0.63&95.77&14.34&0.42&99.88&7.63\\
   \rowcolor{gray}\hline1809.0$^{*}$&103.48&2.96&0.0&46.22&1.36&7.88&57.26\\
   \hline2403.35$^{\dagger}$&16.32&0.47&99.7&16.32&0.48&99.54&0\\
   \rowcolor{lightgray}\hline2749.42&117.88&3.37&0.0&85.62&2.52&0.0&32.26\\
  
       \hline

     \hline

 \end{tabular}

     \caption{Reduced $\chi^2$ of a constant (NDF$=34$) and NFW (NDF$=35$) fit of the form $a\cdot S_{\text{NFW}}+b$, where  $S$ is the DM surface brightness (see Eq.~\ref{eq:Sdm}) and constants $a,b$ are corresponding to scaling $\rho_0$ (see Tab.~\ref{tab:models}) and a displacement. Lines with a star $^{*}$ are prominent astrophysical lines and the ones with a dagger $^{\dagger}$ represent unidentified lines (see Tab.~\ref{tab:lines}). Highlighted lines have $\Delta \chi^2>0$ - Dark grey: lines with $P_{\text{Const}}=0$ and $P_{\text{NFW}}>5\%$ - light grey: lines with $P_{\text{Const}}=0$ only.} 
     \label{tab:line_results_test}
 \end{table*}

\newpage

\def\arraystretch{2}
\begin{center}
\label{fig:lines_profiles}
\begin{figure*}
\captionsetup[subfigure]{labelformat=empty}
\begin{tabular}{cccc}
\subfloat[$22.4\,\text{keV}$]{\includegraphics[width = 1.7in]{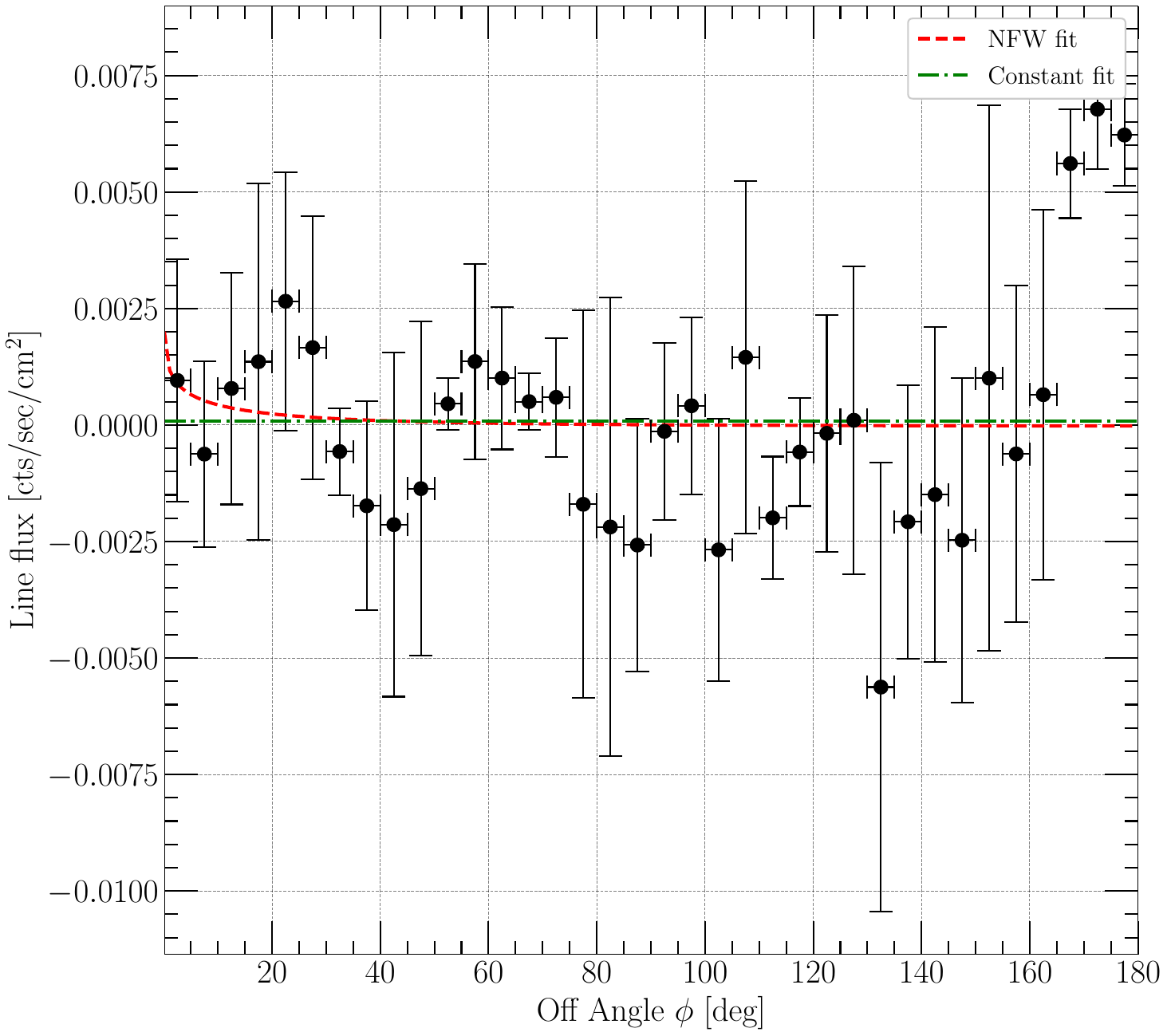}} &
\subfloat[$47.8\,\text{keV}$]{\includegraphics[width = 1.7in]{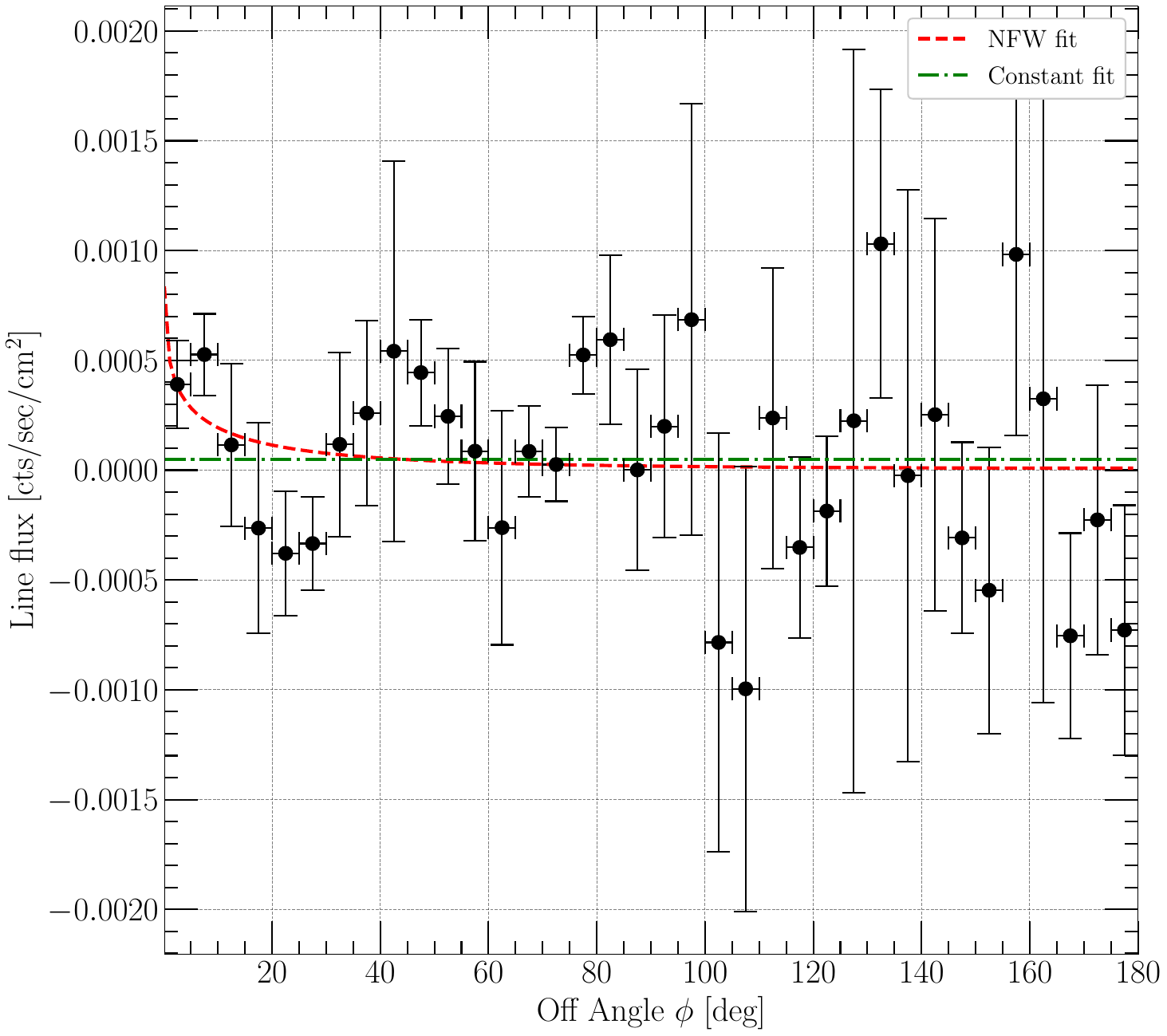}} &
\subfloat[$53.1\,\text{keV}$]{\includegraphics[width = 1.7in]{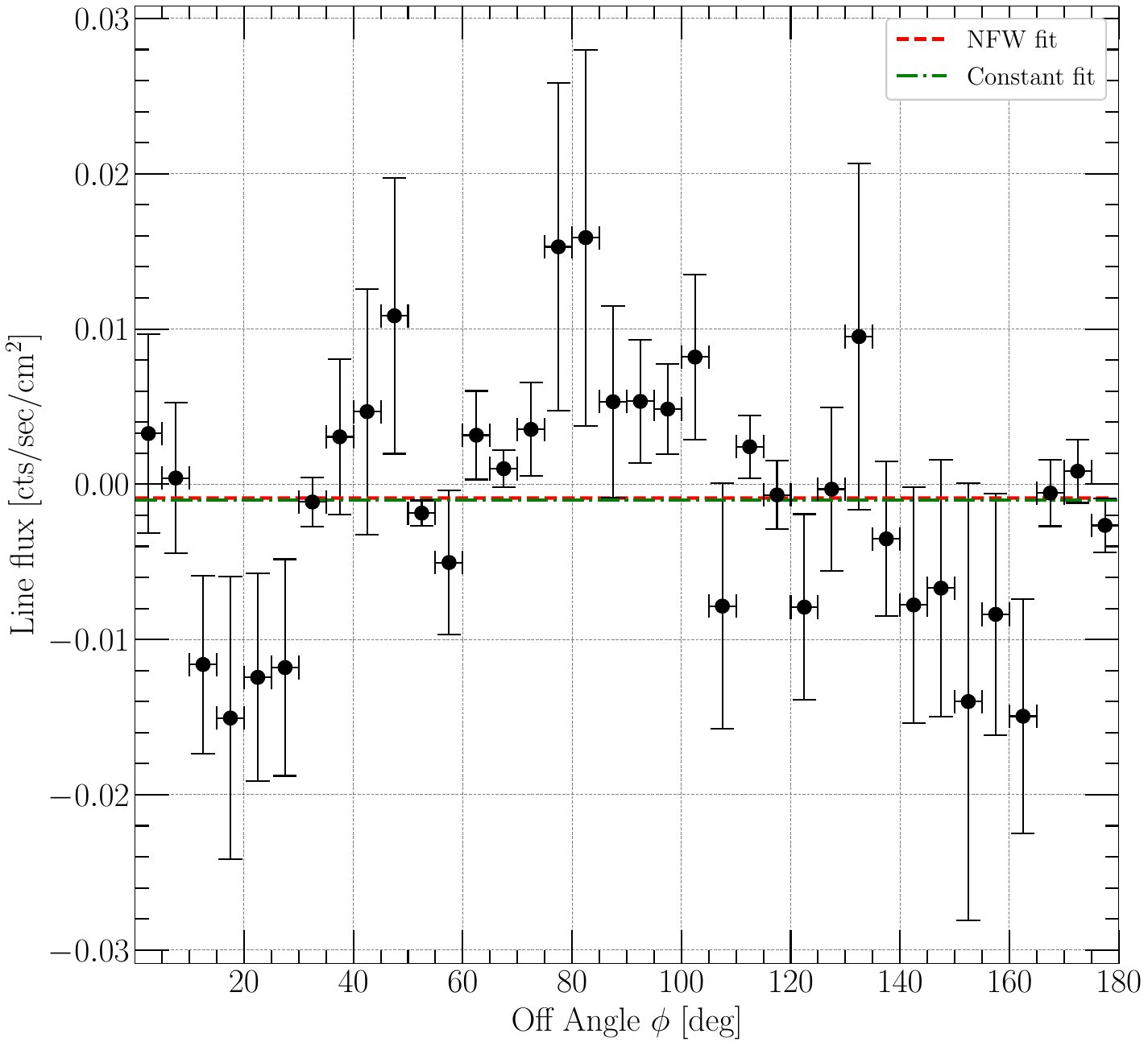}} &
\subfloat[$60.6\,\text{keV}$]{\includegraphics[width = 1.7in]{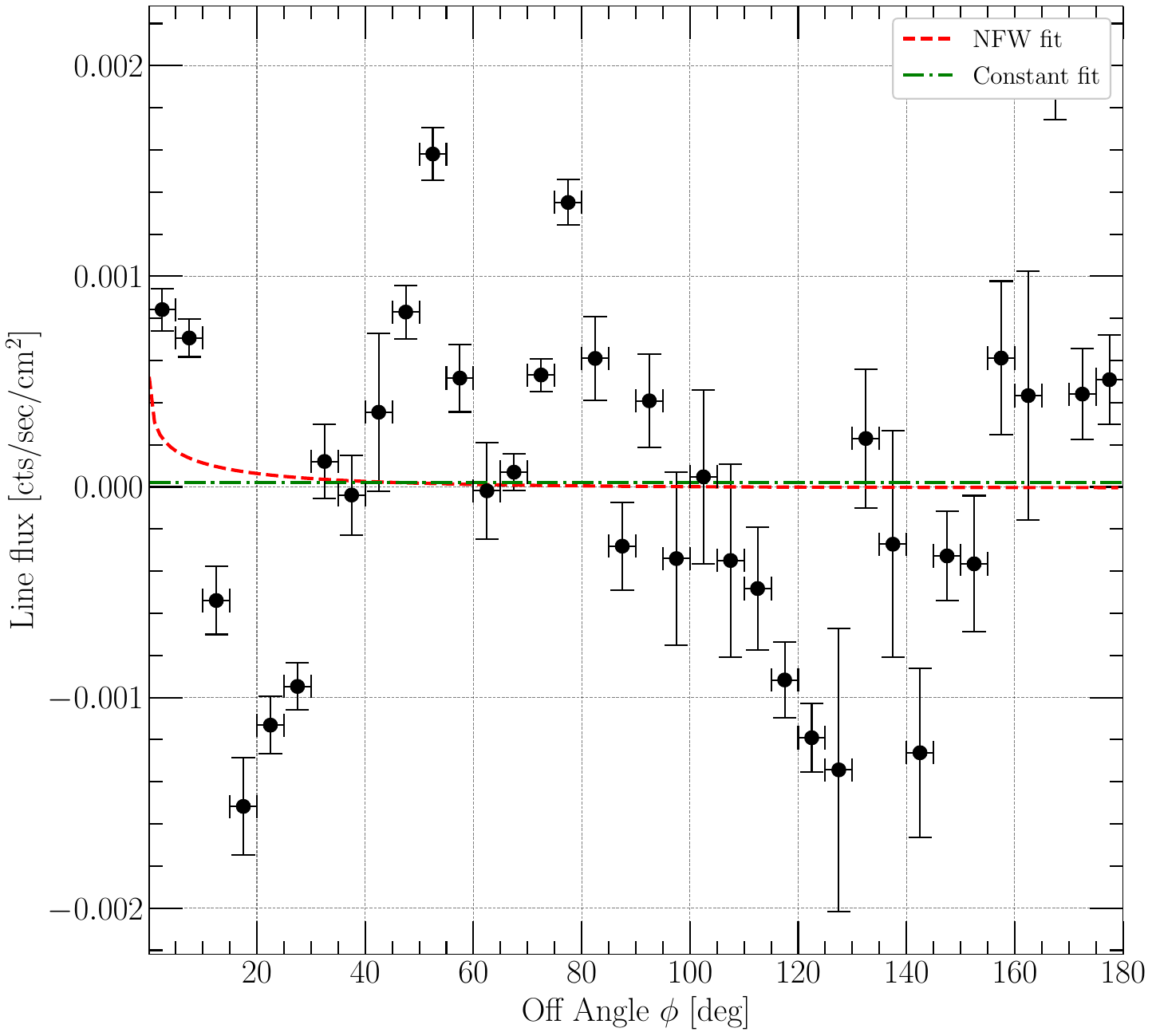}} \\
\subfloat[$65.98\,\text{keV}$]{\includegraphics[width = 1.7in]{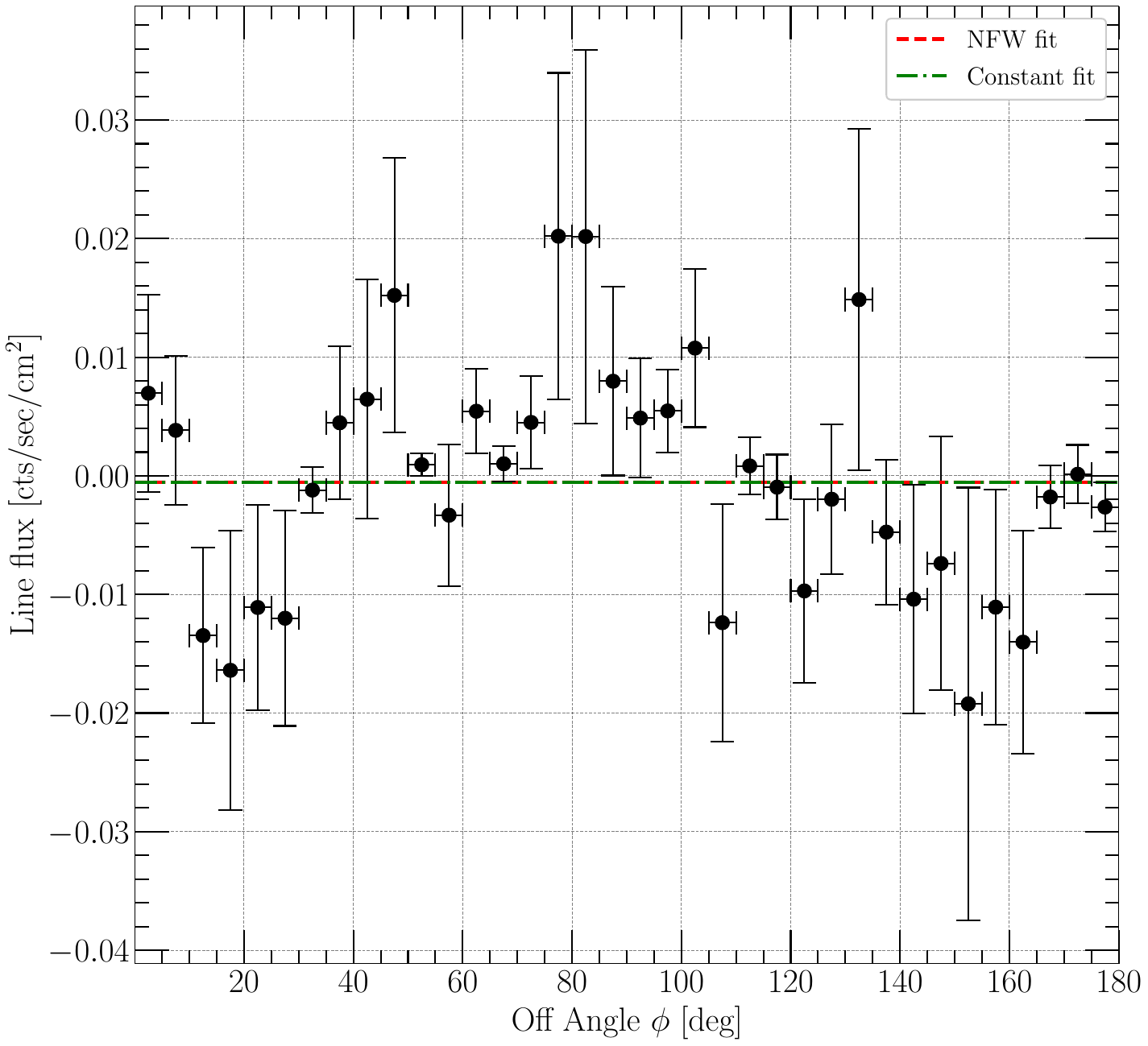}} &
\subfloat[$86.57\,\text{keV}$]{\includegraphics[width = 1.7in]{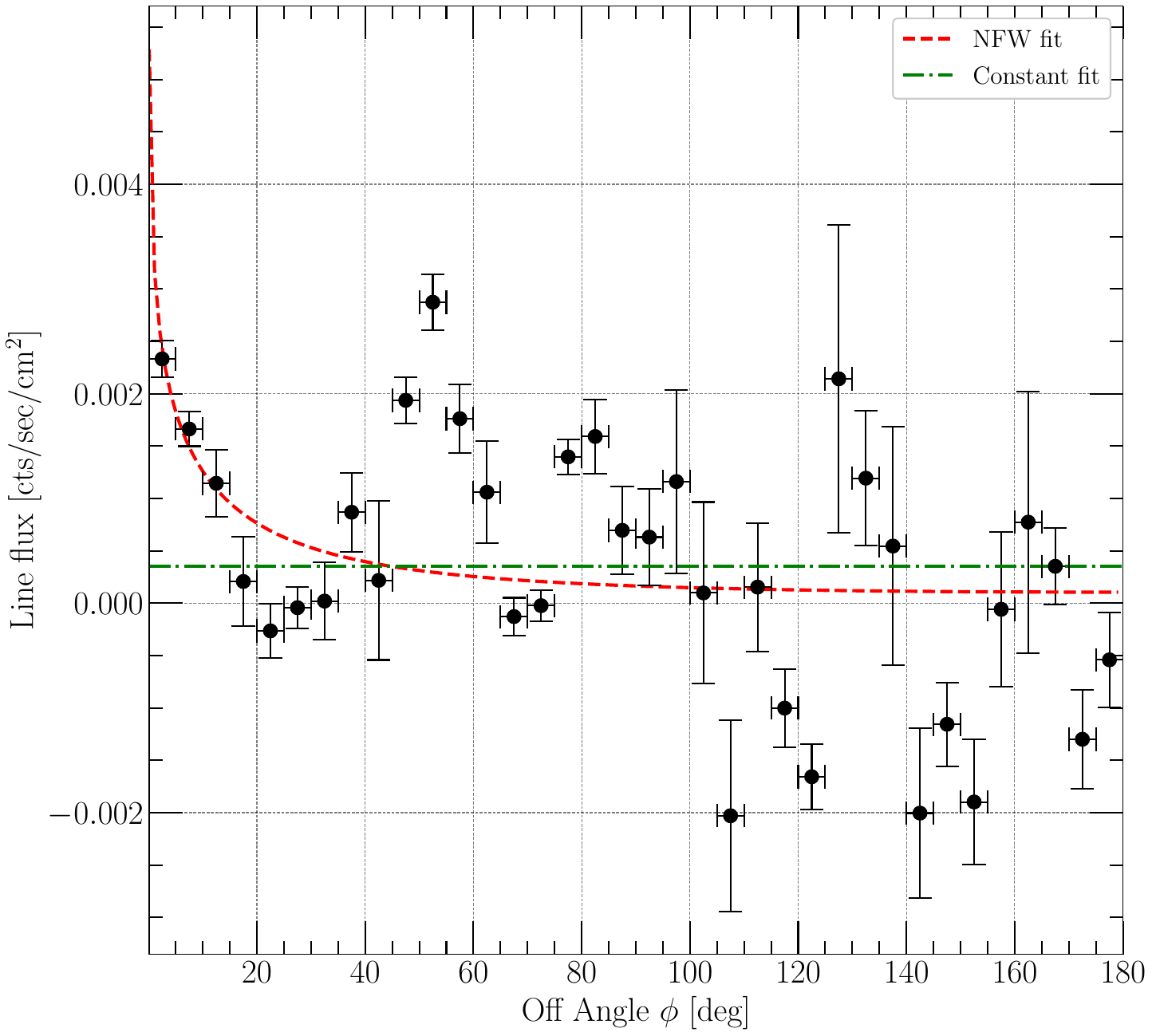}} &
\subfloat[$92.57\,\text{keV}$]{\includegraphics[width = 1.7in]{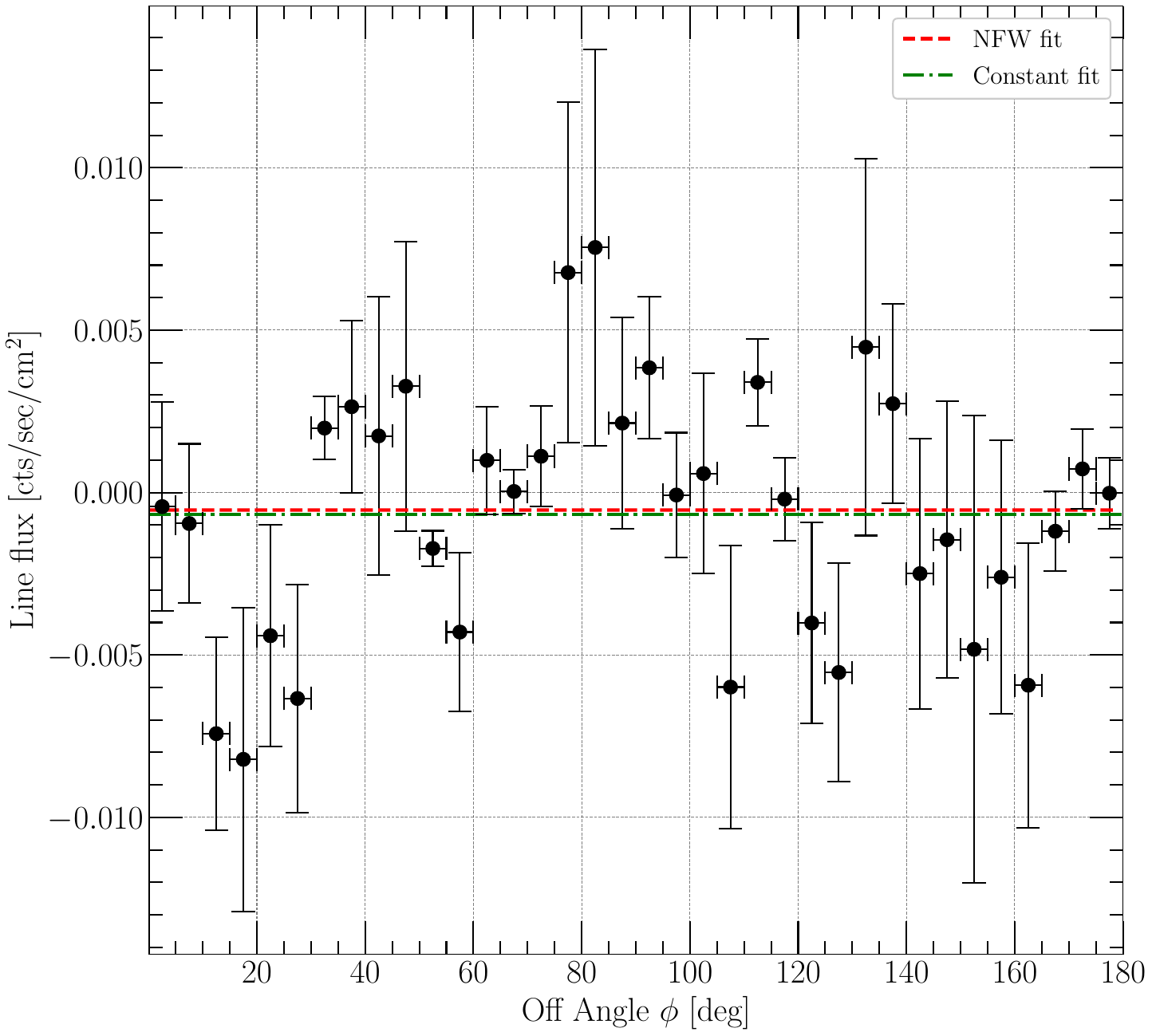}} &
\subfloat[$101.88\,\text{keV}$]{\includegraphics[width = 1.7in]{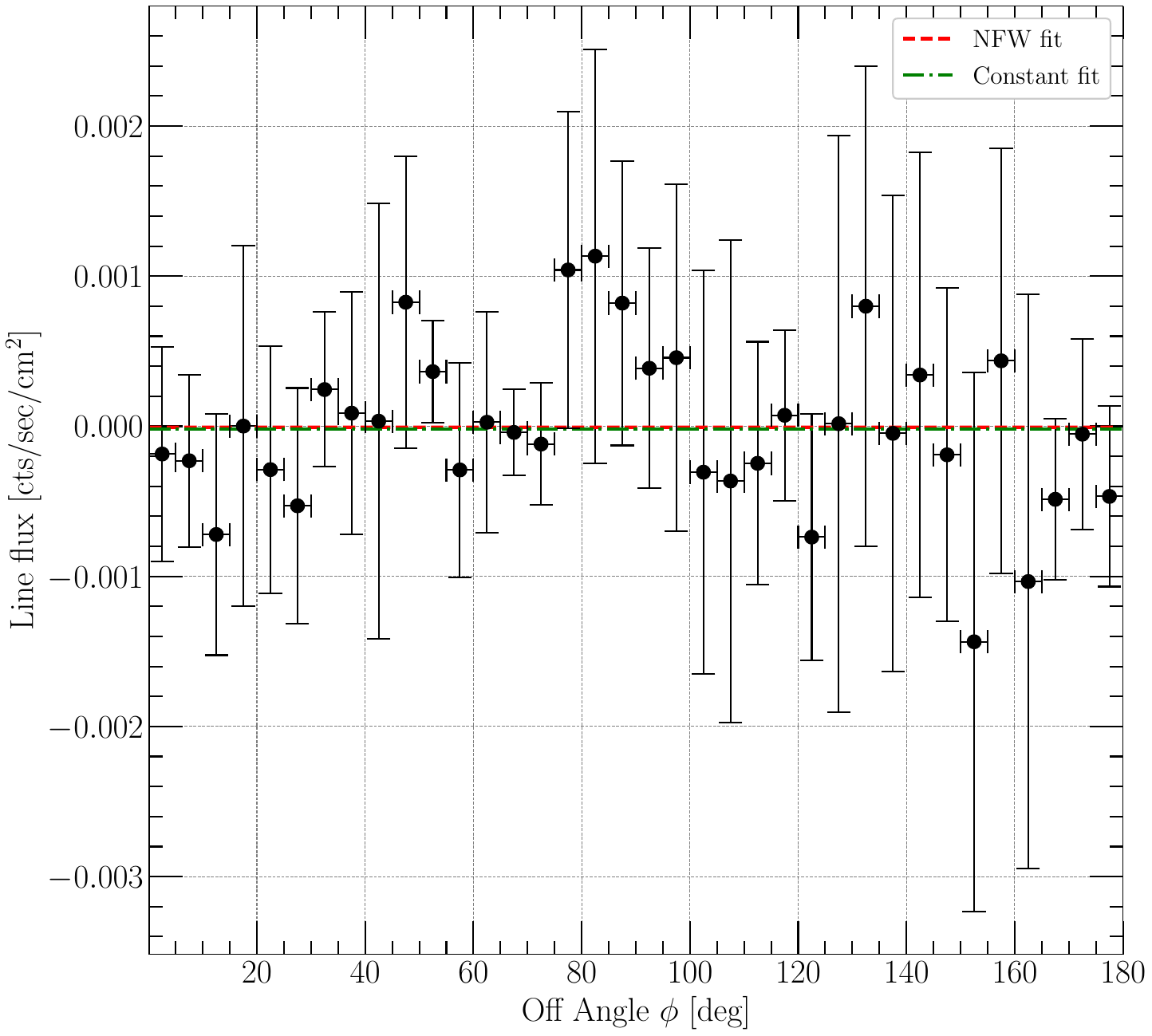}} \\
\subfloat[$138\,\text{keV}$]{\includegraphics[width = 1.7in]{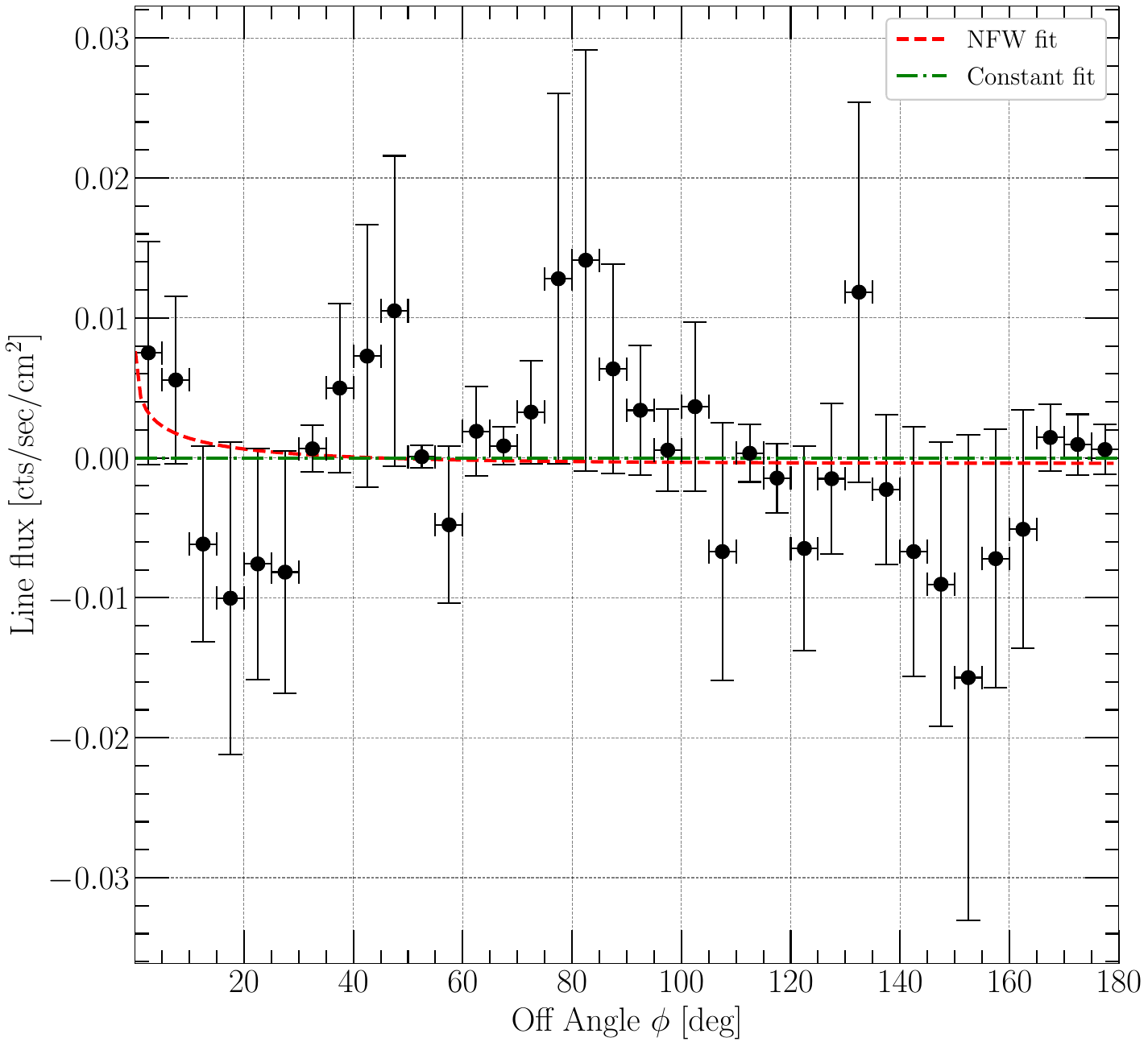}} &
\subfloat[$175.07\,\text{keV}$]{\includegraphics[width = 1.7in]{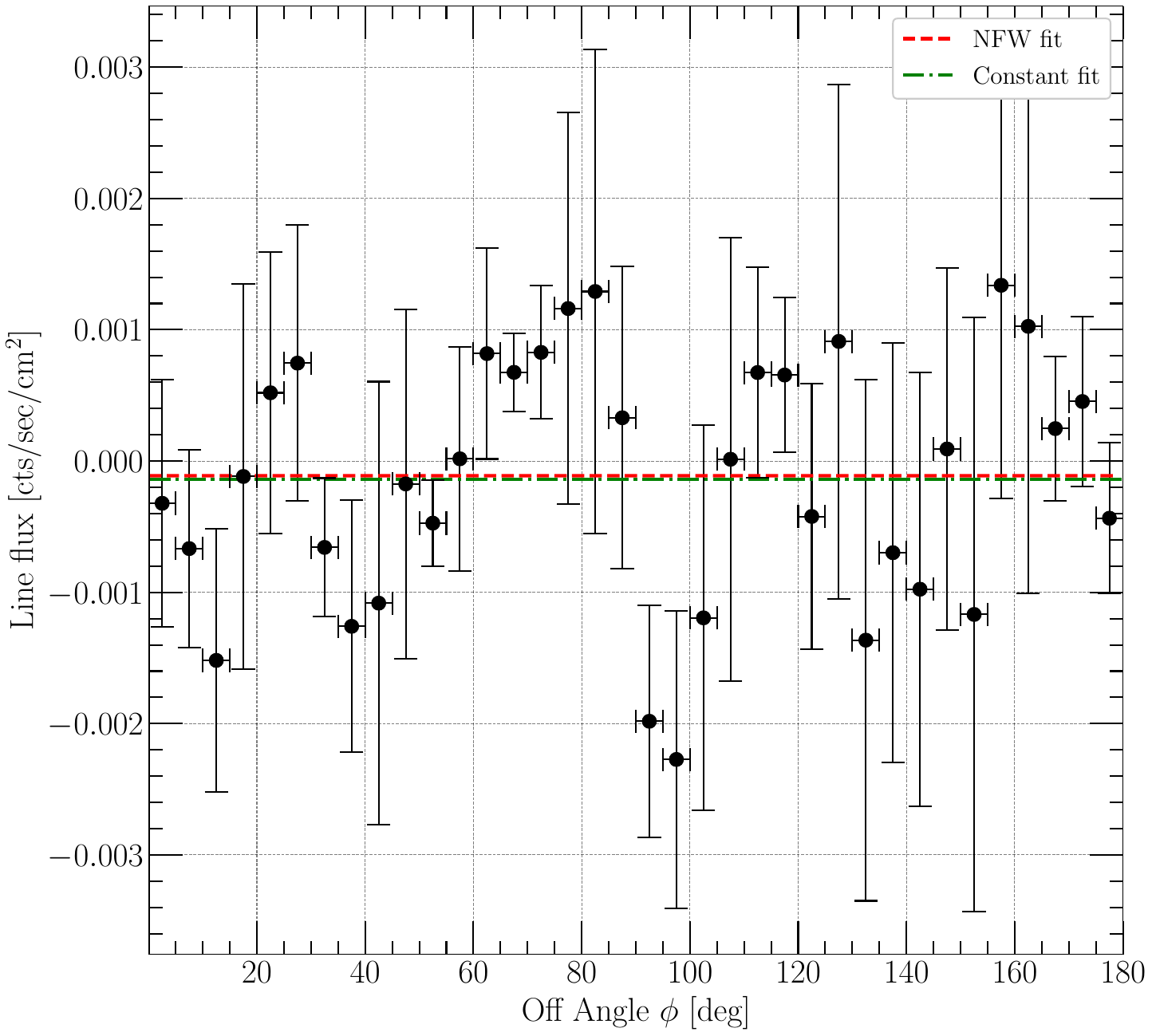}} &
\subfloat[$185.95\,\text{keV}$]{\includegraphics[width = 1.7in]{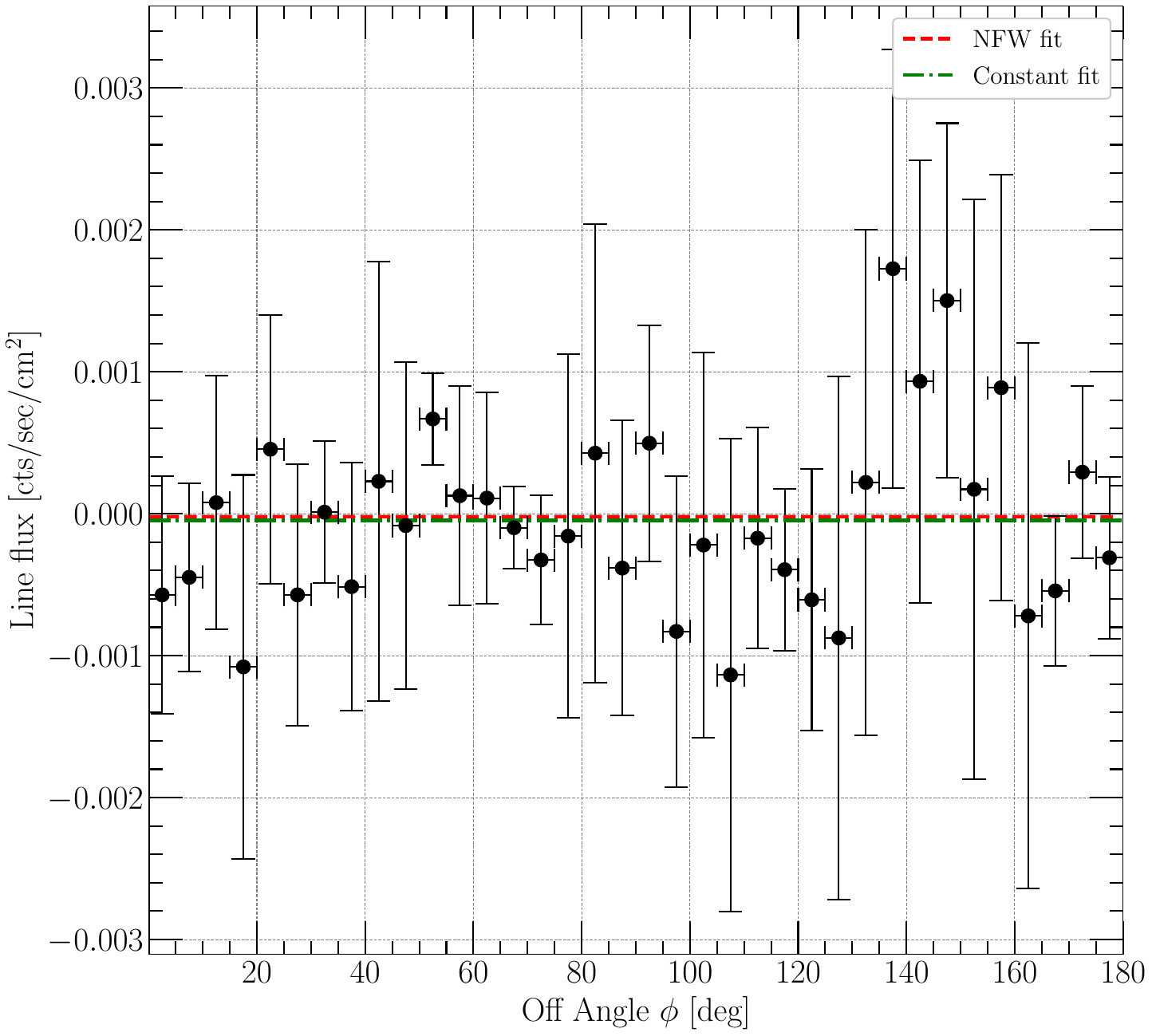}} &
\subfloat[$193.34\,\text{keV}$]{\includegraphics[width = 1.7in]{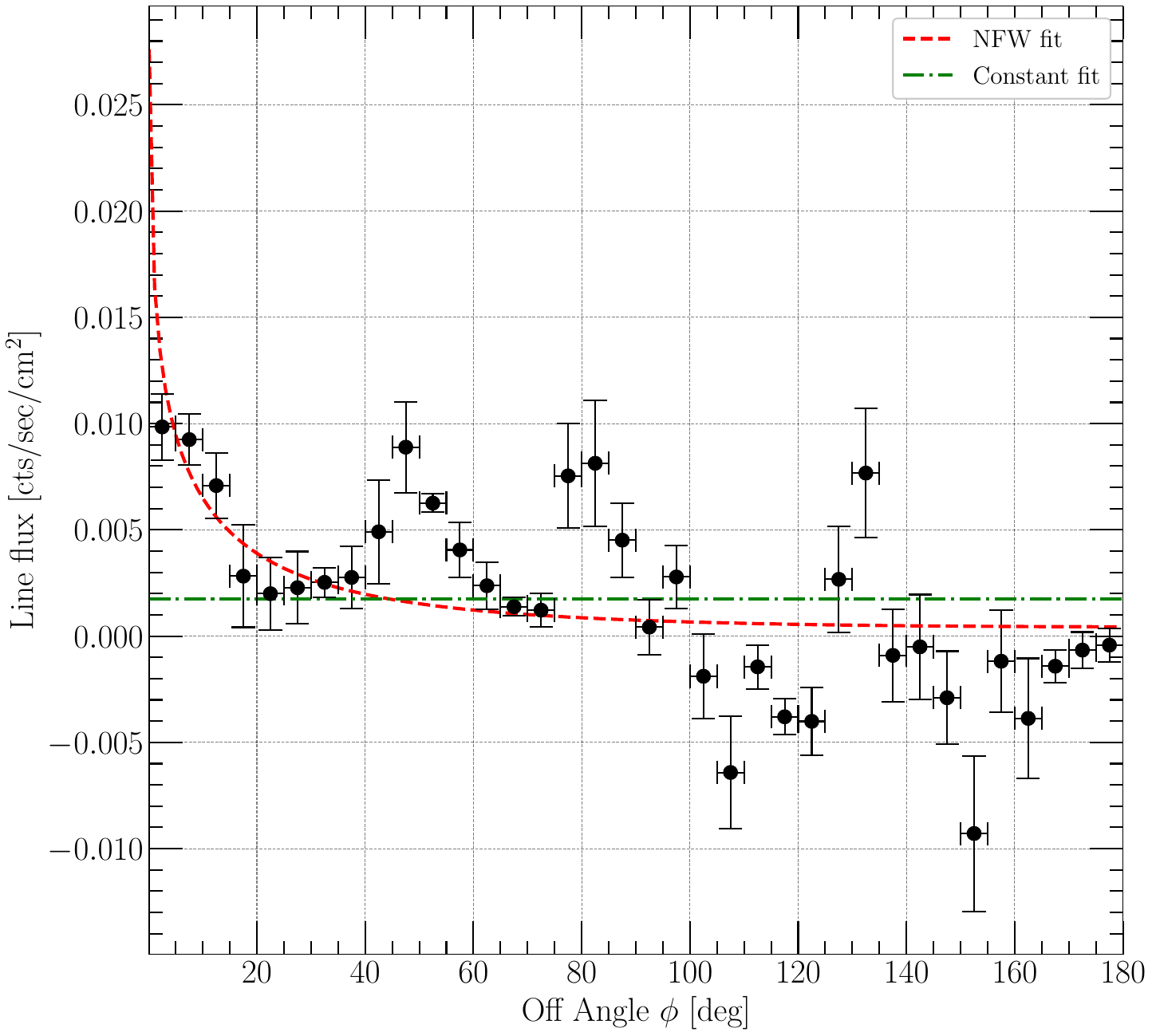}} \\
\subfloat[$197.35\,\text{keV}$]{\includegraphics[width = 1.7in]{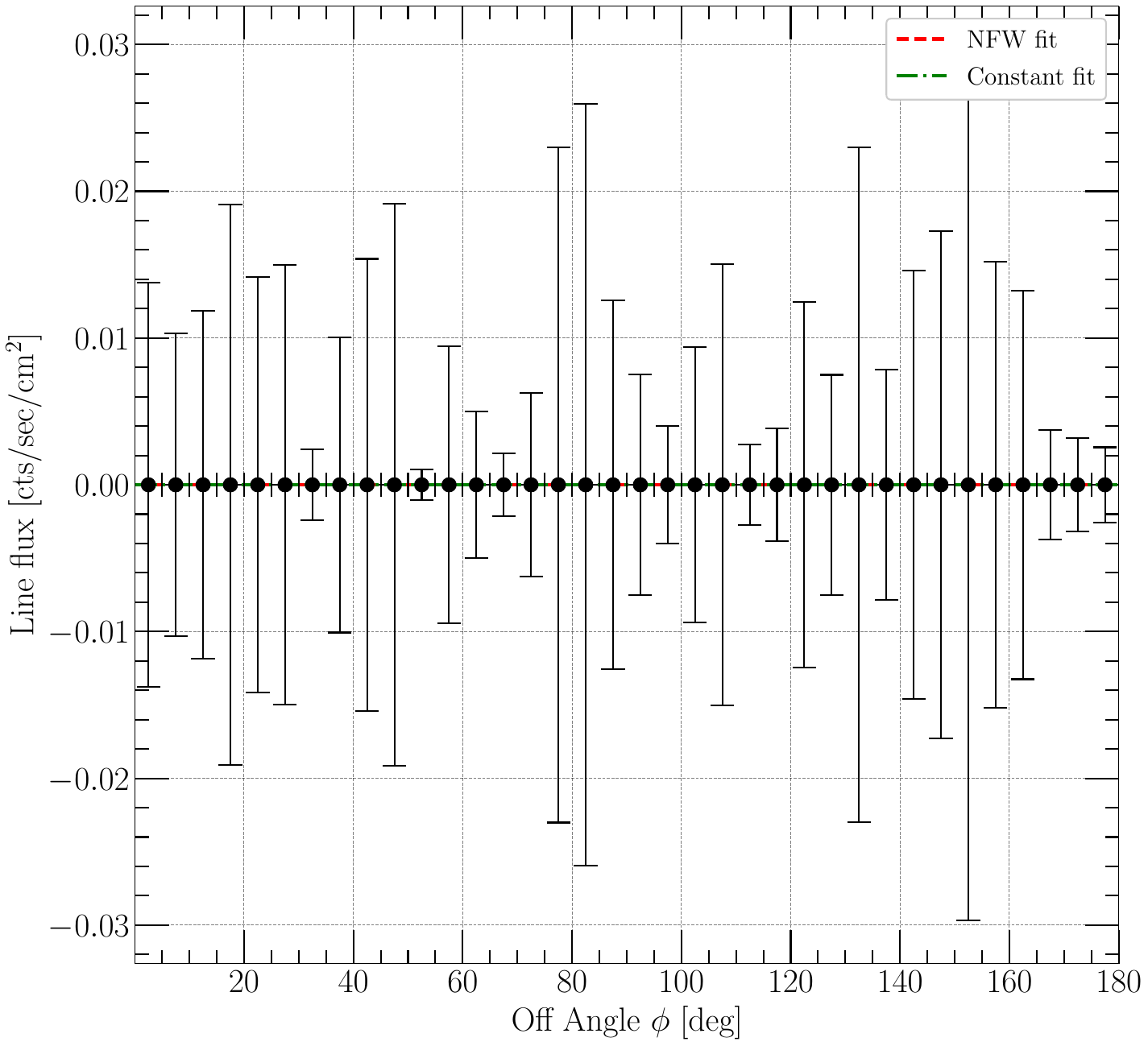}} &
\subfloat[$438.09\,\text{keV}$]{\includegraphics[width = 1.7in]{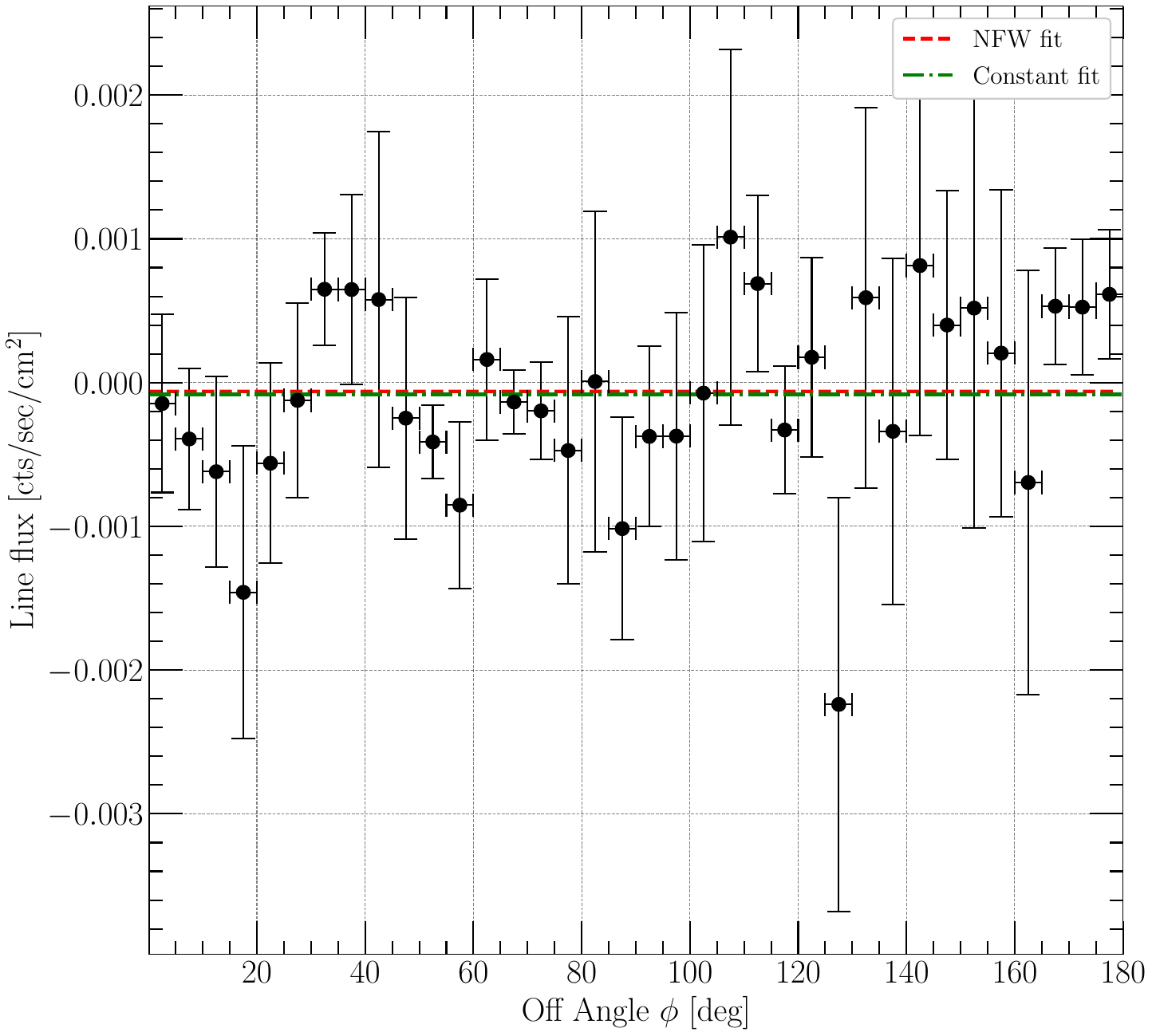}} &
\subfloat[$511\,\text{keV}$]{\includegraphics[width = 1.7in]{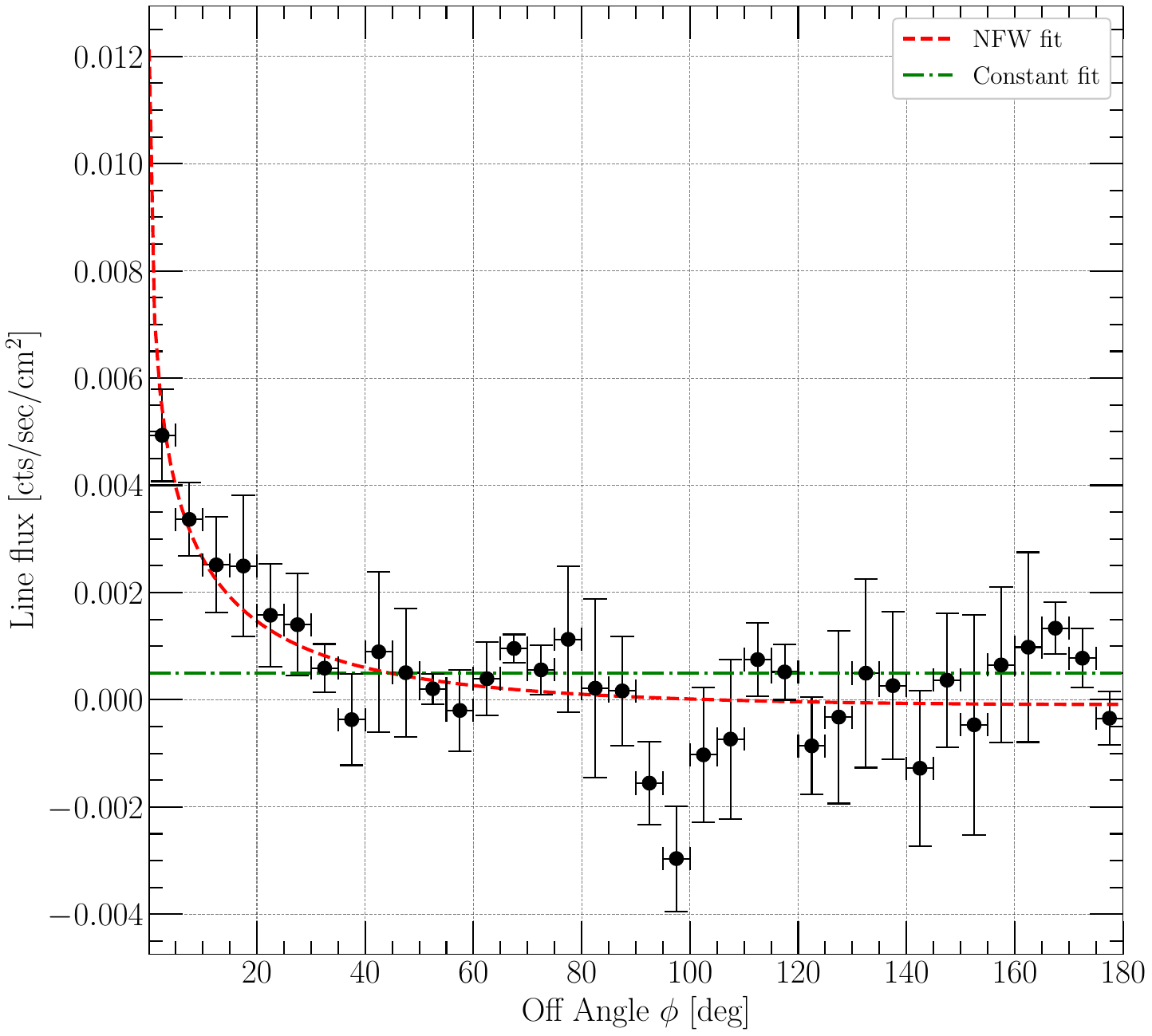}} &
\subfloat[$583.63\,\text{keV}$]{\includegraphics[width = 1.7in]{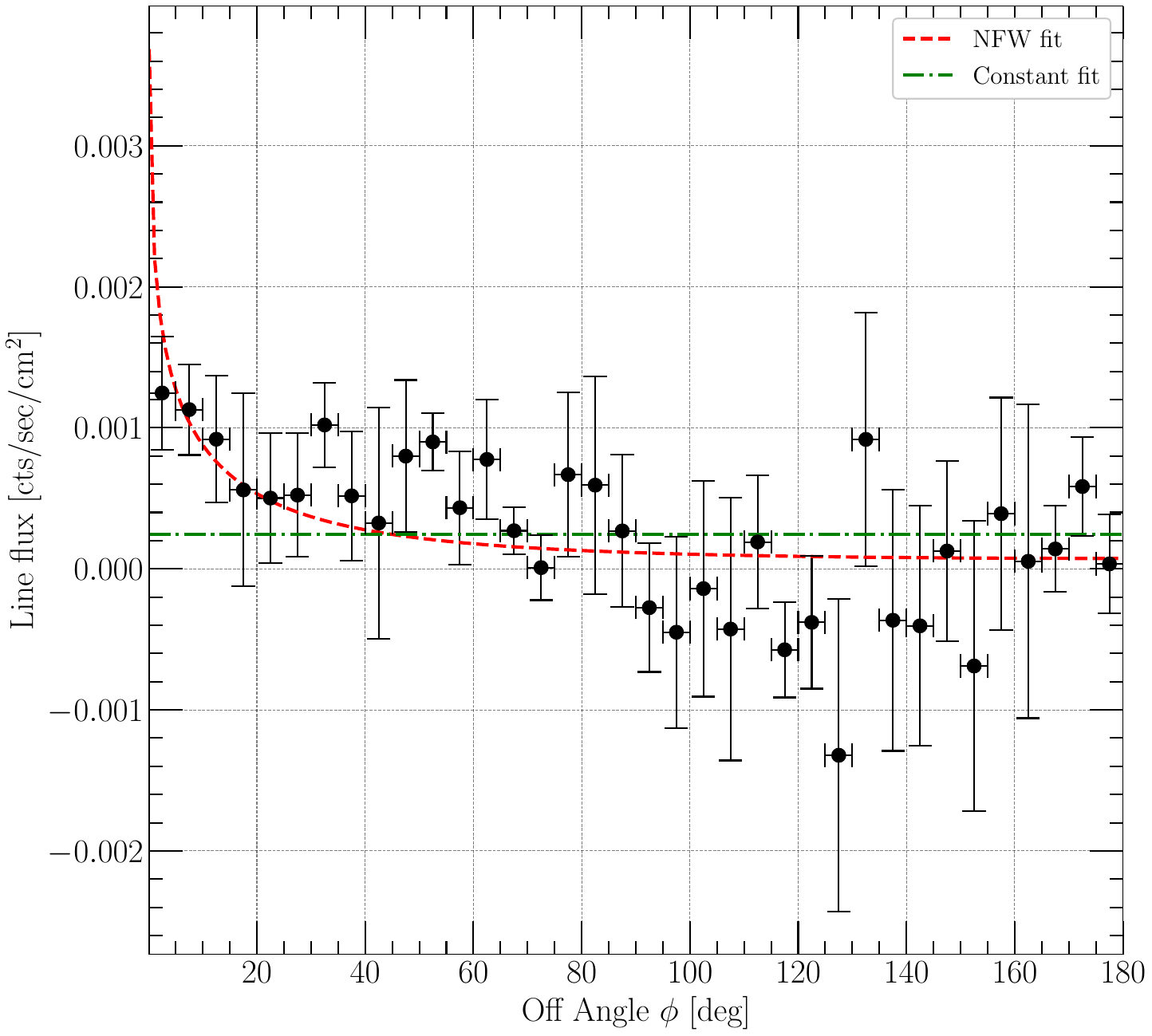}} \\
\subfloat[$699.25\,\text{keV}$]{\includegraphics[width = 1.7in]{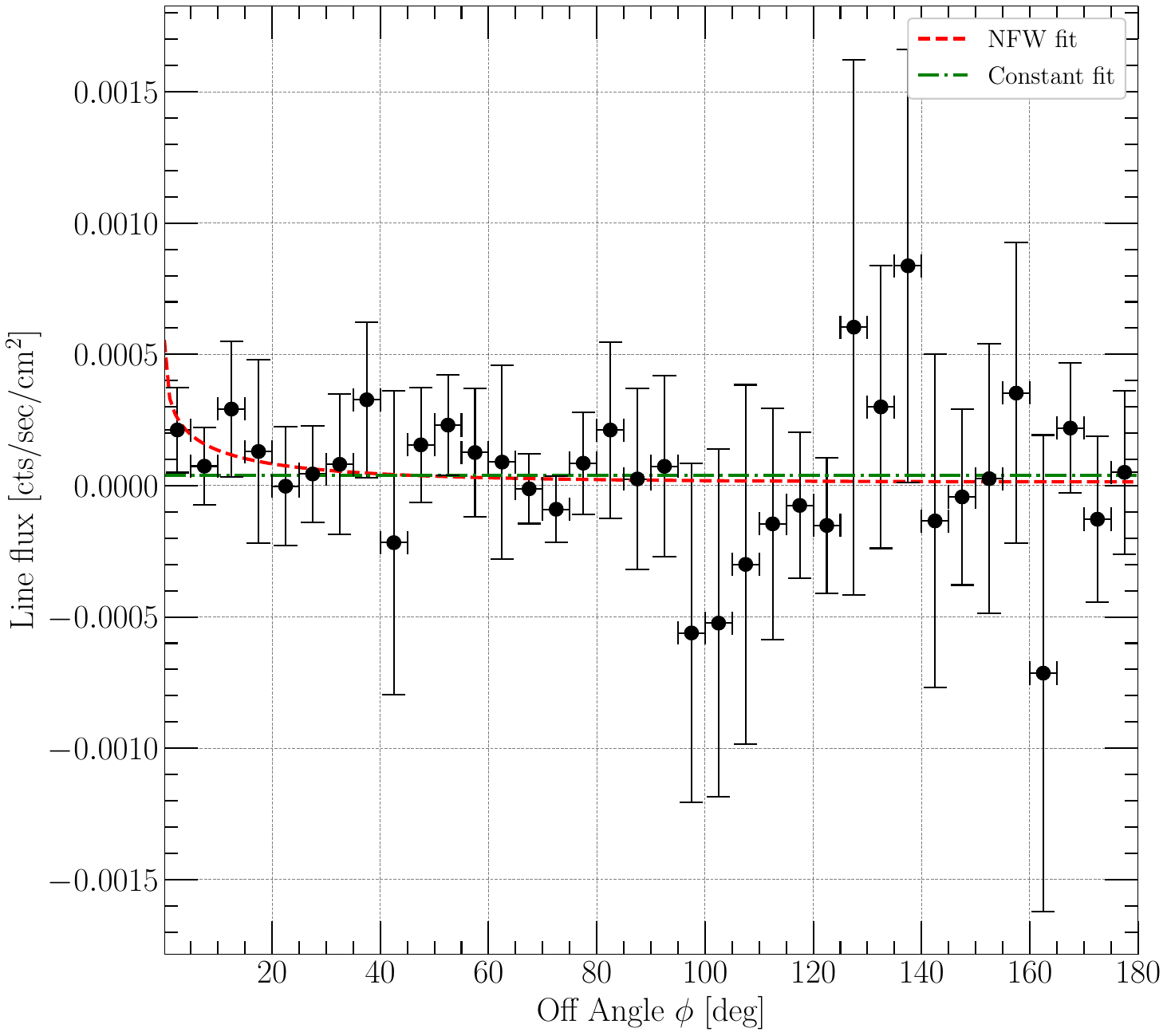}} &
\subfloat[$810.82\,\text{keV}$]{\includegraphics[width = 1.7in]{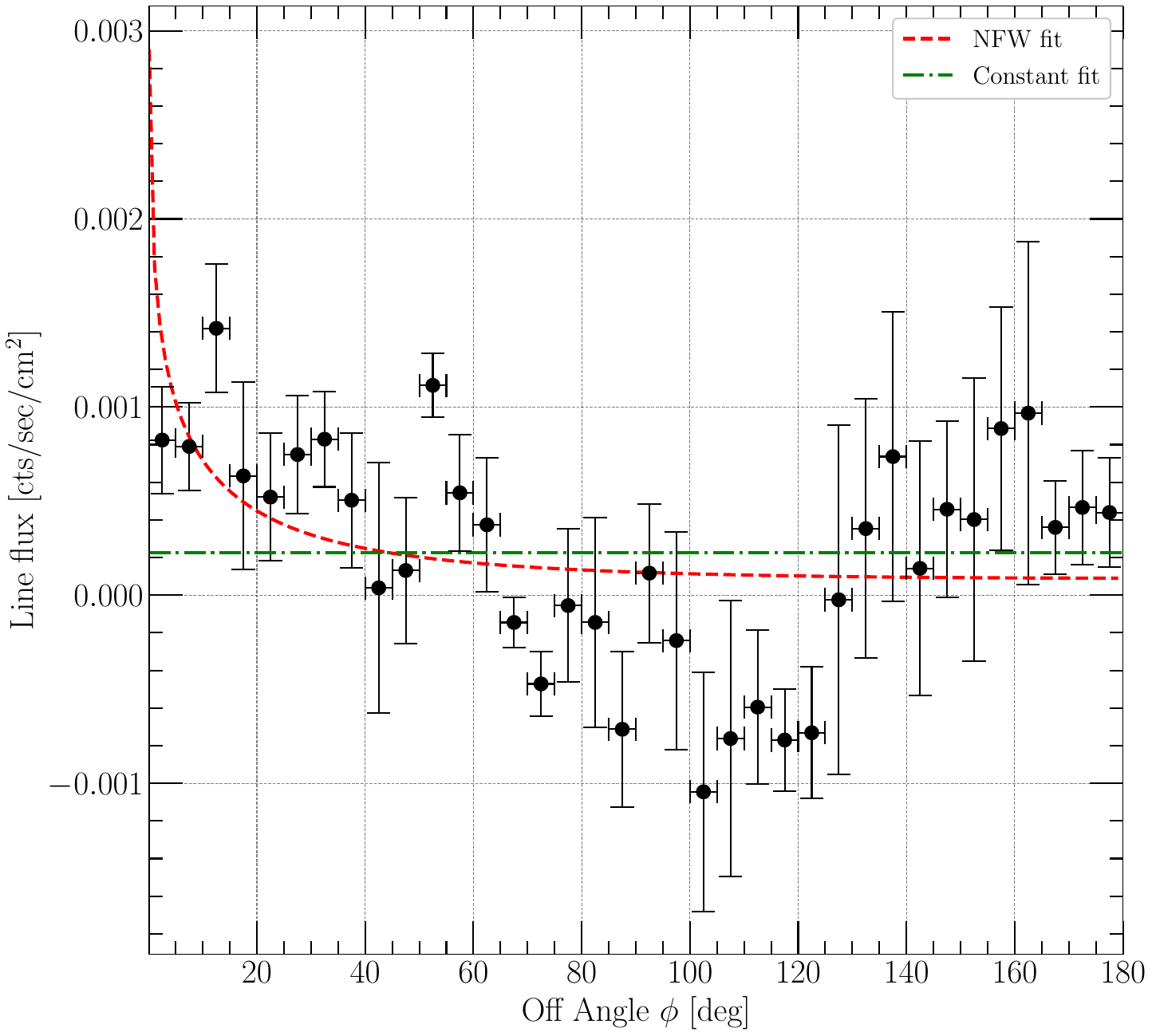}} &
\subfloat[$835.36\,\text{keV}$]{\includegraphics[width = 1.7in]{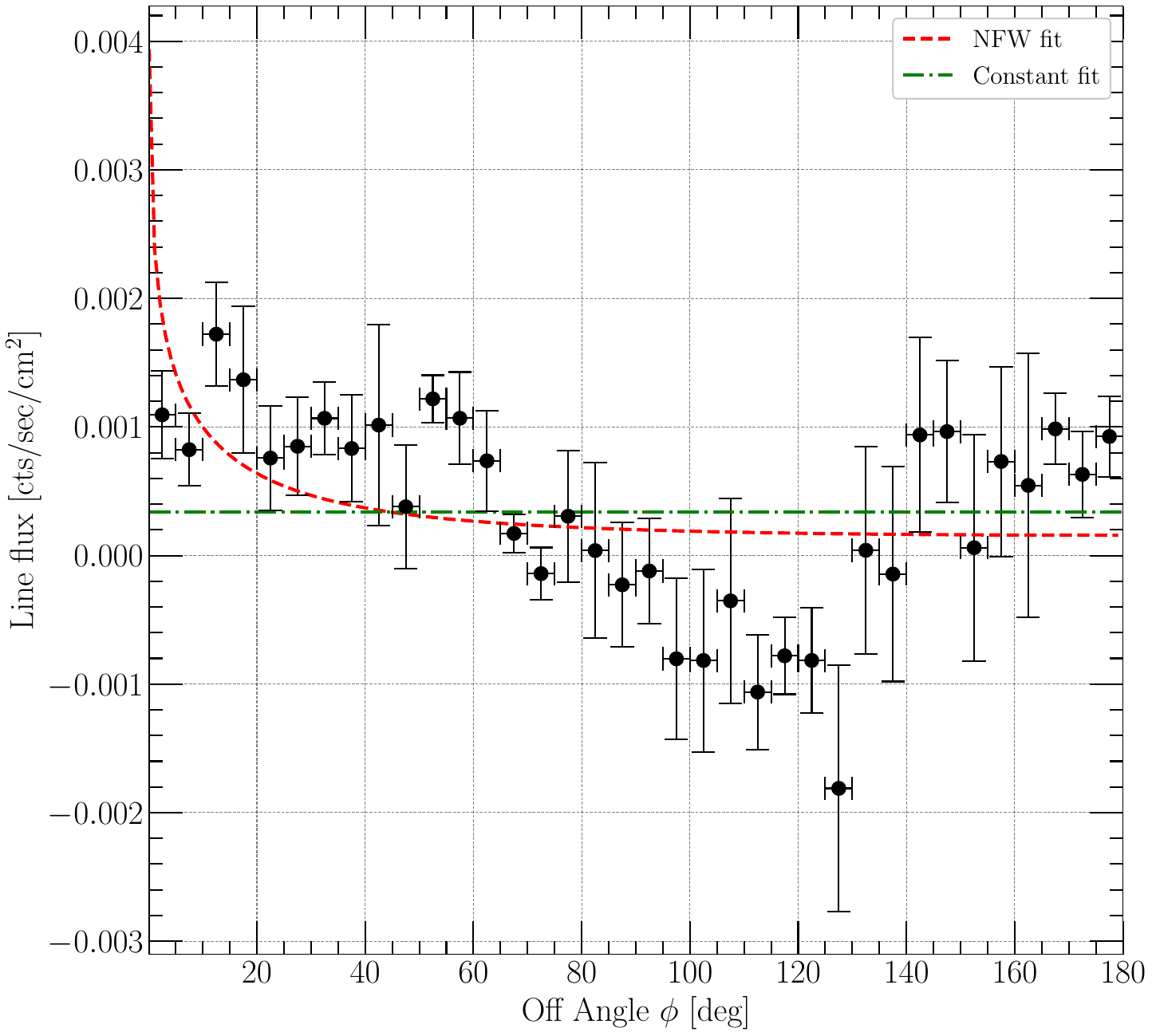}} &
\subfloat[$880.56\,\text{keV}$]{\includegraphics[width = 1.7in]{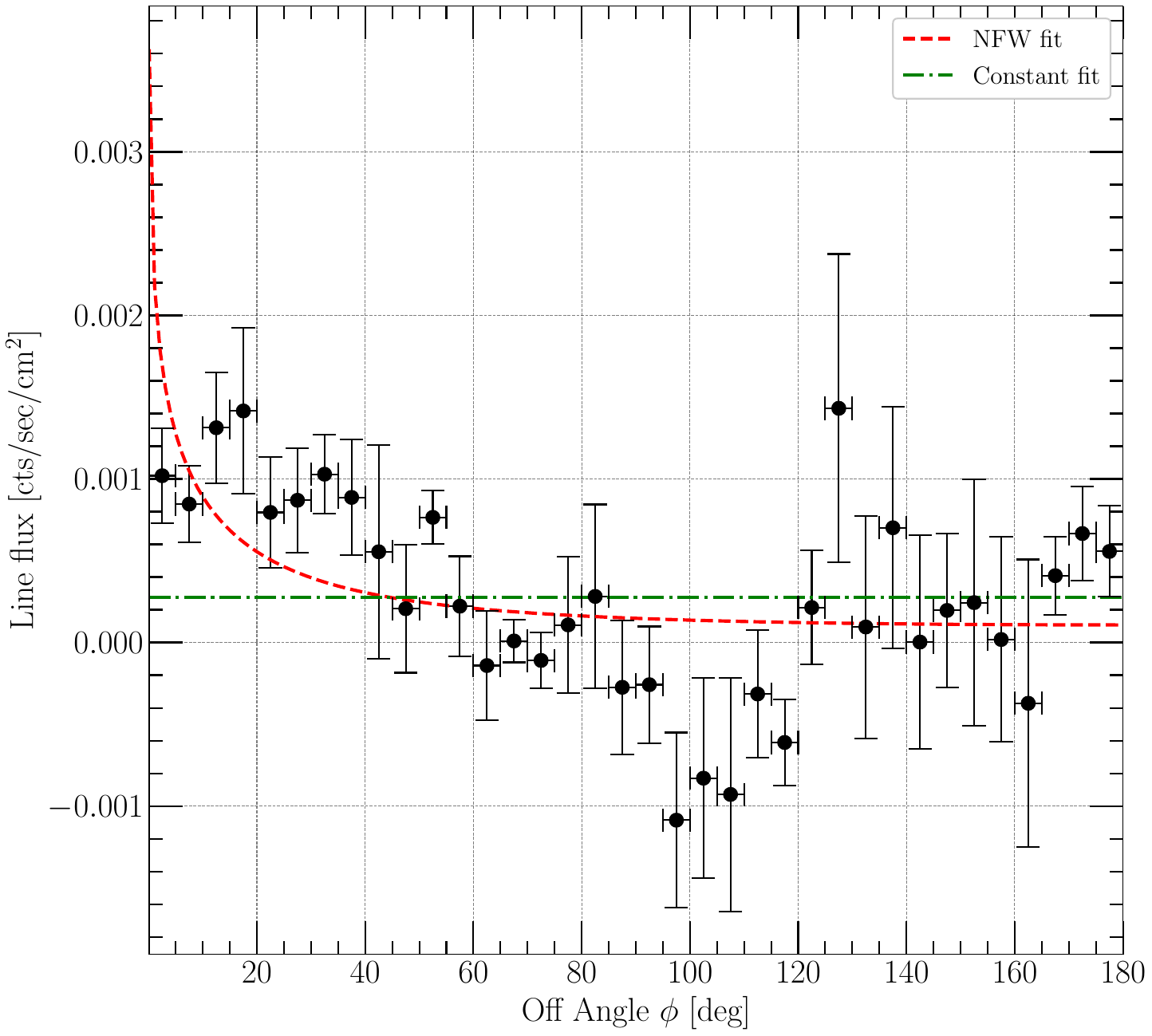}} \\

\captionsetup{labelformat=empty}

\end{tabular}

\end{figure*}
\end{center}

\def\arraystretch{2}

\end{document}